\documentclass[iop, apj, twocolappendix, numberedappendix]{emulateapj}

\usepackage{xspace}
\usepackage{amsmath}
\usepackage{framed} 
\usepackage{txfonts}
\usepackage{epstopdf} 
\usepackage{color}
\usepackage{rotating}
\usepackage{natbib}
\usepackage{ulem}
\usepackage{xspace}
\usepackage[colorlinks=true,urlcolor=black,linkcolor=black,citecolor=blue]{hyperref}

\journalinfo{The Astrophysical Journal, {\rm 789:1 (18pp), 2014 July 1}}
\submitted{Received 2014 January 6; accepted 2014 April 14; published 2014 June 9}

\newcommand{\kpch}{\>{h^{-1}{\rm kpc}}}
\newcommand{\mpch}{\>h^{-1}{\rm {Mpc}}}
\newcommand{\gpch}{\>h^{-1}{\rm {Gpc}}}

\newcommand{\msunh}{\>h^{-1}\rm M_\odot}

\newcommand{\kmsmpc}{\>{\rm km}\,{\rm s}^{-1}\,{\rm Mpc}^{-1}}

\def\LCDM{$\Lambda$CDM\ }

\def\mvir{M_{\rm vir}}
\def\rvir{R_{\rm vir}}
\def\gcm3{\mathrm{g} / \mathrm{cm}^3}

\def\gtsima{$\; \buildrel > \over \sim \;$}
\def\ltsima{$\; \buildrel < \over \sim \;$}
\def\prosima{$\; \buildrel \propto \over \sim \;$}
\def\gsim{\lower.7ex\hbox{\gtsima}}
\def\lsim{\lower.7ex\hbox{\ltsima}}
\def\simgt{\lower.7ex\hbox{\gtsima}}
\def\simlt{\lower.7ex\hbox{\ltsima}}
\def\simpr{\lower.7ex\hbox{\prosima}}

\def\rmb{{\rm b}}

\def\rmm{{\rm m}}

\def\rms{{\rm s}}

\def\rhoc{\rho_{\rm c}}
\def\rhom{\rho_{\rm m}}

\def\cvir{c_{\rm vir}}

\def\mtom{M_{\rm 200m}}
\def\rtom{R_{\rm 200m}}

\def\mtoc{M_{\rm 200c}}
\def\rtoc{R_{\rm 200c}}

\usepackage{etoolbox}
\makeatletter
\patchcmd{\NAT@citex}
  {\@citea\NAT@hyper@{\NAT@nmfmt{\NAT@nm}\NAT@date}}
  {\@citea\NAT@nmfmt{\NAT@nm}\NAT@hyper@{\NAT@date}}
  {}% Do nothing if patch works
  {}% Do nothing if patch fails
\patchcmd{\NAT@citex}
  {\@citea\NAT@hyper@{%
     \NAT@nmfmt{\NAT@nm}%
     \hyper@natlinkbreak{\NAT@aysep\NAT@spacechar}{\@citeb\@extra@b@citeb}%
     \NAT@date}}
  {\@citea\NAT@nmfmt{\NAT@nm}%
   \NAT@aysep\NAT@spacechar%
   \NAT@hyper@{\NAT@date}}
  {}% Do nothing if patch works
  {}% Do nothing if patch fails
\patchcmd{\NAT@citex}
  {\@citea\NAT@hyper@{%
     \NAT@nmfmt{\NAT@nm}%
     \hyper@natlinkbreak{\NAT@spacechar\NAT@@open\if*#1*\else#1\NAT@spacechar\fi}%
       {\@citeb\@extra@b@citeb}%
     \NAT@date}}
  {\@citea\NAT@nmfmt{\NAT@nm}%
   \NAT@spacechar\NAT@@open\if*#1*\else#1\NAT@spacechar\fi%
   \NAT@hyper@{\NAT@date}}
  {}% Do nothing if patch works
  {}% Do nothing if patch fails
\makeatother

\shorttitle{Outer density profiles of halos}
\shortauthors{Diemer \& Kravtsov}
\begin{document}

%%%%%%%%%%%%%%%%%%%%%%%%%%%%%%%%%%%%%%%%%%%%%%%%%%%%%%%%%%%%%%%%%%%%%%%%%%
% EPS OR PDF FIGURES
%%%%%%%%%%%%%%%%%%%%%%%%%%%%%%%%%%%%%%%%%%%%%%%%%%%%%%%%%%%%%%%%%%%%%%%%%%

% PDF FIGURES
\def\figdir{.}
\def\figext{pdf}

% EPS FIGURES
%\def\figdir{FiguresEps}
%\def\figdir{.}
%\def\figext{eps}

\def\panelsize{0.68}
\def\panelsizethree{0.6}

%%%%%%%%%%%%%%%%%%%%%%%%%%%%%%%%%%%%%%%%%%%%%%%%%%%%%%%%%%%%%%%%%%%%%%%%%%
% TITLE ETC
%%%%%%%%%%%%%%%%%%%%%%%%%%%%%%%%%%%%%%%%%%%%%%%%%%%%%%%%%%%%%%%%%%%%%%%%%%

\title{Dependence of the outer density profiles of halos on their mass accretion rate}

\author{Benedikt Diemer \altaffilmark{1,2} and Andrey V. Kravtsov\altaffilmark{1,2,3}}

\affil{
$^1$ Department of Astronomy and Astrophysics, The University of Chicago, Chicago, IL 60637, USA; bdiemer@oddjob.uchicago.edu \\
$^2$ Kavli Institute for Cosmological Physics, The University of Chicago, Chicago, IL 60637, USA \\
$^3$ Enrico Fermi Institute, The University of Chicago, Chicago, IL 60637, USA \\
}

%%%%%%%%%%%%%%%%%%%%%%%%%%%%%%%%%%%%%%%%%%%%%%%%%%%%%%%%%%%%%%%%%%%%%%%%%%
% ABSTRACT
%%%%%%%%%%%%%%%%%%%%%%%%%%%%%%%%%%%%%%%%%%%%%%%%%%%%%%%%%%%%%%%%%%%%%%%%%%

\begin{abstract}
We present a systematic study of the density profiles of $\Lambda$CDM halos, focusing on the outer regions, $0.1<r/\rvir<9$. We show that the median and mean profiles of halo samples of a given peak height exhibit significant deviations from the universal analytic profiles discussed previously in the literature, such as the Navarro-Frenk-White and Einasto profiles, at radii $r\gtrsim0.5\rtom$. In particular, at these radii the logarithmic slope of the median density profiles of massive or rapidly accreting halos steepens more sharply than predicted. The steepest slope of the profiles occurs at $r\approx\rtom$, and its absolute value increases with increasing peak height or mass accretion rate, reaching slopes of $-4$ and steeper. Importantly, we find that the outermost density profiles at $r\gtrsim\rtom$ are remarkably self-similar when radii are rescaled by $\rtom$. This self-similarity indicates that radii defined with respect to the mean density are preferred for describing the structure and evolution of the outer profiles. However, the inner density profiles are most self-similar when radii are rescaled by $\rtoc$. We propose a new fitting formula that describes the median and mean profiles of halo samples selected by their peak height or mass accretion rate with accuracy $\lesssim10\%$ at all radii, redshifts and masses we studied, $r\lesssim9\rvir$, $0<z<6$ and $\mvir>1.7\times10^{10}\msunh$. We discuss observational signatures of the profile features described above, and show that the steepening of the outer profile should be detectable in future weak-lensing analyses of massive clusters. Such observations could be used to estimate the mass accretion rate of cluster halos. 
\end{abstract}

\keywords{cosmology: theory - dark matter - methods: numerical}

%%%%%%%%%%%%%%%%%%%%%%%%%%%%%%%%%%%%%%%%%%%%%%%%%%%%%%%%%%%%%%%%%%%%%%%%%%
% INTRODUCTION
%%%%%%%%%%%%%%%%%%%%%%%%%%%%%%%%%%%%%%%%%%%%%%%%%%%%%%%%%%%%%%%%%%%%%%%%%%

\section{Introduction}
\label{sec:intro}

Theoretical predictions for the structure of dark matter halos forming in the cold dark matter (CDM) scenario play an important role in the interpretation of observations. During the past several decades, a significant effort has been made to understand one of the most basic descriptions of this structure: the spherically averaged, radial density profiles resulting from the gravitational collapse of perturbations in an expanding universe. \citet{gunn_72_sphericalcollapse} made an early prediction for the density profile of collapsed halos based on the spherical top hat model. Subsequently, \citet{fillmore_84} showed that the spherically symmetric radial collapse of a perturbation with an initial density profile $\delta_{\rm i} \propto r^{-\gamma}$ results in a power-law density profile, $\rho \propto r^{-g}$, where $g = 2$ for $\gamma < 2$ and $g = 3 \gamma / (1+\gamma)$ for $\gamma \geq 2$. Thus, for example, secondary collapse onto a preexisting point perturbation ($\delta_{\rm i} \propto r^{-3}$) results in a $\rho \propto r^{-9/4}$ profile \citep[cf. also,][]{gott_75,bertschinger_85}. 

The collapse of peaks in the initial Gaussian density perturbation field is generally expected to be triaxial and significantly more complicated than envisioned in the spherical collapse model \citep[e.g.,][]{doroshkevich_70, bond_96_peakpatch, bond_96_filaments}, a picture confirmed by cosmological simulations \citep[e.g.,][]{klypin_83, miller_83, davis_85_clustering}.  Early simulations of halos showed that their profiles were roughly consistent with isothermal profiles, $\rho \propto r^{-2}$, required to explain the flat rotation curves of galaxies \citep{frenk_88_haloformation}. Higher resolution simulations, however, showed that in general profiles of halos forming in the hierarchical structure scenario are not well described by a single power law. Thus, \citet{dubinski_91} modeled the collapse of individual halos in the CDM model and showed that the \citet{hernquist_90} profile, in which the slope changes from $-1$ at small radii to $-4$ at large radii, provides a good description of the collapsed halos in their simulations. \citet[][hereafter NFW, see also \citealt{cole_96_halostructure}]{navarro_95, navarro_96, navarro_97_nfw} proposed a similar form of the density profile with an inner asymptotic slope of $-1$ and an outer slope of $-3$. These authors did not focus on the structure of the outer density profile, however, and the outer slope was shown to exhibit significant halo-to-halo scatter \citep{avilareese_99}. Subsequent studies have confirmed that the profiles of halos resulting from the cold collapse of a wide variety of initial conditions are described by profiles that gradually steepen with increasing radius \citep[e.g.,][for a recent theoretical explanation of this behavior see \citealt{lithwick_11}]{huss_99}. However, they showed that the profiles are  more accurately described by the \citet{einasto_65, einasto_69} functional form \citep{navarro_04, graham_06, merritt_06, gao_08, stadel_09, navarro_10, ludlow_11}. 

The main focus of most of the studies of halo density profiles has been on the innermost regions \citep[e.g.,][]{moore_99, navarro_04, navarro_10, stadel_09}, which are critical for understanding the observed distribution of mass within the visible regions of galaxies. The outer regions, however, are increasingly being probed by X-ray and Sunyaev--Zel'dovich effect observations of clusters of galaxies \citep[e.g.,][]{reiprich_13} and weak-lensing analyses \citep[e.g.,][]{mandelbaum_06_group_profiles, umetsu_11}. It is important to understand theoretical expectations for the outer density profiles in order to interpret such observations properly. For example, \citet{becker_11_bias} showed that typical cluster-sized halos exhibit deviations from the NFW form, and that NFW profile fits to shear profiles extended to large radii can result in sizeable systematic bias in weak-lensing mass measurements \citep[see also][]{oguri_11}.  Although a number of recent studies have considered the overall shape of the density profiles at large radii \citep{prada_06_outerregions, betancortrijo_06, tavio_08, cuesta_08_infall, oguri_11} and proposed analytic profiles to describe them, it is not yet clear whether the shape is universal for halos in different stages of their evolution.

In this paper, we present a systematic study of the outer density profiles of halos, focusing specifically on the dependence of the profiles on the evolutionary stage of halos and their mass accretion rate. We report significant deviations from previously proposed fitting formulae at radii $r \gtrsim 0.5 \rtom$. Specifically, we show that halos that rapidly accrete mass exhibit a sharp steepening of their profile slope at $r \gtrsim 0.5 \rtom$, with the maximum absolute value of the slope increasing with increasing mass accretion rate. We propose a new fitting formula that accounts for this behavior and report best-fit parameters for the outer profiles as a function of halo peak height and mass accretion rate. 

The paper is organized as follows. In Section \ref{sec:numerical} we describe the numerical simulations used as well as the selection criteria and relevant definitions of mass, radius, and other quantities. In Section \ref{sec:results} we present our main results, while in Section \ref{sec:discussion} we discuss their interpretation and implications for observational analyses. Finally, we summarize our main results and conclusions in Section \ref{sec:conclusion}.

%%%%%%%%%%%%%%%%%%%%%%%%%%%%%%%%%%%%%%%%%%%%%%%%%%%%%%%%%%%%%%%%%%%%%%%%%%
% SIMULATIONS 
%%%%%%%%%%%%%%%%%%%%%%%%%%%%%%%%%%%%%%%%%%%%%%%%%%%%%%%%%%%%%%%%%%%%%%%%%%

\section{Numerical Simulations and Methods}
\label{sec:numerical}

In this section we describe the cosmological simulations used in our study, halo identification and construction of the halo density profiles, and the relevant mass and radius definitions.

\subsection{Cosmological $N$-body Simulations}
\label{sec:numerical:sims}

\begin{deluxetable}{lccccc}
\tablecaption{N-body Simulations
\label{table:sims}}
\tablewidth{0pt}
\tablehead{
\colhead{Box} &
\colhead{$L (\mpch)$} &
\colhead{$N^3$} &
\colhead{$m_{\rm p} (\msunh)$} &
\colhead{$\epsilon (\kpch)$} &
\colhead{$\epsilon / (L / N)$}
}
\startdata
L1000 & $1000$ & $1024^3$ & $7.0 \times 10^{10}$ & $33.0$ & $1/30$ \\
L0500 & $500$  & $1024^3$ & $8.7 \times 10^{9}$  & $14.0$ & $1/35$  \\
L0250 & $250$  & $1024^3$ & $1.1 \times 10^{9}$  & $5.8$  & $1/42$  \\
L0125 & $125$  & $1024^3$ & $1.4 \times 10^{8}$  & $2.4$  & $1/51$  \\
L0063 & $62.5$ & $1024^3$ & $1.7 \times 10^{7}$  & $1.0$  & $1/60$ 
\enddata
\tablecomments{Numerical parameters of the five N-body simulations used in this paper. $L$ denotes the box size, $N^3$ the number of particles, $m_{\rm p}$ the particle mass, and $\epsilon$ the force softening. All simulations were started at an initial redshift of $49$, and run with a GADGET2 timestep parameter of $\eta = 0.025$.}
\end{deluxetable}

To investigate halos across a wide range of masses and redshifts, we use a suite of dissipationless \LCDM  simulations of different box sizes (Table \ref{table:sims}). The largest simulation, L1000, was introduced in \citet{diemer_13_scalingrel}. All simulations use the same cosmological parameters, initial redshift, and number of particles. We adopt the cosmological parameters of the Bolshoi simulation \citep{klypin_11_bolshoi}: a flat $\Lambda$CDM model with $\Omega_\rmm = 1 - \Omega_\Lambda = 0.27$, $\Omega_\rmb = 0.0469$, $h = H_0 / (100\kmsmpc) = 0.7$, $\sigma_8 = 0.82$ and $n_\rms = 0.95$. These parameters are compatible with constraints from a combination of WMAP5, baryon acoustic oscillations and Type Ia supernovae \citep{komatsu_etal11, jarosik_11_wmap7}, X-Ray cluster abundance evolution \citep{vikhlinin_09_clusters}, and observations of the clustering of galaxies and galaxy--galaxy/cluster weak lensing \citep[see, e.g.,][]{tinker_12_galcosmo, cacciato_13}. The same cosmology was used for all calculations in this paper, such as peak height. The initial conditions for the simulations were generated using a second-order Lagrangian perturbation theory code \citep[2LPTic;][]{crocce_06_2lptic}. The simulations were started at redshift $z=49$, which has been shown to be sufficiently high to avoid transient effects \citep{crocce_06_2lptic}. The simulations were run using the publicly available code Gadget2 \citep{springel_05_gadget2}. Each run followed $1024^3$ dark matter particles, corresponding to particle masses between $1.7 \times 10^{7} \msunh$ and $7.0 \times 10^{10} \msunh$ (Table \ref{table:sims}). 

Given that we focus on the outer density profile, we set the force resolution in such a way that the smallest halos that can be used for profile analysis are sufficiently resolved. Specifically, we set the force softening to a quarter of the scale radius expected for a halo with $\mvir = 1000 m_{\rm p}$, using the concentration--mass relation of \citet{zhao_09_mah}. According to this criterion, a force softening of $\epsilon \approx 1/30 \times L/N$ is appropriate for large box sizes such as $1 \gpch$, while for the smallest box $\epsilon \approx 1/60 \times L/N$. 

\subsection{Halo Samples and Resolution Limits}
\label{sec:numerical:halos}

We used the phase--space--based halo finder Rockstar \citep{behroozi_13_rockstar} to extract all isolated halos and subhalos from the 100 snapshots of each simulation. A halo is deemed to be isolated if its center does not lie inside $\rvir$ of another, larger halo, where $\rvir$ is the radius enclosing the ``virial'' overdensity implied by the spherical collapse model \citep{bryan_98_virial}. We derived merger trees from the halo catalogs using the code of \citet{behroozi_13_trees}. Whenever we refer to the progenitor of a halo, we mean the halo along its most massive progenitor branch at each redshift. We use the merger trees to identify halos with recent major mergers and to estimate the mass accretion rates using the masses of the main progenitors over a particular redshift interval.

We extracted spherically averaged density profiles of halos in 80 logarithmically spaced bins between $0.05 \rvir$ and $10 \rvir$. We are agnostic as to which of the simulations in Table \ref{table:sims} a halo profile originated from, and instead we combine all profiles in order to access a large range of masses and redshifts. As a check, we compared the density profiles to a set that was extracted from the Bolshoi simulation \citep{klypin_11_bolshoi} using a different code and found excellent agreement. Furthermore, we only consider isolated halos, as the density profiles of subhalos often contain a dominant contribution from their host halo. We do not, however, attempt to remove the contribution of subhalos to the density profiles of their host halos, because it is often ambiguous whether a particle belongs to the host or subhalo, and because such a procedure cannot be replicated in observations.

Any $N$-body simulation has limited mass and force resolution, and these limitations need to be taken into account when analyzing the structure of halos. We test for resolution effects by comparing halo samples of the same mass range from different simulation boxes (corresponding to different mass and force resolutions; see Table \ref{table:sims}). We find that the mean and median profiles of halos with $N_{\rm p} \geq 1000$ particles within $\rvir$ differ by less than $5\%$ for the entire radial range $0.1 \rvir < r < 9 \rvir$, with a typical difference of $\approx 3\%$ at most radii. The differences are random and do not exhibit any systematic trend with mass or redshift for all masses and redshifts used in our analyses. Given that the simulations were started from different initial conditions, the mean and median profiles of halos of the same mass may differ somewhat due to sample variance or Poisson fluctuations. Such random differences can therefore be expected and are sufficiently small not to affect our conclusions.

We conclude that the profiles of halos with $N_{\rm p} \geq 1000$ particles within $\rvir$ have converged to better than $5\%$ in the radial range $0.1 \rvir < r < 9 \rvir$, and we adopt $N_{\rm p} = 1000$ as the lower limit for our halo samples, corresponding to a mass limit of $\mvir \geq 1.7 \times 10^{10} \msunh$ in the smallest simulation box. The limit was relaxed to $N_{\rm p} = 200$ for the progenitors of halos that were used to estimate the mass accretion rate. The profiles of these progenitor halos were not used for any analyses, however.

\subsection{Mass and Radius Definitions}
\label{sec:numerical:definitions}

\begin{figure}
\centering
\includegraphics[trim = 4mm 8mm 4mm 4mm, clip, scale=0.72]{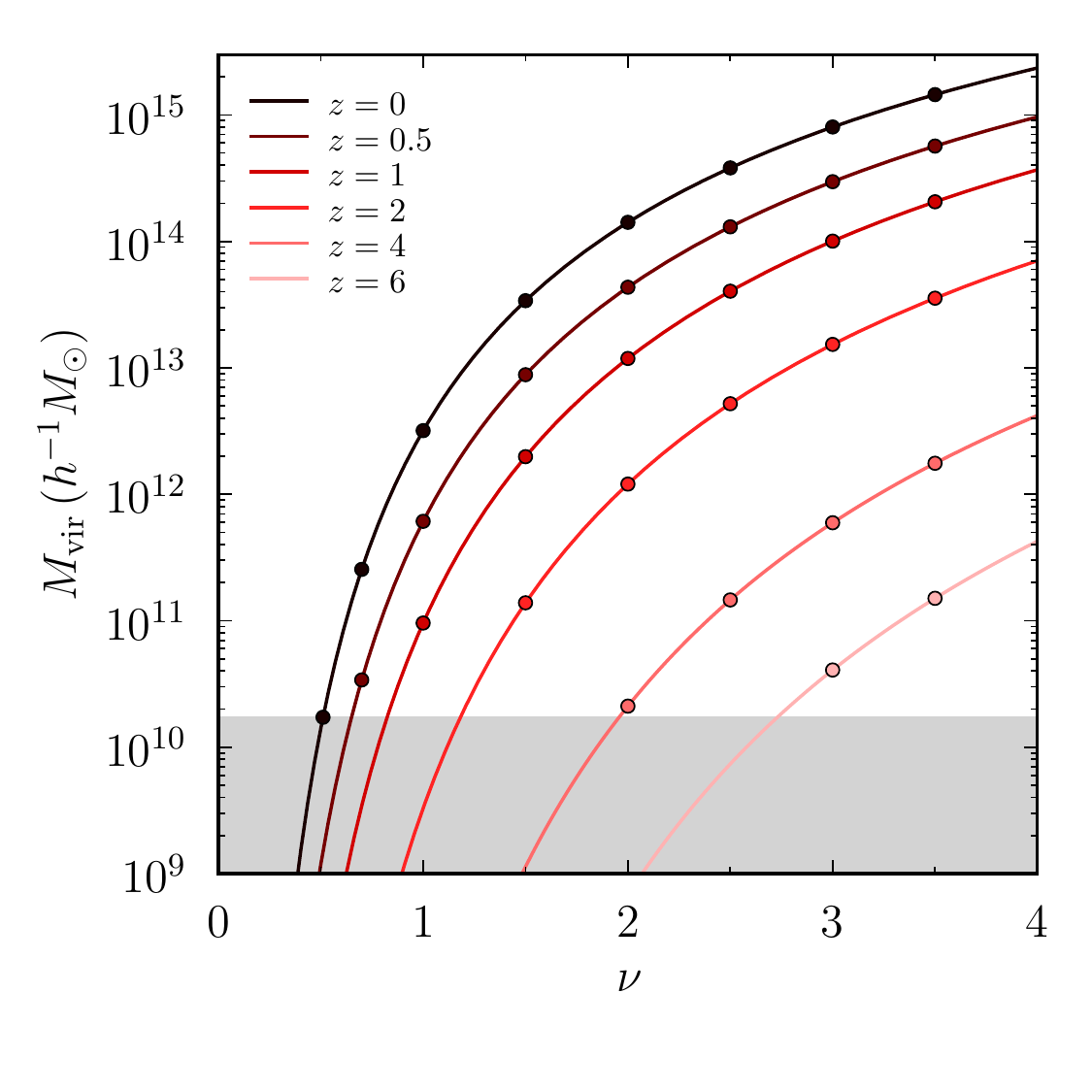}
\caption{Virial mass of halos as a function of their peak height, $\nu$, at different redshifts. The circles mark the edges of the $\nu$ bins used in our analyses. The gray shaded area at the bottom indicates the mass range beyond the resolution limit of our simulations ($1000 m_{\rm p}$ in the smallest simulation box, or $1.7 \times 10^{10} \msunh$). }
\label{fig:nu_M}
\end{figure}

Throughout the paper, we denote the three-dimensional halo-centric radius as $r$, reserving capital $R$ for specific radii used to define halo mass. We denote the mean matter density of the universe $\rho_{\rm m}$, and the critical density $\rho_{\rm c}$. Spherical overdensity mass definitions referring to $\rho_{\rm m}$ or $\rho_{\rm c}$ are understood to have a fixed overdensity $\Delta$, and are denoted $M_{\Delta \rm m} = M(<R_{\Delta \rm m})$, such as $\mtom$, or $M_{\Delta \rm c} = M(<R_{\Delta \rm c})$, such as $\mtoc$. The labels $\mvir$ and $\rvir$ are reserved for a varying overdensity $\Delta(z)$ with respect to the matter density, where $\Delta_{\rm vir}(z=0) \approx 358$ and $\Delta_{\rm vir}(z>2) \approx 180$ for the cosmology assumed in this paper \citep[e.g.,][]{bryan_98_virial}.

\begin{figure*}
\centering
\includegraphics[trim = 5mm 6mm 3mm 0mm, clip, scale=\panelsize]{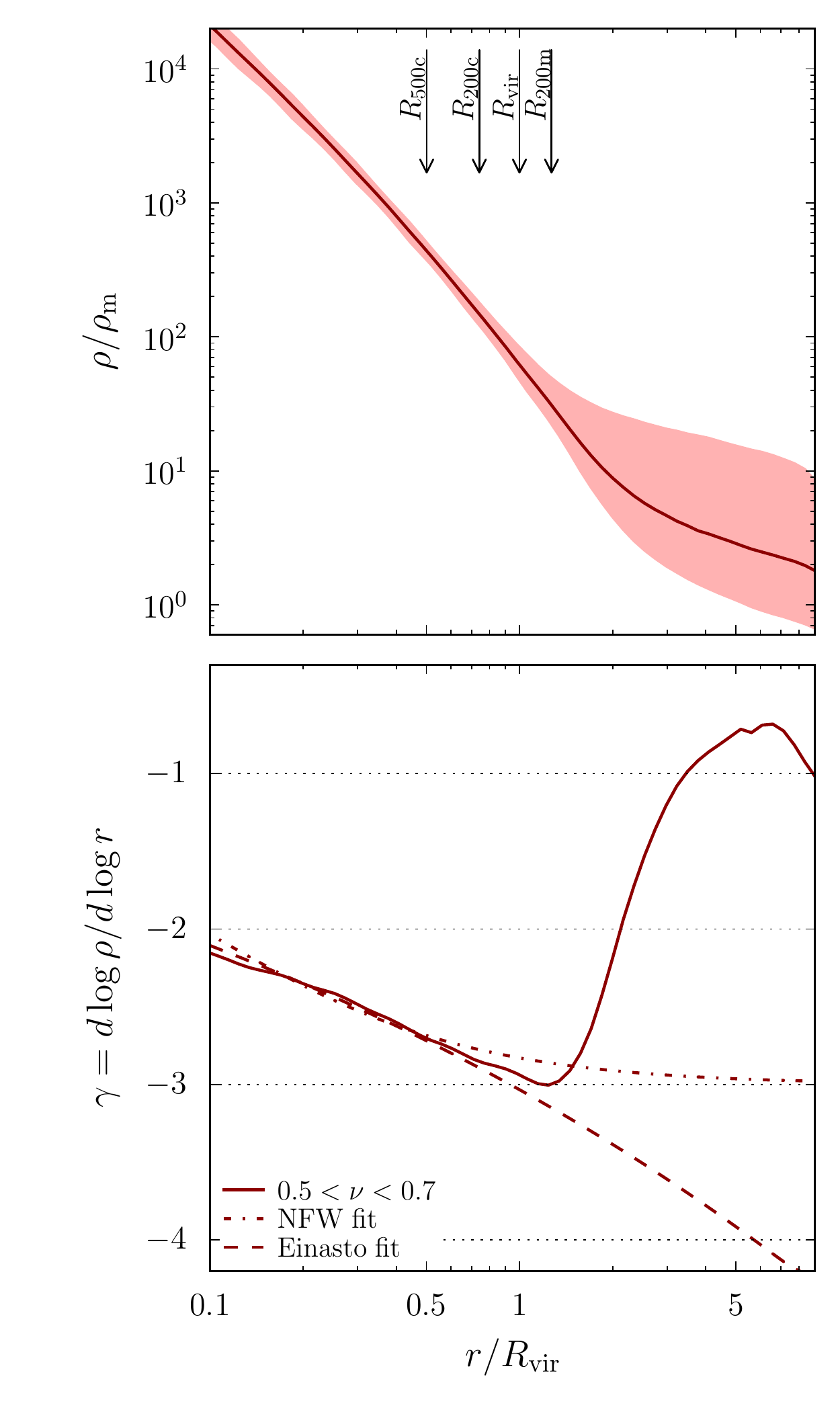}
\includegraphics[trim = 5mm 6mm 3mm 0mm, clip, scale=\panelsize]{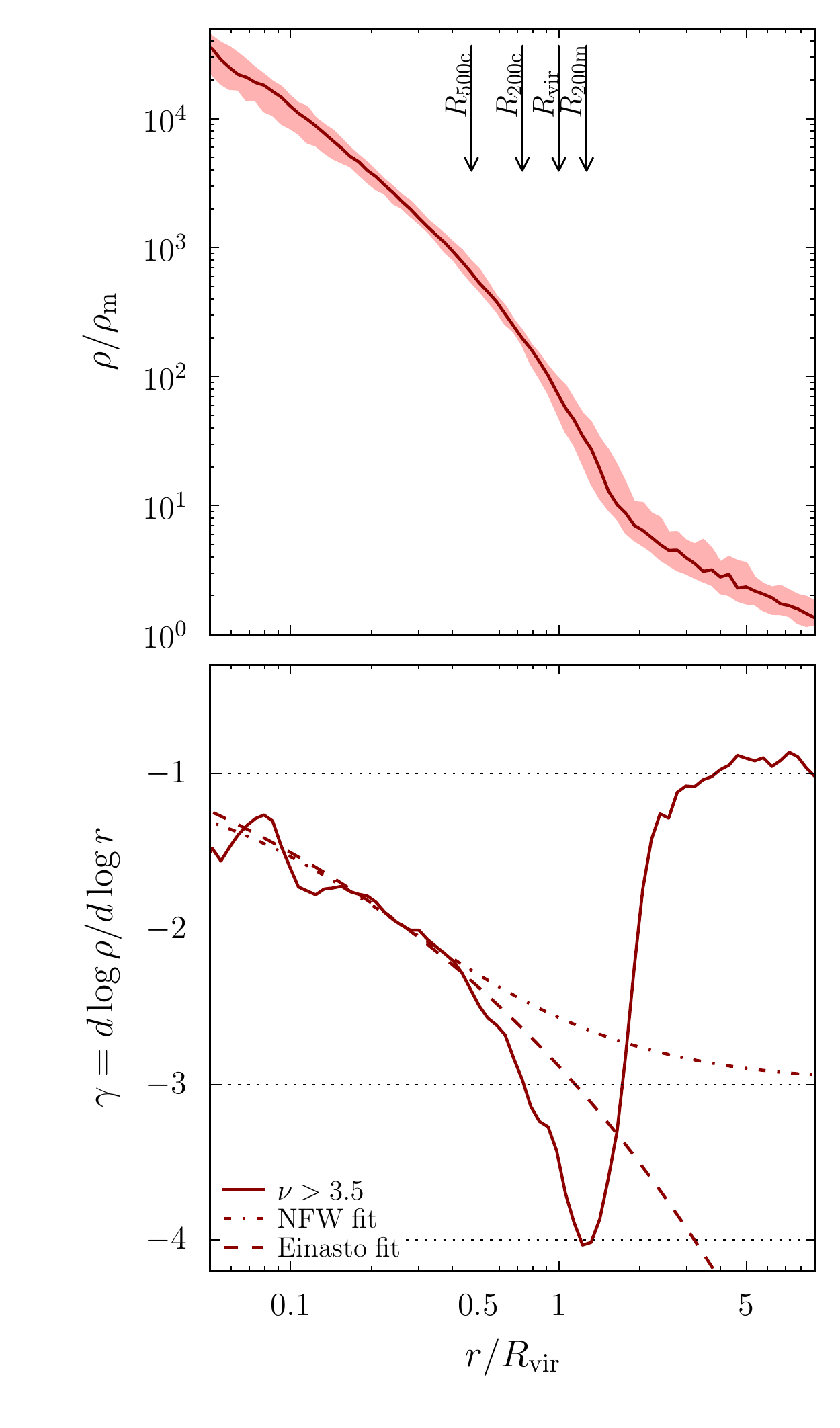}
\caption{Median density profiles of low-mass (top left panel) and very massive (top right panel) halos at $z = 0$. The shaded bands show the interval around the median that contains 68\% of the individual halo profiles in the corresponding $\nu$ bin. The plots include somewhat smaller radii for the high-$\nu$ sample compared to the low-$\nu$ sample due to the different resolution limits of the simulations from which the profiles were extracted. The shapes of the high- and low-mass profiles are noticeably different: the slope of the high-$\nu$ profile steepens sharply at $r \gtrsim 0.5 \rvir$, while the profile of the low-$\nu$ sample changes slope gradually until $r \approx 1.5 \rvir$, where the profiles of both samples flatten significantly. The sharp steepening of the outer profile of the high-$\nu$ sample cannot be described by the NFW or Einasto profiles, as is evident in the bottom panels. The bottom panels show the logarithmic slope profile of the median density profiles in the top panels, as well as the corresponding slope profiles for the best-fit NFW (dot-dashed) and Einasto (dashed) profiles. To avoid crowding, we only show the NFW and Einasto fits in the bottom panels where the differences can be seen more clearly. The vertical arrows indicate the position of various radius definitions, evaluated for the median mass profile.}
\label{fig:prof_simple}
\end{figure*}

We bin halos by peak height, $\nu$, rather than mass, because halo properties are expected to be similar across redshifts for a fixed value of $\nu$. The peak height is defined as
\begin{equation}
\nu \equiv \frac{\delta_{\rm c}}{\sigma(M, z)} = \frac{\delta_{\rm c}}{\sigma(M, z = 0) \times D_+(z)},
\end{equation}
where $\delta_{\rm c} = 1.686$ is the critical overdensity for collapse derived from the spherical top hat collapse model \citep[][we ignore a weak dependence of $\delta_{\rm c}$ on cosmology and redshift]{gunn_72_sphericalcollapse}, and $D_+(z)$ is the linear growth factor normalized to unity at $z=0$. Here $\sigma$ is the rms density fluctuation in a sphere of radius $R$,
\begin{equation}
\sigma^2(R) = \frac{1}{2 \pi^2} \int_0^{\infty} k^2 P(k) |\tilde{W}(kR)|^2 dk
\end{equation}
where $\tilde{W}(kR)$ is the Fourier transform of the spherical top hat filter function, and $P(k)$ is the linear power spectrum. We approximate $P(k)$ using the formula of \citet{eisenstein_98}, normalized such that $\sigma(8 \mpch) = \sigma_8$ = $0.82$. The variance of a certain mass is defined as $\sigma(M) = \sigma(R[M])$ where $M = (4 \pi/3) \rho_{\rm m}(z=0) R^3$. To compute $\nu$ we use $M = \mvir$. Figure \ref{fig:nu_M} shows the halo masses corresponding to the peak height bins used in this paper. We note that by splitting halo samples into equal bins in peak height we emphasize large halo masses. We use $\rvir$ to translate halo masses into peak heights, as it corresponds to the largest radius where the scatter in the density profiles of a given mass is still relatively small, whereas scatter quickly increases at $r \gtrsim \rvir$ (see Figure~\ref{fig:prof_simple}). For the same reason, we use $\rvir$ rather than $\rtom$ when we estimate the mass accretion rate between two redshifts (see Section \ref{sec:results}). We have verified that the choice of mass definition does not qualitatively influence our results and conclusions.

\subsection{Other Numerical Aspects}
\label{sec:numerical:other}

Whenever we show the mean or median profiles in rescaled radial units, such as $r / R_{\Delta}$, we first rescale each individual halo profile using the halo's $R_{\Delta}$, and then construct the mean and median from the rescaled profiles. We compute the slope profiles using the fourth-order Savitzky--Golay smoothing algorithm over the 15 nearest bins \citep{savitzky_64}. This algorithm is designed to smooth out noise in the profiles without affecting the actual values of the slope. We found 15 bins to be the optimal window size to smooth out random fluctuations without introducing artificial steepening or other features compared to the unsmoothed slope profile. Due to this large window size, the method fails for the seven innermost and outermost bins, where we replace it with the algorithm described in the Appendix of \citet{churazov_10}.

All functional fits are performed using the Levenberg-Marquart minimization algorithm. The merit function that is minimized is the sum of the square differences in units of $r^2 \rho$ rather than just $\rho$, as the numerical value of $\rho$ decreases by many orders of magnitude between the inner and outer radii. The $r^2 \rho$ metric provides a more balanced indicator of goodness of fit across the radial range we are fitting. We exclude the outer radii ($r > 0.5 \rvir$) when fitting functions that are not designed to fit the outer halo profile, for example, the NFW and Einasto profiles. If larger radii are included in the fit, the shapes of the outer profiles drag the fit away from the values suggested by the central region. Due to the potential force resolution issues discussed in Section \ref{sec:numerical:halos}, we do not attempt to fit for both the scale radius and steepness parameter $\alpha$ of the Einasto profile. Instead, we use the relation of \citet{gao_08} to fix $\alpha$ as a function of $\nu$.

In this paper, we use median profiles for most of our analyses. The median profile is an approximation of the most typical profile for a given halo sample, and is thus well suited for studying trends in the density profiles. However, for certain purposes, the mean profile may be more applicable. For example, in weak-lensing analyses using stacked shear maps of many galaxies or galaxy clusters, the derived density profile may correspond more closely to the mean profile of a sample. Our conclusions described below hold for the mean density profiles as well, and the fitting formula we devise in Section \ref{sec:results:formula} is valid for both the median and mean profiles.

%%%%%%%%%%%%%%%%%%%%%%%%%%%%%%%%%%%%%%%%%%%%%%%%%%%%%%%%%%%%%%%%%%%%%%%%%%
% RESULTS
%%%%%%%%%%%%%%%%%%%%%%%%%%%%%%%%%%%%%%%%%%%%%%%%%%%%%%%%%%%%%%%%%%%%%%%%%%

\section{Results}
\label{sec:results}

We use the simulations and halo samples described in the previous section to construct the median and mean density profiles of halos binned by peak height, redshift, and mass accretion rate. In this section, we explore the variation of the profiles with these properties.  

\subsection{Density Profiles as a Function of Peak Height}
\label{sec:results:nu}

Figure \ref{fig:prof_simple} shows the median density profiles at $z = 0$ of two halo samples representing extremes of the range of halo peak heights, and the corresponding profiles of the logarithmic slope, $\gamma(r) \equiv d \log \rho / d \log r$. The low-mass sample (left panels) corresponds to the peak height range of $0.5 < \nu < 0.7$ (see Figure \ref{fig:nu_M} for the respective mass range), while the high-mass sample corresponds to $\nu > 3.5$. We also show the interval containing 68\% of the individual profiles with a shaded band. 

\begin{figure*}
\centering
\includegraphics[trim = 5mm 1mm 118mm 0mm, clip, scale=\panelsize]{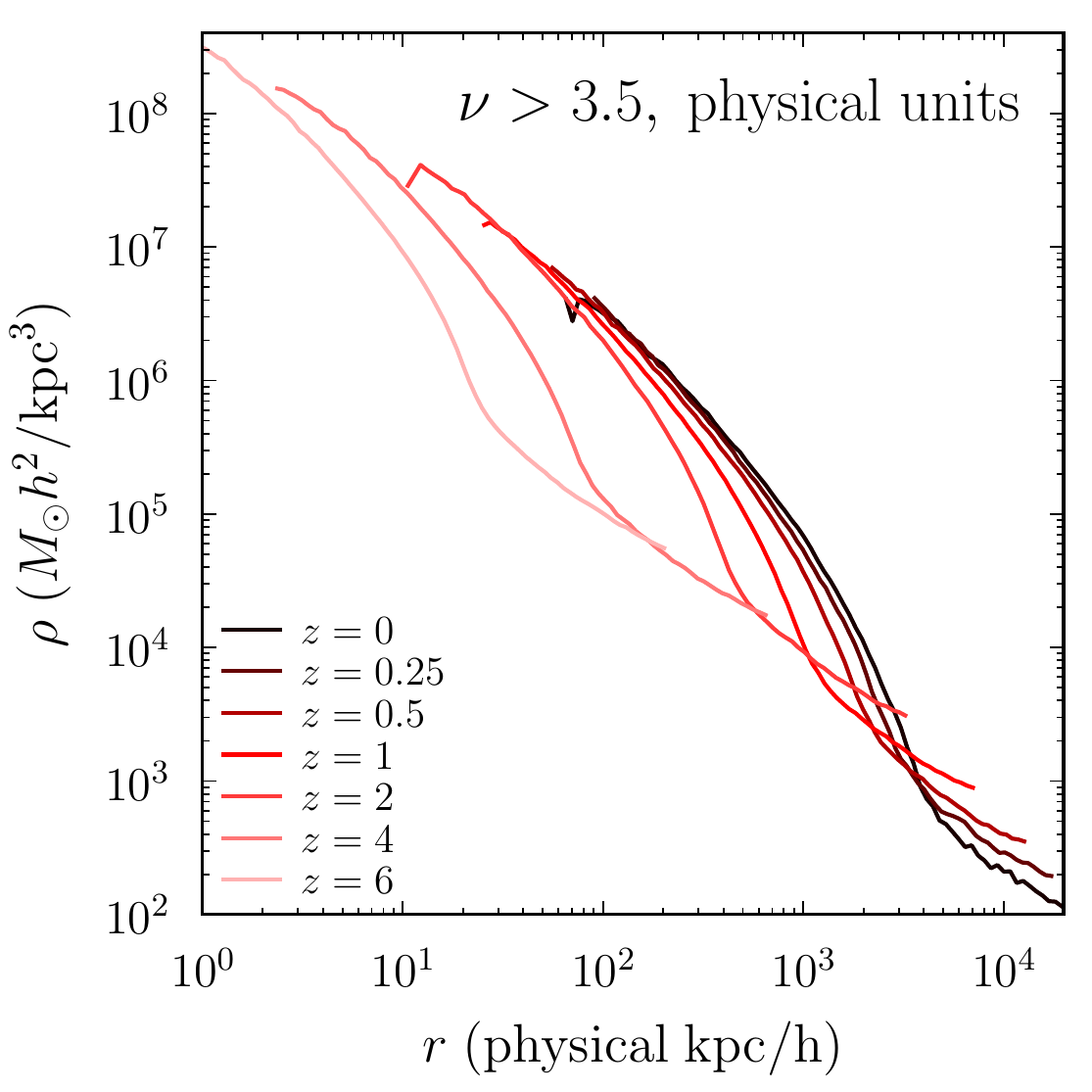}
\includegraphics[trim = 5mm 1mm 118mm 0mm, clip, scale=\panelsize]{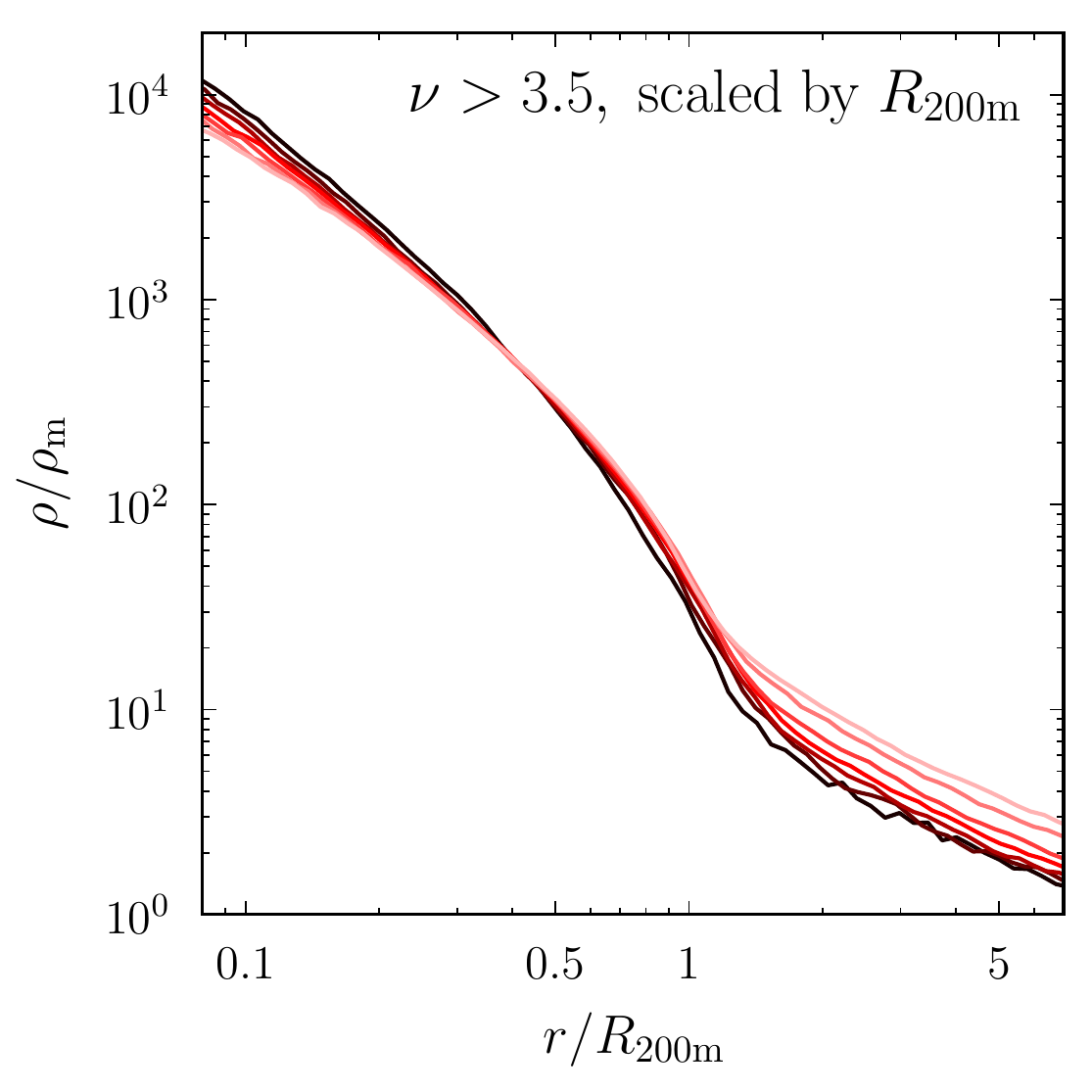}
\includegraphics[trim = 5mm 1mm 118mm 0mm, clip, scale=\panelsize]{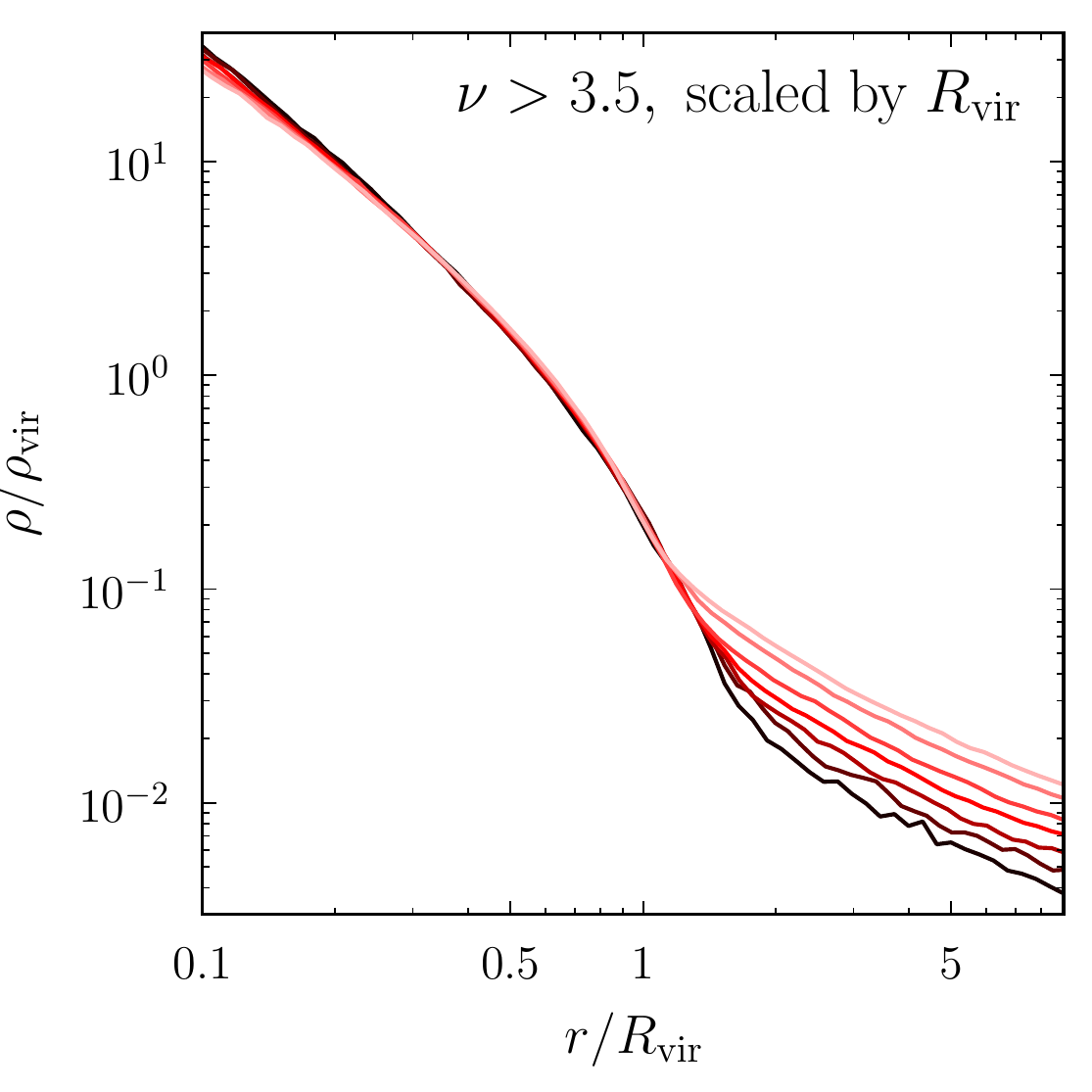}
\includegraphics[trim = 5mm 1mm 118mm 0mm, clip, scale=\panelsize]{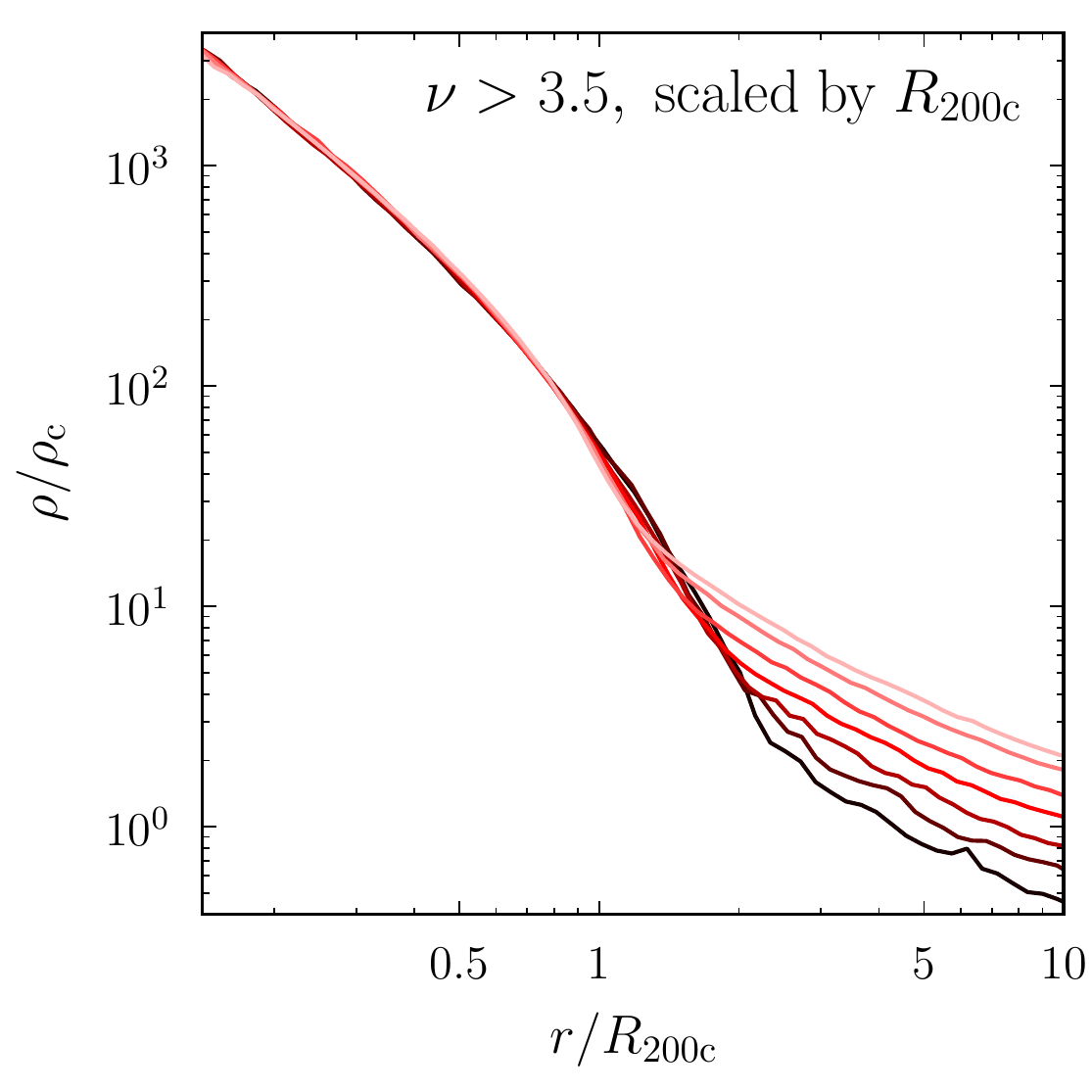}
\caption{Self-similarity of the redshift evolution of density profiles. The top left panel shows the redshift evolution of the median density profiles of the highest peak halos, $\nu>3.5$, as a function of proper radius (the results for lower-$\nu$ halos are similar). The rest of the panels show the same profiles as the top left panel, but rescaled by $\rtoc$, $\rvir$, and $\rtom$, with density rescaled correspondingly by $\rhoc$, $\rho_{\rm vir}$, and $\rhom$. The plots demonstrate that the structure of halos of a given $\nu$ is nearly self-similar when rescaled by any $R_{\Delta}$. However, they also reveal that the inner structure of halos is most self-similar when radii and densities are rescaled by $\rtoc$ and $\rhoc$, while the outer profiles are most self-similar when rescaled by $\rtom$ and $\rhom$. See also Figure~\ref{fig:prof_by_z} where we show the slope profiles of the scaled profiles.}
\label{fig:prof_physical}
\end{figure*}

It is clear that the profiles of the two samples in Figure \ref{fig:prof_simple} are quite different. The median profile of the low-$\nu$ sample has a slowly changing slope out to $r \gsim \rvir$ and large scatter around the flattening at larger radii. The high-$\nu$ sample, on the other hand, has a sharply steepening profile at $r \gtrsim 0.5 \rvir$ with the slope changing from $-2$ to $-4$ over a range of only $\approx 4$ in radius, as can be seen in the slope profiles (bottom panels). For comparison, the slope of an NFW profile is expected to change by only $\approx 0.6$ over the same radial range for typical concentrations. The slope profiles show that although the NFW and Einasto profiles provide a reasonable description to the profiles of the low-$\nu$ sample out to $r \approx \rvir$, they fail to describe the rapid steepening of the slope in the high-$\nu$ sample. Clearly, the functional form of the high-$\nu$ profiles differs from the fit at large radii, implying that the outer density profiles of halos cannot be universally described by a single NFW or Einasto profile. We note that these fitting functions were not designed to match profiles outside $r \approx \rvir$, but the deviations from the NFW and Einasto profiles in high-$\nu$ halos begin at smaller radii, $r \approx 0.5 \rvir$ \citep[see also][]{meneghetti_13, balmes_14_sparsity}. In Section \ref{sec:results:formula} and the Appendix we present a more flexible functional form that can describe the profiles of halos of different peak heights.

We note that the profiles of both the low-$\nu$ and high-$\nu$ samples flatten to a slope of $\approx -1$ at $r \gtrsim 2 \rvir$, as the profile approaches the 2-halo term of the halo--matter correlation function \citep[see, e.g.,][]{hayashi_08}. However, the scatter around the median profiles is much larger for low-$\nu$ halos, even though such halos form earlier and are thus more relaxed on average. The reason for the increased scatter is that some of the low-$\nu$ halos are located in crowded environments near massive neighbors, while others are relatively isolated. High-$\nu$ halos are massive and rare, and their environments are much more uniform.

\begin{figure*}
\centering
\includegraphics[trim = 123mm 2mm 2mm 0mm, clip, scale=\panelsizethree]{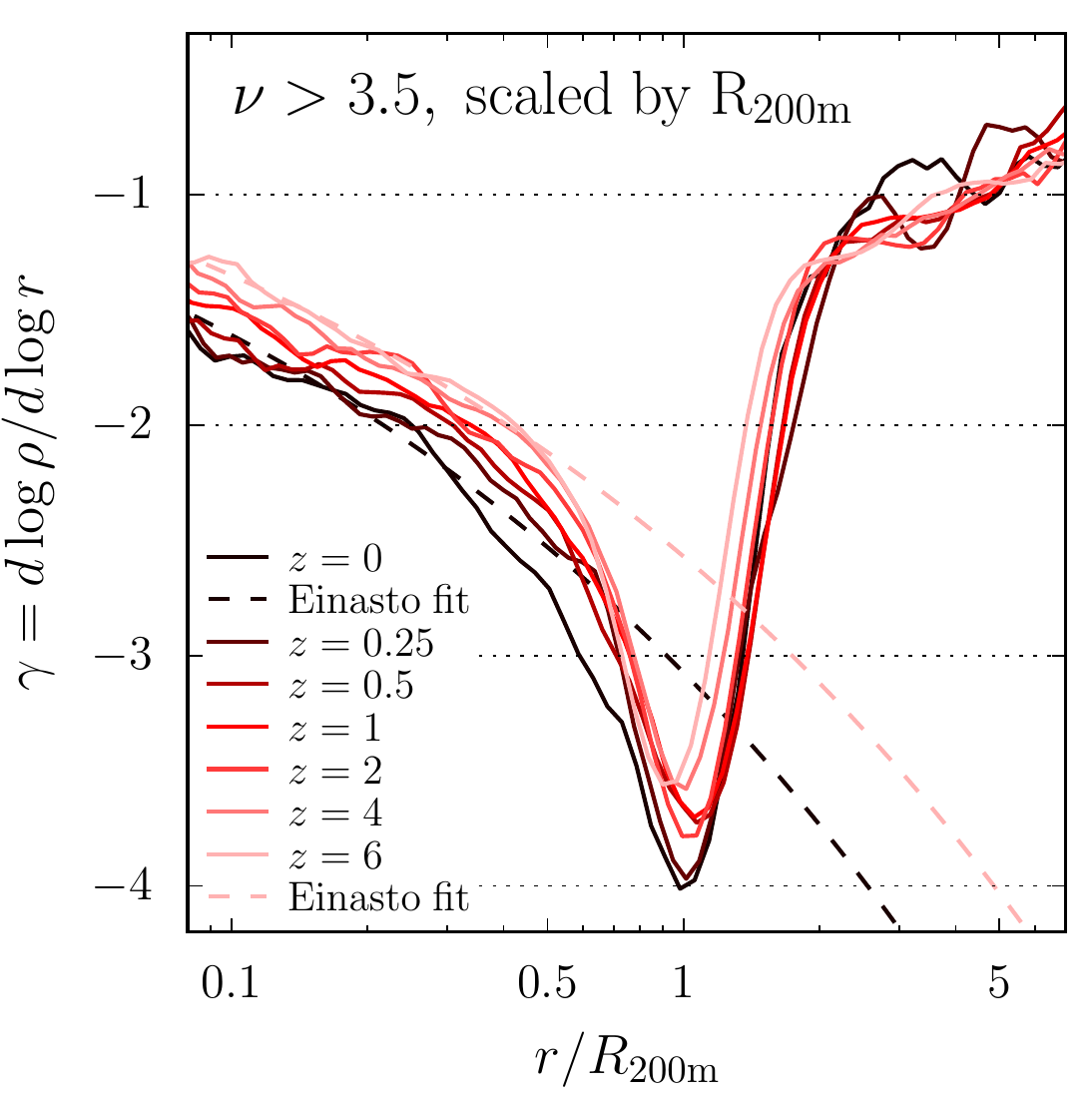}
\includegraphics[trim = 141mm 2mm 2mm 0mm, clip, scale=\panelsizethree]{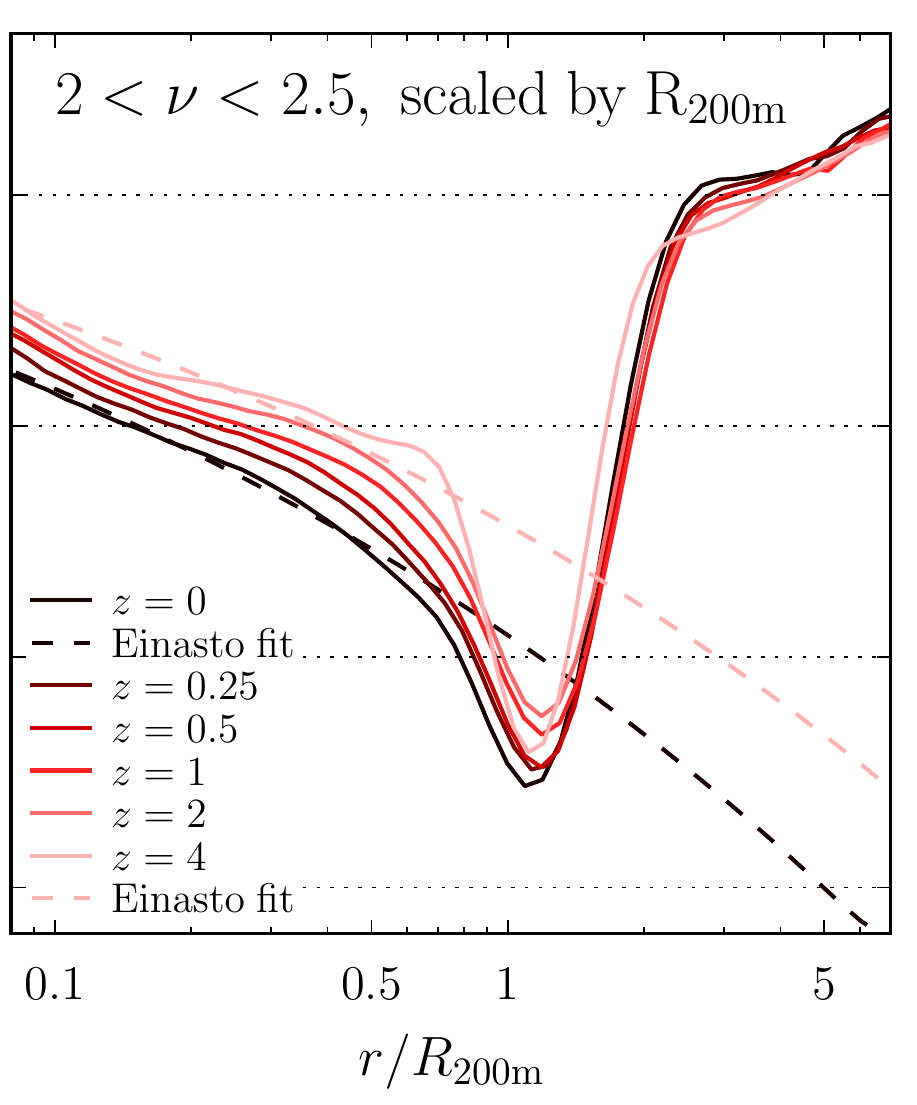}
\includegraphics[trim = 141mm 2mm 2mm 0mm, clip, scale=\panelsizethree]{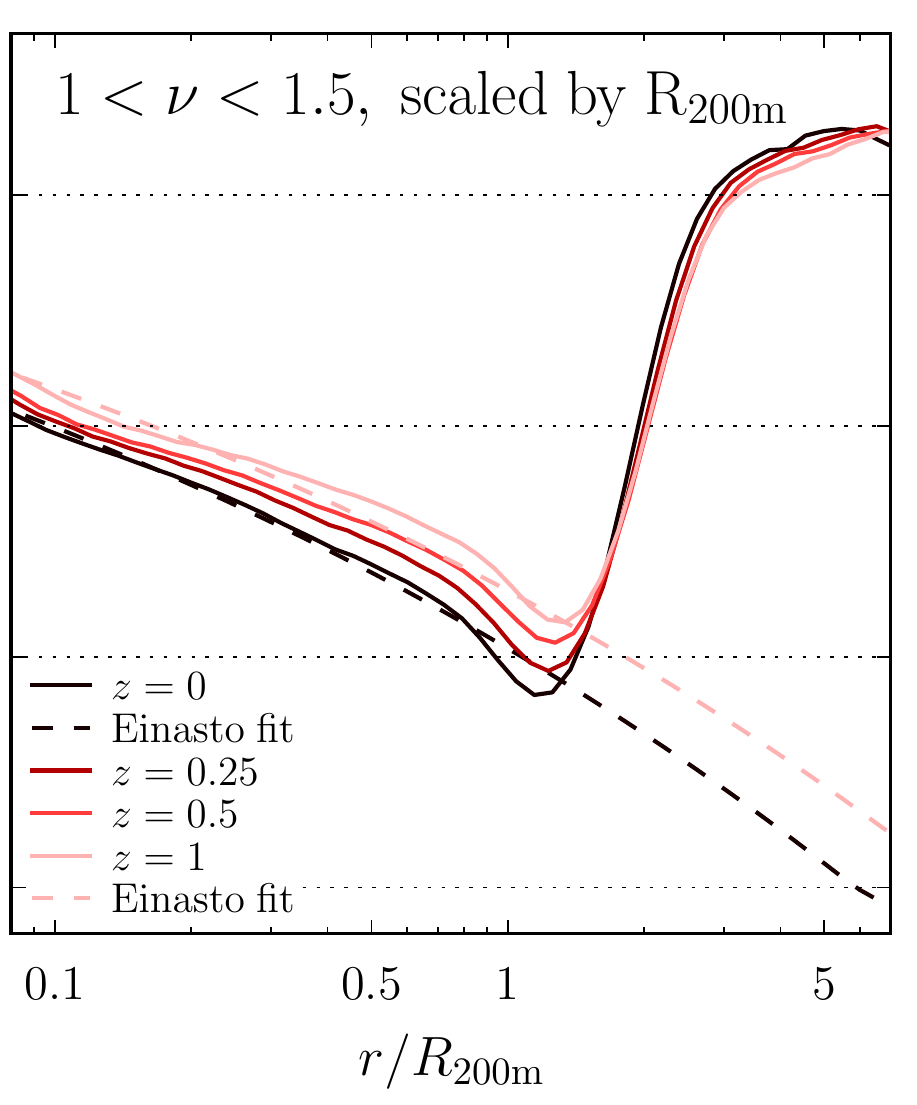}
\includegraphics[trim = 123mm 2mm 2mm 0mm, clip, scale=\panelsizethree]{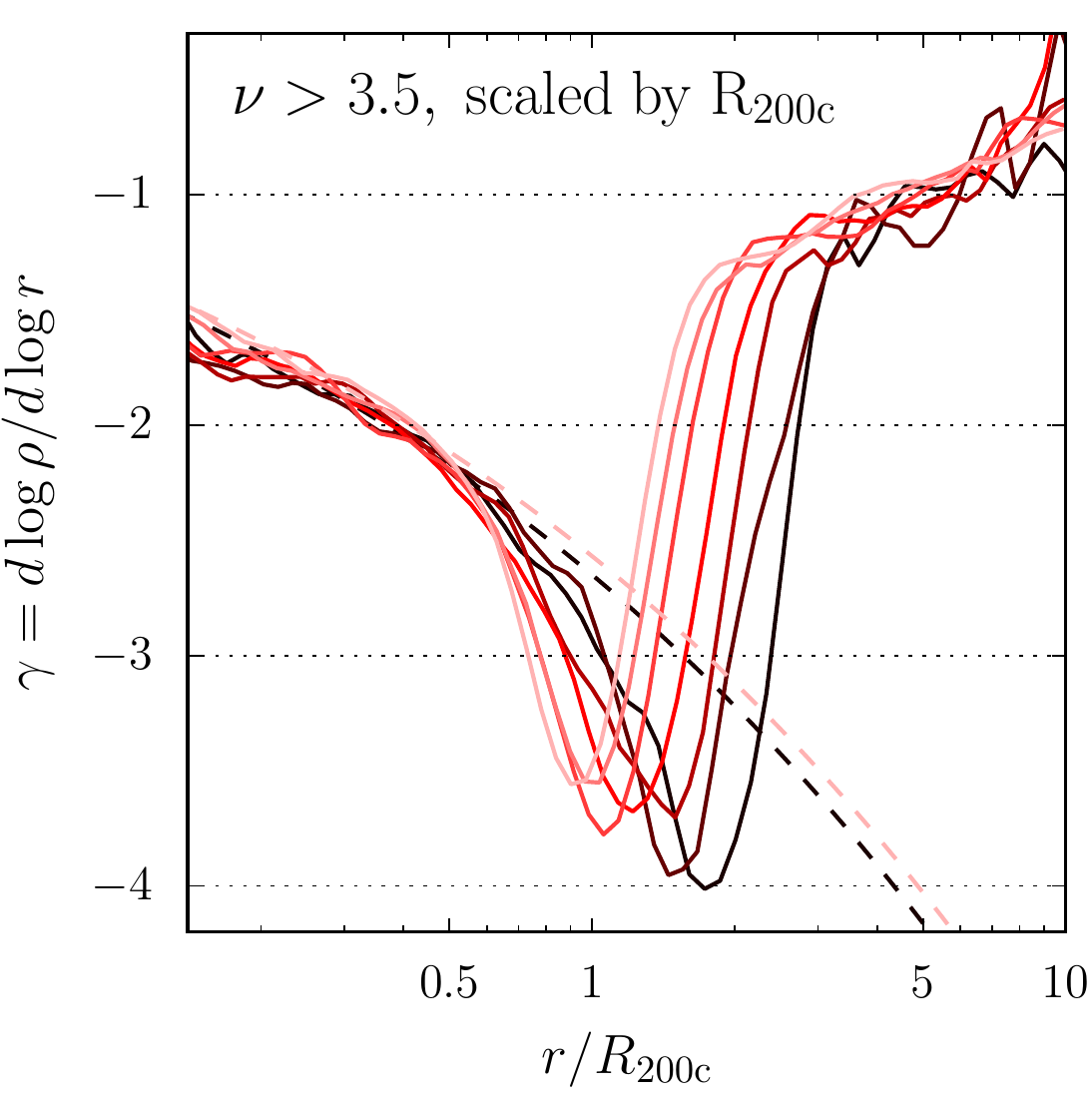}
\includegraphics[trim = 141mm 2mm 2mm 0mm, clip, scale=\panelsizethree]{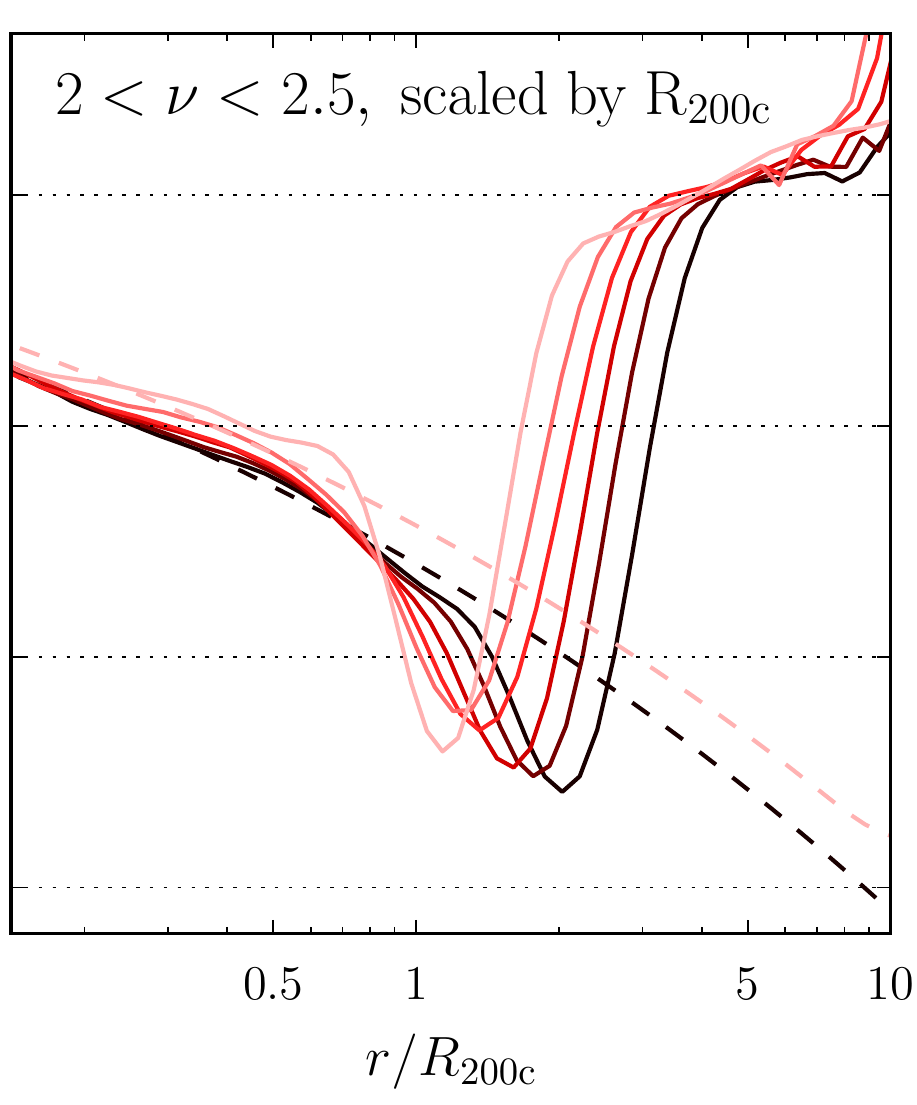}
\includegraphics[trim = 141mm 2mm 2mm 0mm, clip, scale=\panelsizethree]{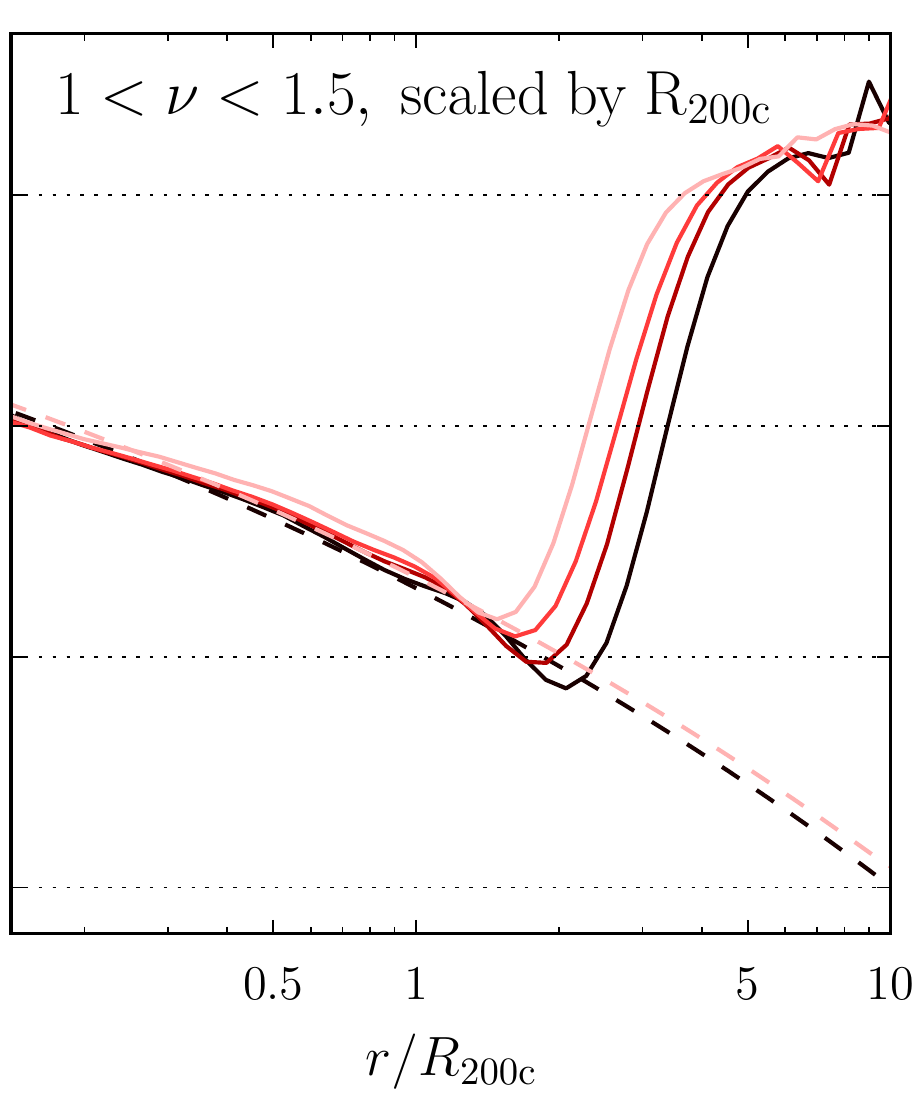}
\caption{Slope profiles of the three $\nu$ bins shown in Figure \ref{fig:prof_by_nu}, at different redshifts. The left panels refer to the $\nu > 3.5$ sample shown in Figure \ref{fig:prof_physical}. For the lower-$\nu$ bins (center and right panels), fewer redshift bins are accessible with our simulations. In the top panels radii are rescaled by $\rtom$, in the bottom panels by $\rtoc$. The slope profiles confirm the results of Figure \ref{fig:prof_physical} that the outer profiles at $r \gtrsim \rtom$ are most self-similar when radii are rescaled by $\rtom$, with the steepest slope reached at $r \approx \rtom$ regardless of redshift. The inner profiles at $r\lesssim 0.6 \rtoc$, however, are most self-similar when rescaled by $\rtoc$. We note that at $z \gtrsim 2$ the difference between $\rtom$ and $\rtoc$ becomes negligible. The $2 < \nu< 2.5$ bin at $z = 4$ (lightest red line in the center panels) exhibits a slightly different shape than the other redshift bins, possibly due to sample variance as almost all halos in this bin originate from the smallest simulation box, L0063.}
\label{fig:prof_by_z}
\end{figure*}

Figure~\ref{fig:prof_simple} shows the profiles of a given $\nu$ bin only at $z=0$. However, we can in general expect that profiles of halos of a given $\nu$ are self-similar in shape, as long as the density and radii are properly rescaled. However, it is not clear a priori what radii and characteristic densities should be used for such rescaling, leading us to investigate several choices.

The top left panel of Figure \ref{fig:prof_physical} shows a sequence of profiles of the highest-$\nu$ bin at different redshifts in proper units (physical density and radius). We stress that we compare the median profiles of halos of similar peak heights, not the profiles of progenitor and descendant halos. The peak height bin $\nu > 3.5$ corresponds to halos of very different mass at different redshifts, from $\mvir > 1.4 \times 10^{15} \msunh$ at $z = 0$ to $\mvir > 1.5 \times 10^{11} \msunh$ at $z = 6$ (see Figure~\ref{fig:nu_M}). Their virial radii span over two orders of magnitude over this redshift interval. The other panels of Figure \ref{fig:prof_physical} show the same profiles, but rescaled by $\rtoc$, $\rvir$, and $\rtom$, with the densities rescaled correspondingly by $\rhoc$, $\rho_{\rm vir}$, and $\rho_{\rm m}$. These panels demonstrate that the structure of halos of a given $\nu$ is nearly self-similar when rescaled by any $R_{\Delta}$ in a reasonable range. However, they also reveal that the inner structure of halos is most self-similar when radii and densities are rescaled by $\rhoc$ and $\rtoc$, while the outer profiles are most self-similar when rescaled by $\rtom$ and $\rhom$. A scaling with $\rho_{\rm vir}$ and $\rvir$ produces intermediate results.

The degree of self-similarity can be assessed more robustly in profiles of the logarithmic slope, which show particularly clearly at which radii the profiles undergo rapid changes in slope. Figure \ref{fig:prof_by_z} shows the slope profiles for three $\nu$ bins, rescaled by $\rtom$ (top row) and $\rtoc$ (bottom row). The sharp steepening of the profile and subsequent sharp flattening occur at the same radii in units of $\rtom$, and the radius of the steepest slope occurs at $\approx 1-1.2\rtom$ for all $\nu$ and redshifts. At $r < \rtom$, however, the slopes of the profiles at a given $r / \rtom$ vary for different $\nu$ and $z$. The opposite is true when the densities and radii are rescaled by $\rho_c$ and $\rtoc$. In particular, at $r \lesssim 0.8-1\rtoc$, the slopes at a given $r / \rtoc$ agree for halos of the same $\nu$ at different $z$. Although the shapes of the low-$\nu$ and high-$\nu$ profiles are different, with the former exhibiting a slower change of slope, they exhibit a similarly remarkable degree of uniformity at $r > \rtom$ when rescaled by $\rtom$, and at $r < \rtoc$ when rescaled by $\rtoc$.

Our results thus lead to the conclusion that {\it the inner, most relaxed regions of halo profiles are self-similar in units of $r / \rtoc$, while the outer profiles are self-similar in units of $r / \rtom$.} This conclusion would of course hold for any radius definition using a fixed overdensity relative to the mean and critical density within a reasonable range of overdensities. This observation implies that the concentration of halos should be more universal as a function of $\nu$ when one uses a radius definition tied to the critical density. On the other hand, for modeling the transition radius between the 1-halo and 2-halo terms in the halo model, the use of radii tied to the mean density may be preferable. Given that we focus on the outer profiles in this study, we will scale the profiles at different redshifts using $\rhom$ and $\rtom$ in the subsequent analyses. We further discuss the self-similarity of the profiles in Section \ref{sec:discussion:selfsimilarity}.

\begin{figure}
\centering
\includegraphics[trim = 12mm 6mm 3mm 3mm, clip, scale=\panelsize]{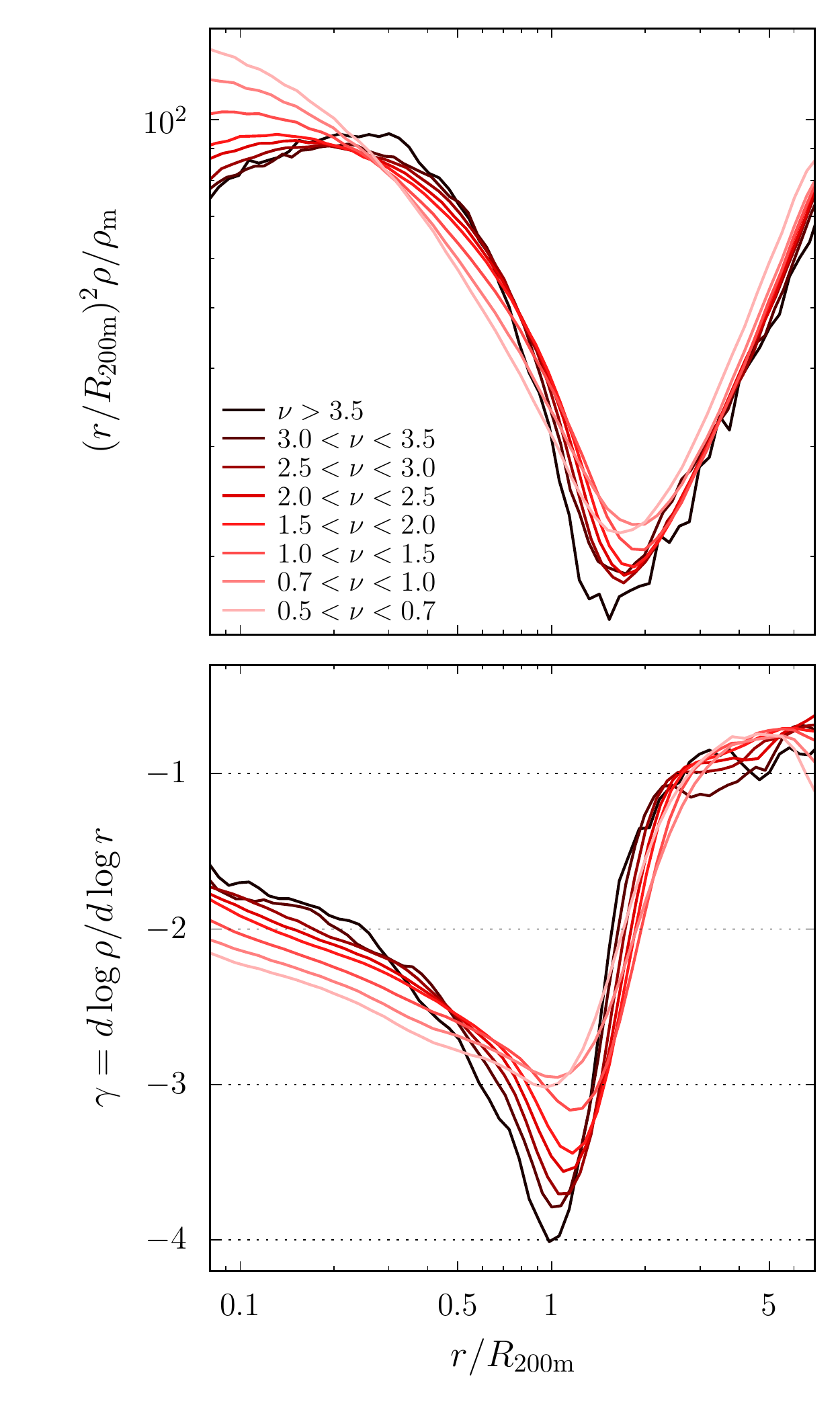}
\caption{Median density profiles (top panel) and their logarithmic slopes (bottom panel) for various bins in peak height, $\nu$, at $z = 0$. For clarity, the density is plotted in units of $\rho r^2$, which makes it easier to see differences between profiles. The $\nu$ bins range from small peaks ($\nu = 0.5$, $\mvir = 1.4 \times 10^{10} \msunh$) to rare peaks ($\nu > 3.5$, $\mvir > 1.4 \times 10^{15} \msunh$).  The steepest slope of the profiles increases with peak height, but all profiles of samples with $\nu > 1$ reach slopes steeper than $-3$.}
\label{fig:prof_by_nu}
\end{figure}

\begin{figure}
\centering
\includegraphics[trim = 0mm 2mm 0mm 0mm, clip, scale=0.75]{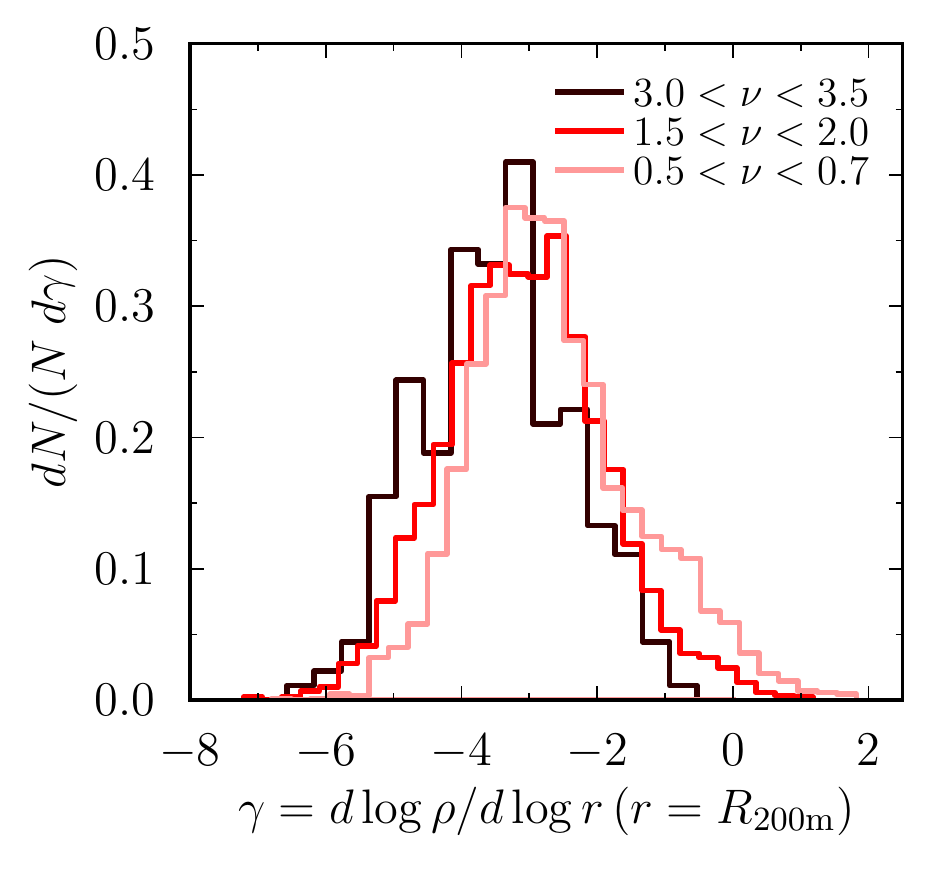}
\caption{Distribution of the logarithmic slope $\gamma \equiv d \log \rho / d \log r$ at $\rtom$, for three bins in peak height. The slope is measured for about $3000$ individual halo profiles in the lower-$\nu$ bins, and about $220$ in the highest-$\nu$ bin. The slopes span a wide range: some halos have outer slopes as steep as $-6$ or $-7$, while other halos have flat or even positive slopes. The latter halos likely have nearby massive neighbors, while the former halos accrete mass at a high rate, as we will show in Section \ref{sec:results:accrate}.}
\label{fig:slopedist}
\end{figure}

Finally, we investigate whether the shape of the profiles follows a continuous function of peak height, as indicated by the trend with $\nu$ in Figure \ref{fig:prof_by_z}. Figure \ref{fig:prof_by_nu} shows the density profiles (in units of $\rho(r) r^2$ to minimize the dynamic range) and corresponding slope profiles for a range of peak heights spanning five orders of magnitude in mass. As the peak height increases, the slope of the profiles becomes shallower at $r \lesssim 0.5 \rvir$, but steeper at $0.5 \lesssim r \lesssim 1.5 \rtom$. At $r \gtrsim 1.5 \rtom$ the profiles are remarkably self-similar for halos of different $\nu$ when rescaled using $\rtom$. 

Although the shape of the median profiles follows a continuous trend with $\nu$, the scatter of the individual profiles around the median of each $\nu$ sample is substantial. Figure \ref{fig:slopedist} shows the distribution of slopes at $\rtom$ for three of the $\nu$ bins shown in Figure \ref{fig:prof_by_nu}. The distributions are quite broad, with particularly long tails toward shallower, or even positive, slopes. On the other hand, the tails toward very steep slopes indicate that the steepening demonstrated in Figure \ref{fig:prof_by_nu} can actually be even more pronounced for individual halos, as many halos have slopes significantly steeper than $\gamma \approx -4$. We have verified this observation by examining individual profiles. The radii of the steepest slope, however, do not exactly overlap, and are thus smoothed out in the median profiles.

Figure \ref{fig:slopedist} also demonstrates why we chose to investigate the median rather than mean profiles in this section. The distributions of slopes are not symmetric and have long tails that strongly influence the mean, but not the median. Furthermore, we find that the mean and median of the slope distribution can differ from the slope of the mean and median profile. We will return to this issue when considering individual halo profiles in Section \ref{sec:discussion:observable}.

A similarly large scatter in the outer profiles was reported by \citet{prada_06_outerregions}, who also showed that the mean outer profile depends on how subhalos are excluded from the sample. For example, if one uses a larger radius to define the halo boundary and define subhalos, this lowers the averaged outer profile of the isolated halo sample because it lowers the fraction of halos located right next to a larger, isolated halo.  

\subsection{Dependence on the Mass Accretion Rate}
\label{sec:results:accrate}

\begin{figure}
\centering
\includegraphics[trim = 118mm 17mm 2mm 2mm, clip, scale=\panelsize]{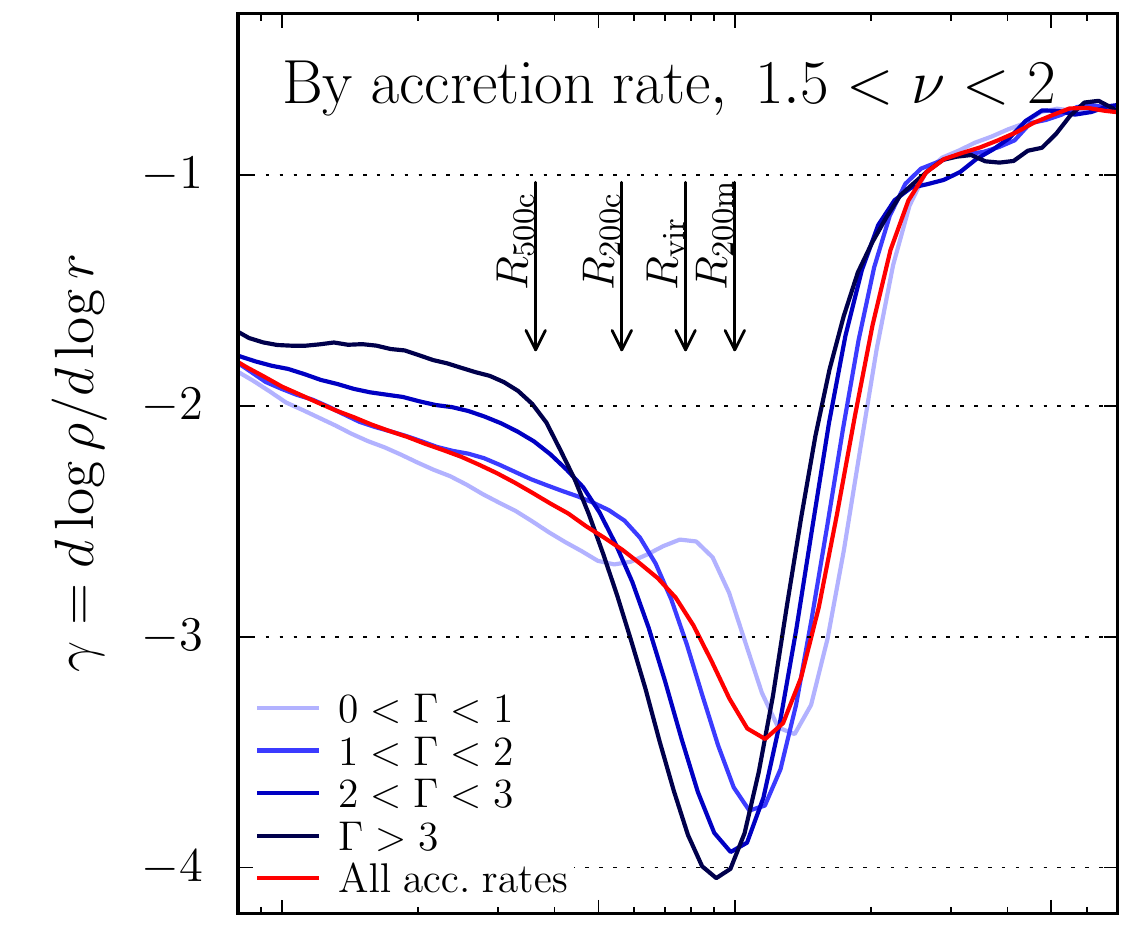}
\includegraphics[trim = 118mm 3mm 2mm 2mm, clip, scale=\panelsize]{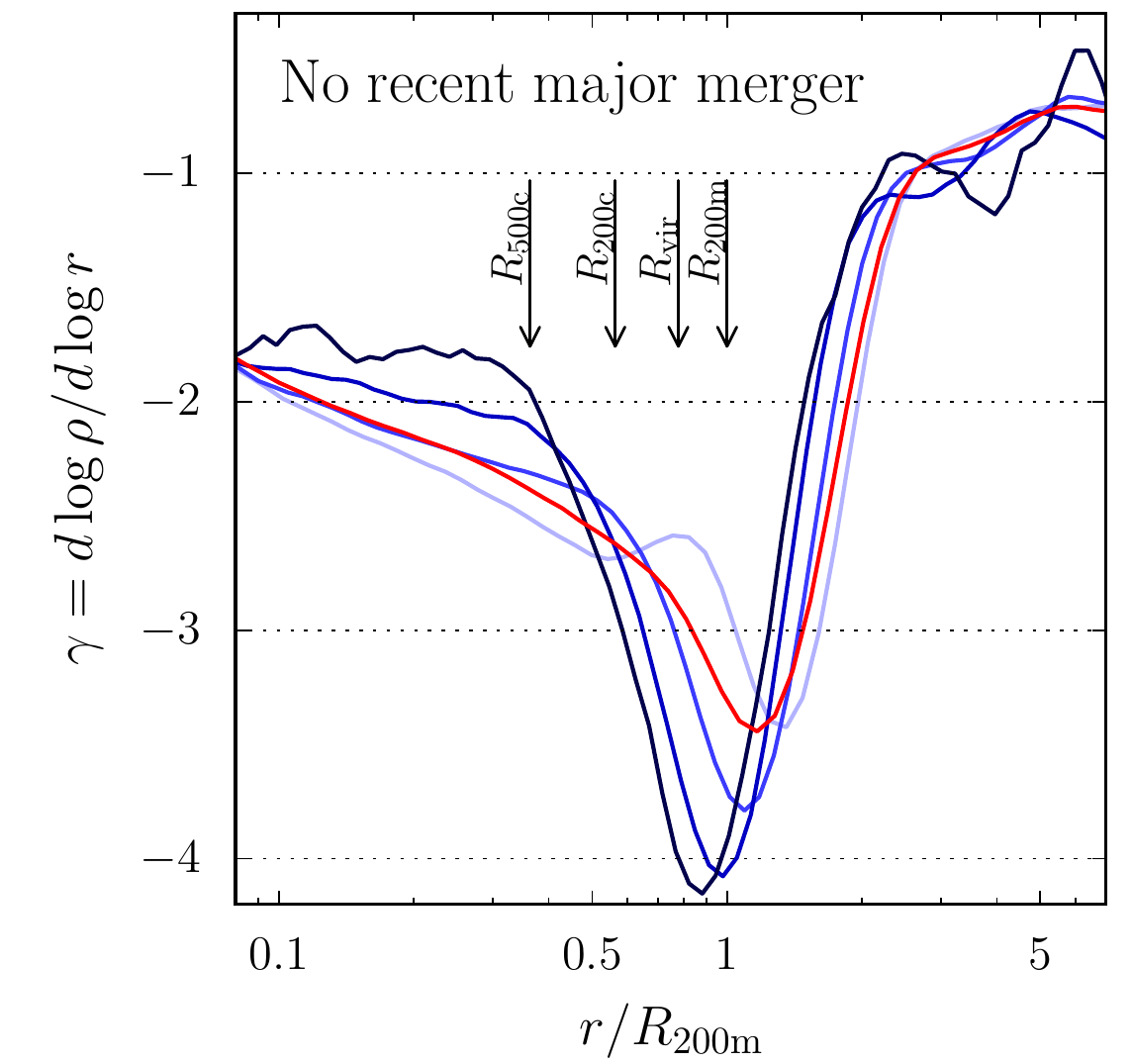}
\caption{Dependence of the slope profiles on the mass accretion rate and occurrence of a recent major merger. In both panels, the red line shows the median density profile of all halos in the peak height range $1.5 < \nu < 2$ at $z = 0$, previously shown in Figure \ref{fig:prof_by_nu}. In the top panel, the sample is further split by accretion rate, measured as the logarithmic change in halo mass, $\Gamma \equiv \Delta\log(\mvir) / \Delta\log (a)$, with differences evaluated for the main progenitor and descendant halo at $z = 0.5$ and $z=0$. Halos with high mass accretion rates exhibit very different median profiles compared to their slowly accreting counterparts. The bottom panel shows the same samples, but with the additional condition that the halos have not undergone a major merger since $z = 0.5$. The profiles are very similar to those in the top panel, which demonstrates that systematic deviations in the shape of the outer profile correlate with the overall mass accretion rate rather than a sharp increase of mass due to a recent major merger.}
\label{fig:prof_accrate}
\end{figure}

\begin{figure}
\centering
\includegraphics[trim = 1mm 5mm 0mm 1mm, clip, scale=0.8]{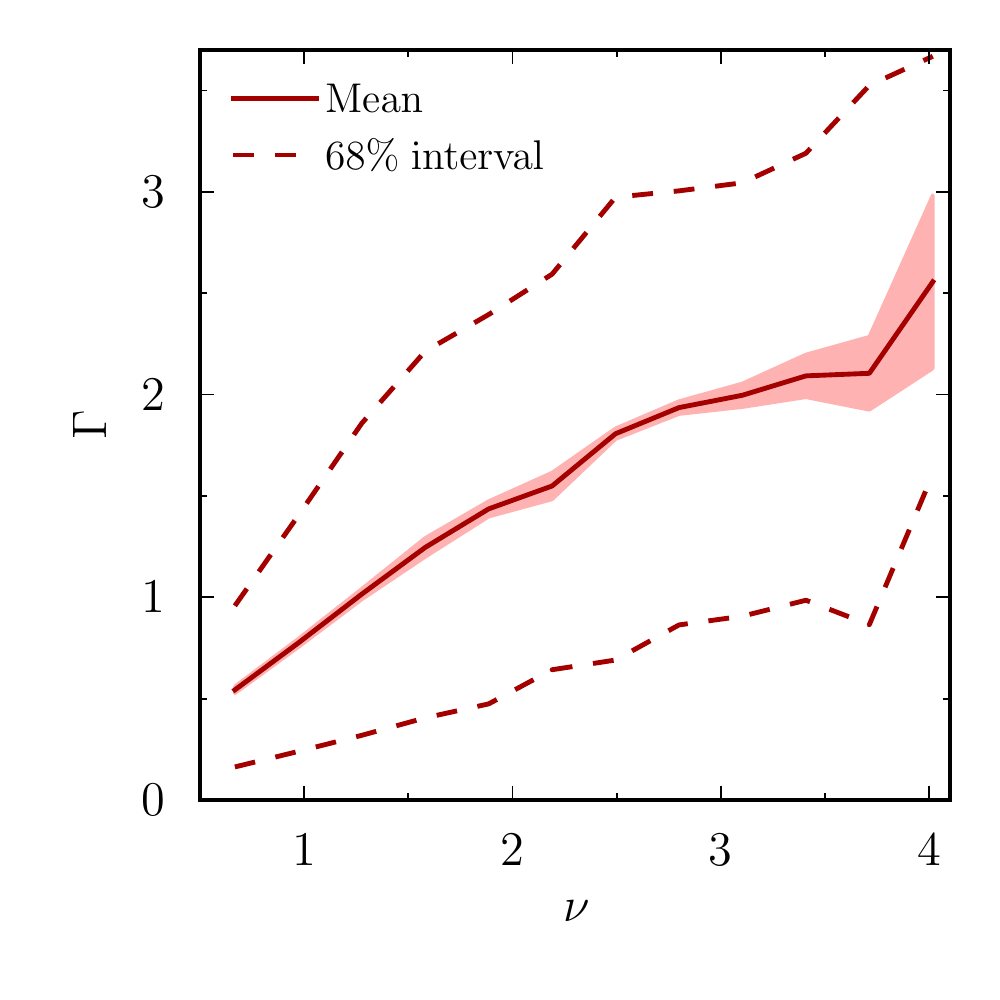}
\caption{Mean mass accretion rate, $\Gamma$, as a function of peak height, $\nu$. The shaded contour indicates the uncertainty on the mean, whereas the dashed lines show the $68\%$ interval. The median $\Gamma$ is slightly lower than the mean at all $\nu$. The dependence of $\Gamma$ on $\nu$ explains why high-$\nu$ halo samples have similar profiles as samples selected by a high accretion rate (Figures \ref{fig:prof_by_nu} and \ref{fig:prof_accrate}).}
\label{fig:nu_gamma}
\end{figure}

\begin{figure}
\centering
\includegraphics[trim = 5mm 17mm 120mm 2mm, clip, scale=\panelsize]{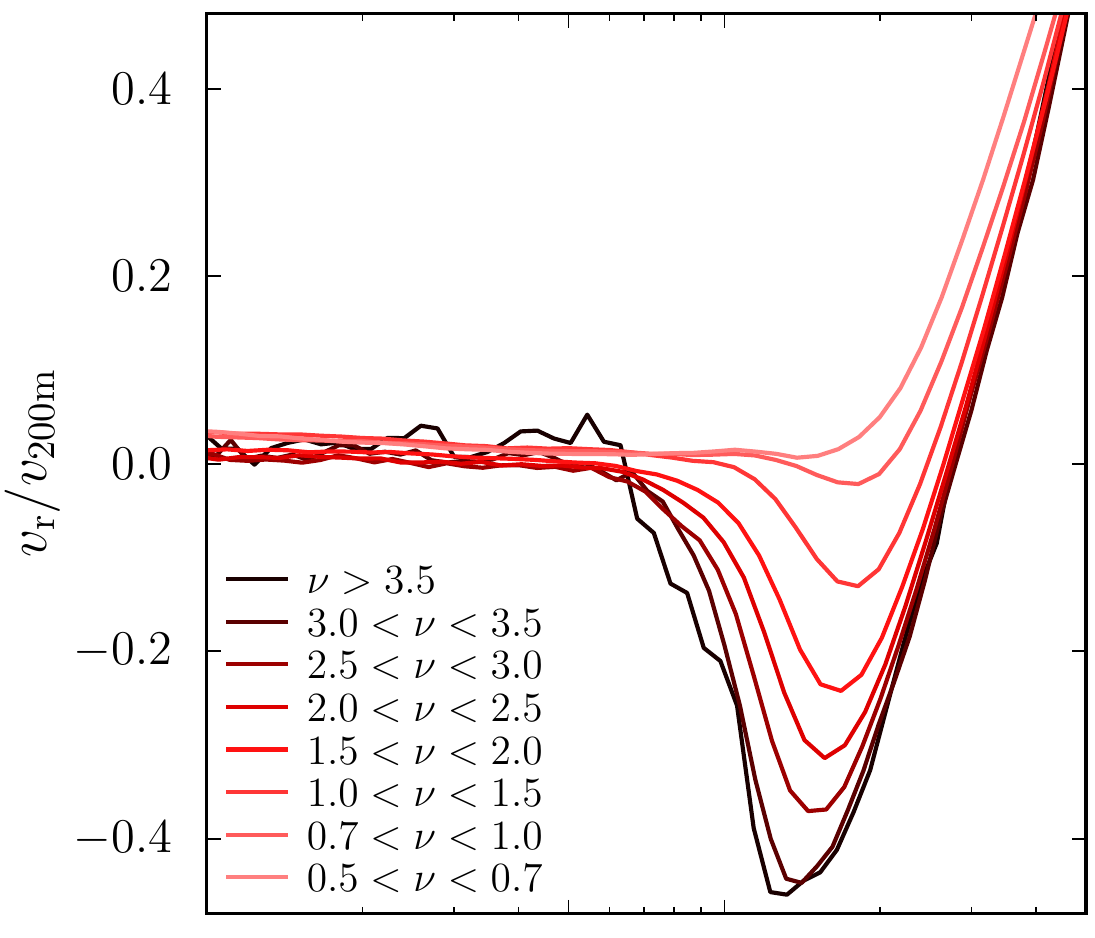}
\includegraphics[trim = 5mm 3mm 120mm 2mm, clip, scale=\panelsize]{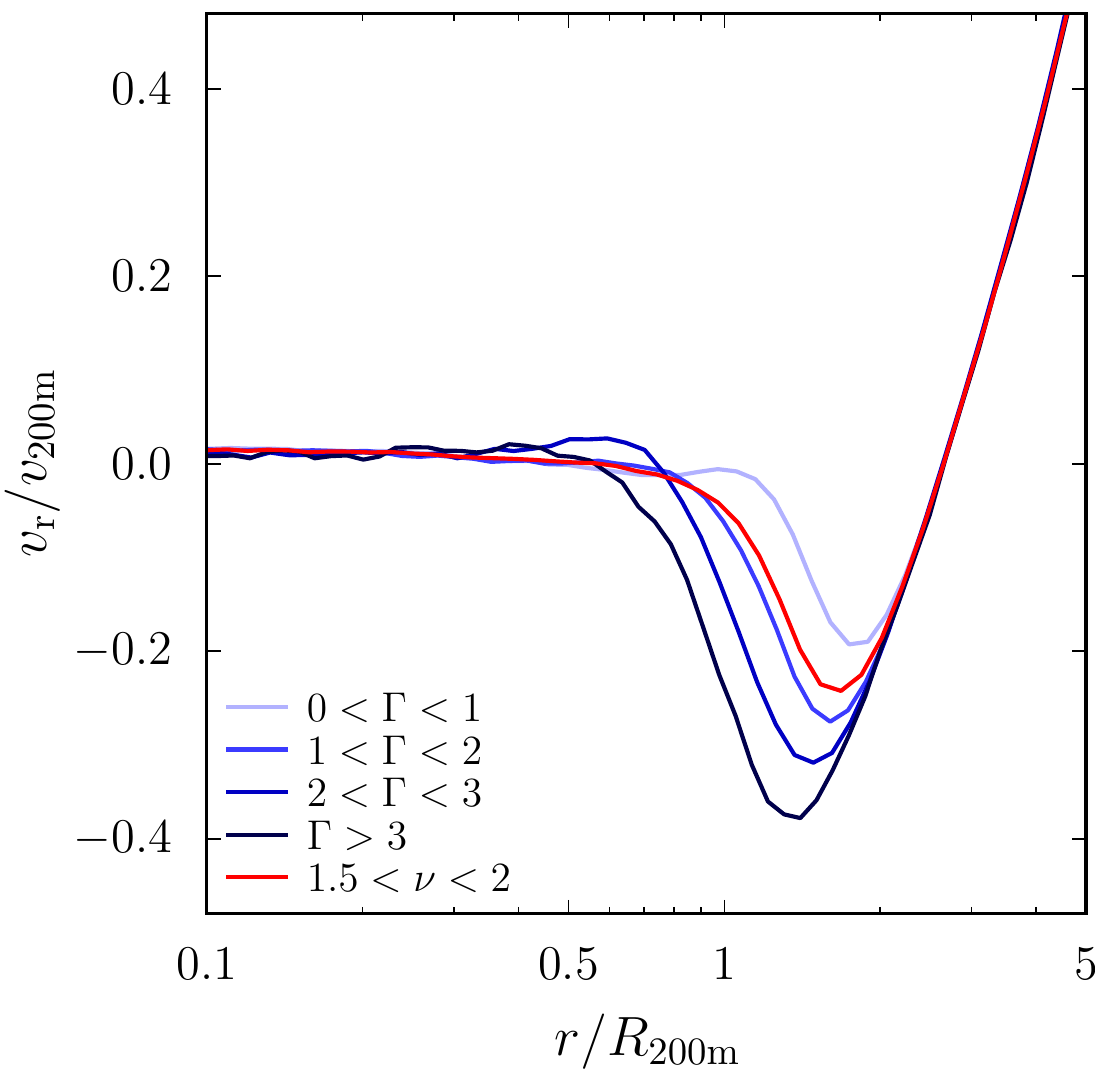}
\caption{Median radial velocity profiles of halos. Top panel: profiles of the same $\nu$ bins as in Figure \ref{fig:prof_by_nu}, at $z = 0$. As expected, the high-$\nu$ bins have much higher infall velocities, even when rescaled by their $v_{\rm 200m}$. Bottom panel: halos from the $1.5 < \nu < 2$ bin, split according to their accretion rate as in the top panel of Figure \ref{fig:prof_accrate}, with the red line showing the median profile of the entire $1.5<\nu<2.0$ sample. The radius where the infall velocity is most negative shows a similar evolution with $\Gamma$ as the radius of the steepest slope in Figure \ref{fig:prof_accrate}.}
\label{fig:prof_vr}
\end{figure}

\begin{figure*}
\centering
\includegraphics[trim = 118mm 17mm 2mm 0mm, clip, scale=\panelsize]{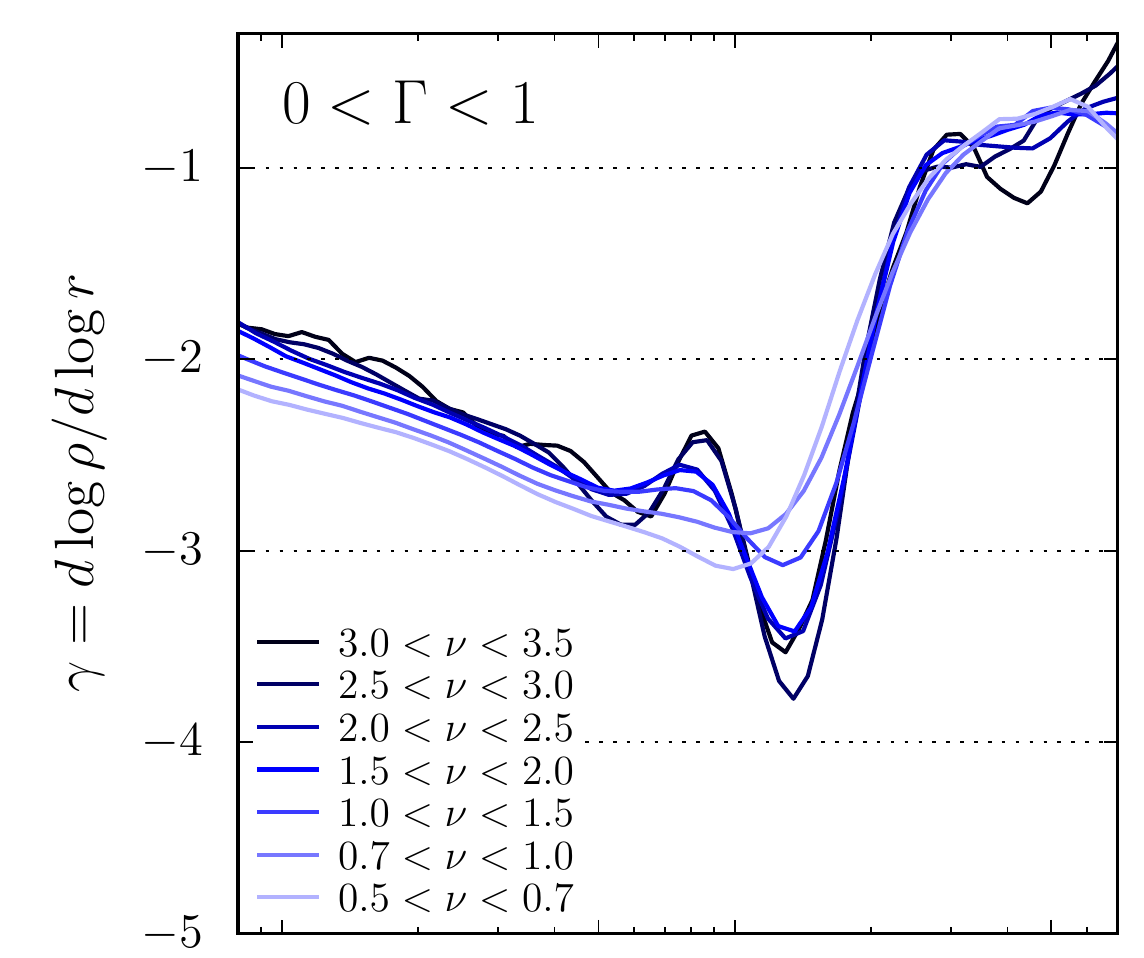}
\includegraphics[trim = 139mm 17mm 2mm 0mm, clip, scale=\panelsize]{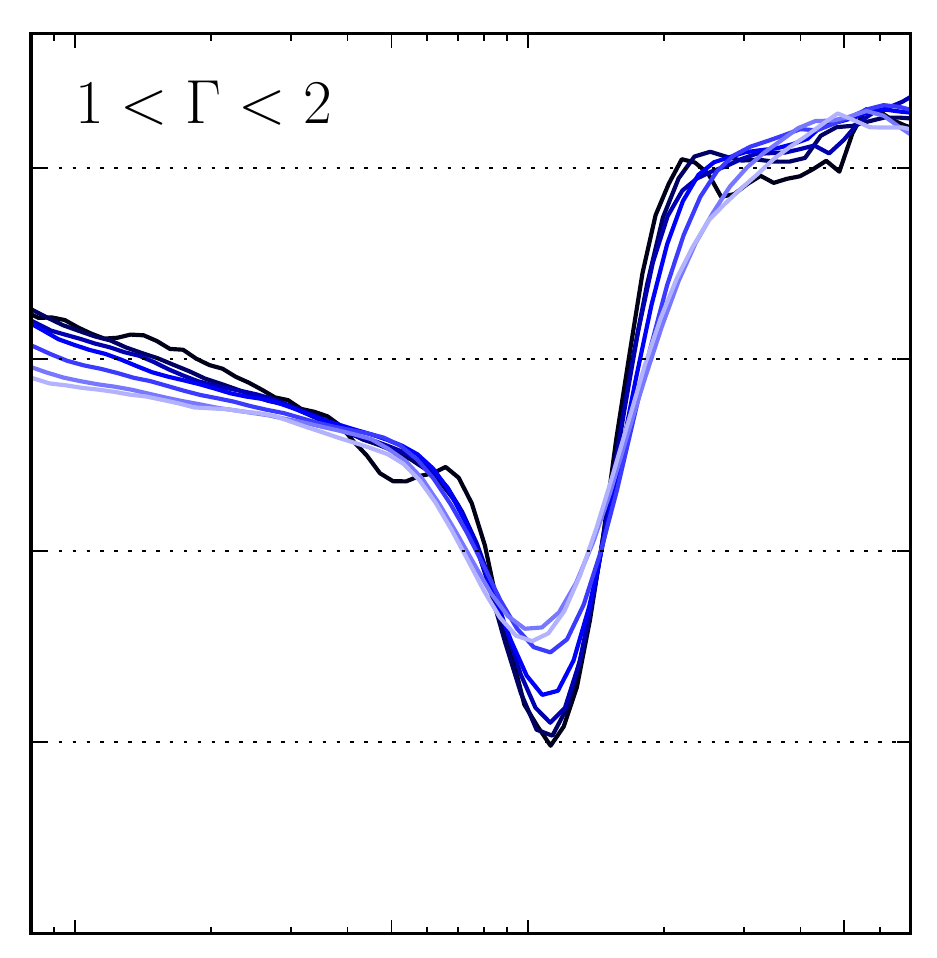}
\includegraphics[trim = 118mm 3mm 2mm 0mm, clip, scale=\panelsize]{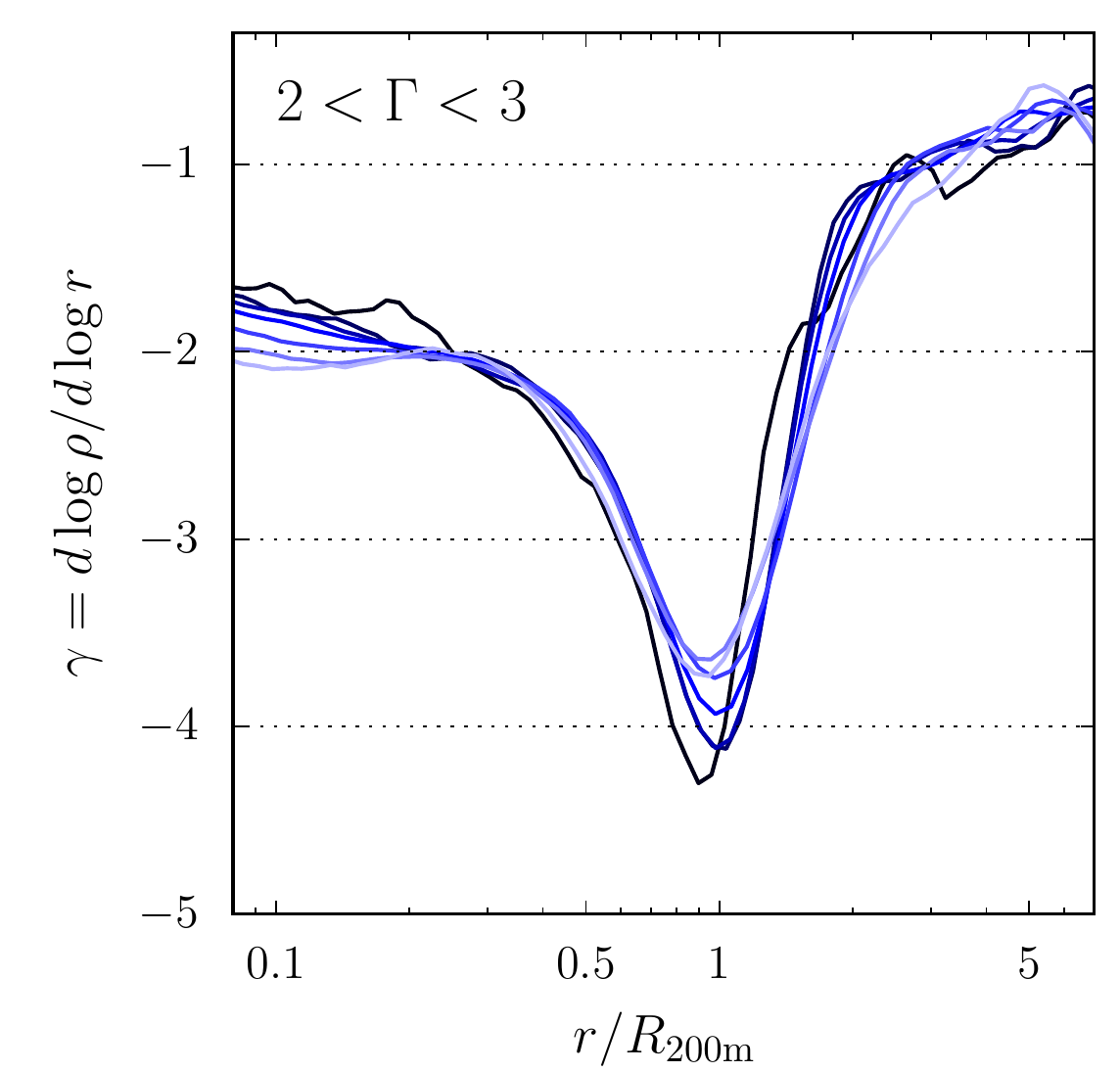}
\includegraphics[trim = 139mm 3mm 2mm 0mm, clip, scale=\panelsize]{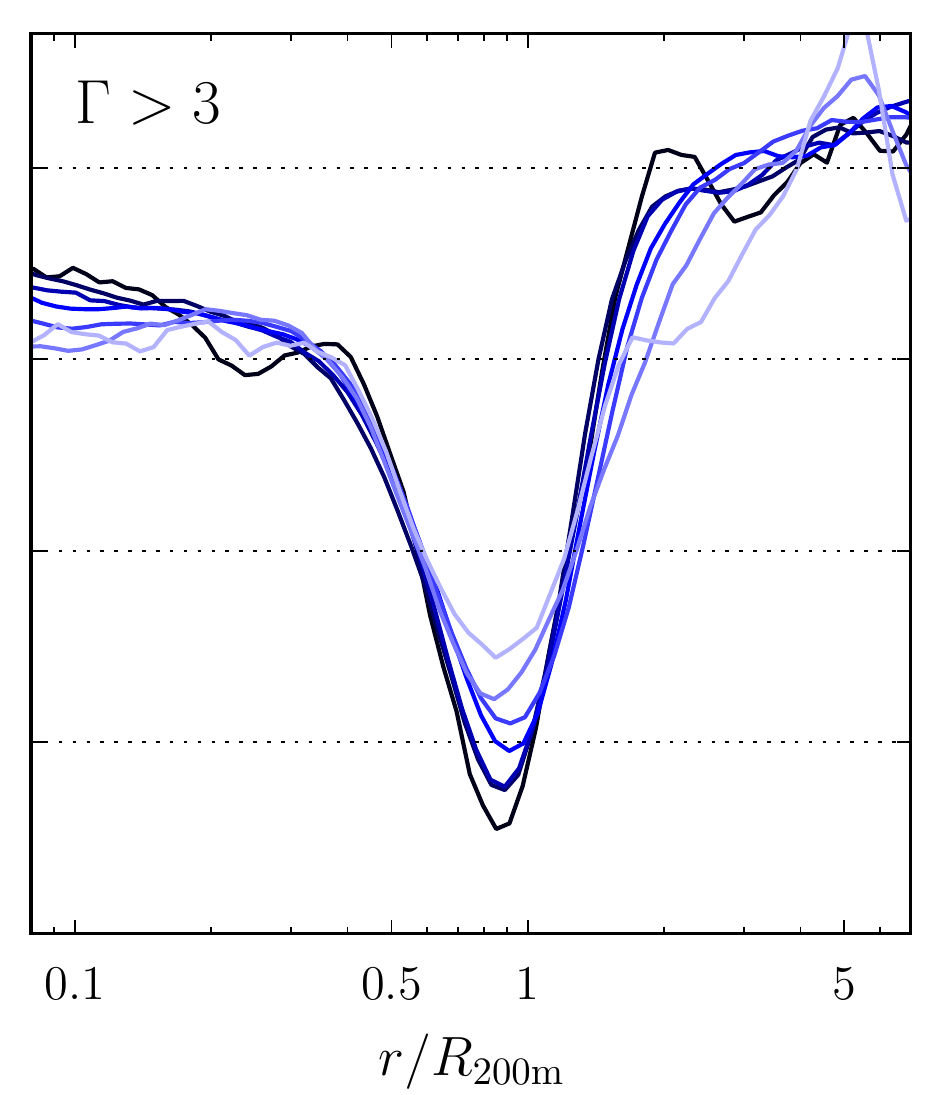}
\caption{Median slope profiles of halo samples with different mass accretion rates $\Gamma = \Delta \log(\mvir) / \Delta \log (a)$; each range of $\Gamma$ is further split into subsamples of different $\nu$. The figure shows that for $\nu \gtrsim 1.5$ the profiles at a given $\Gamma$ become more or less independent of $\nu$. This independence illustrates that the primary cause of the variation in the shape of the outer profiles is a variation in the mass accretion rate. The profiles of the $\nu < 1.5$ halos do show some residual dependence on $\nu$, which we attribute to environment variations around halos of lower peak heights.}
\label{fig:prof_accrate2}
\end{figure*}

In the previous section, we showed that the outer profiles of halos exhibit systematic variations, with their logarithmic slope at $r\approx 0.5-1\rtom$ becoming steeper with increasing peak height, independent of redshift. To understand the origin of this trend we must seek the corresponding physical property of halos that shapes the profiles. One of the most salient differences between halos of different peak height is the degree to which they dominate their environment, and are capable of accreting matter. To this end, we examine the median profiles of halos binned by their mass accretion rate, which we define as
\begin{equation}
\label{eq:gamma}
\Gamma\equiv \Delta\log(\mvir) / \Delta\log(a),
\end{equation}
using the masses of the main progenitor at $z = 0.5$ and its descendant at $z = 0$. We note that halo masses change both due to actual physical accretion and due to changes of the reference density with respect to which the halo radius is defined. The accretion rate $\Gamma$ is thus the sum of the real physical accretion and the so-called pseudo-evolution of mass \citep[][]{diemer_13_pe}. However, for our current purposes we are interested not in the absolute value of the accretion rate but in its relative differences between halos. The contribution to $\Gamma$ due to pseudo-evolution is similar for all halos independent of mass, meaning that a higher $\Gamma$ still implies a higher rate of physical accretion. Thus, the simple definition of $\Gamma$ in Equation (\ref{eq:gamma}) is sufficient for our purposes. We have verified that using an estimate of the physical accretion (based on the minimum estimator of pseudo-evolution defined in \citealt{diemer_13_pe}) leads to qualitatively similar results. 

The top panel of Figure \ref{fig:prof_accrate} shows the median profile of the $1.5 < \nu < 2$ halo sample at $z = 0$. This sample is further split by the accretion rate of halos, $\Gamma$, as indicated in the legend. The figure shows a strikingly clear correlation between mass accretion rate and the steepness of the median outer profile: rapidly accreting halos exhibit steepest slopes as steep as those observed in the highest-$\nu$ bin in Figure \ref{fig:prof_by_nu}, whereas slowly accreting halos reach slopes comparable to those of the median profile of the overall $\nu$ sample. We can also see that the radius at which the steepest slope is reached decreases with increasing accretion rate, although the variation occurs in a rather narrow range around $\rtom$. These differences demonstrate that the median profiles for a given range of $\nu$ are not representative of all halos in that range. Instead, the outer profiles depend on the mass accretion rate. The correlation of the profile shape with $\nu$ is secondary and arises because higher-$\nu$ halos tend to dominate their environment and thus generally have larger mass accretion rates, as shown in Figure \ref{fig:nu_gamma}.

Furthermore, the bottom panel of Figure \ref{fig:prof_accrate} shows the same halo samples as the top panel but excluding halos that underwent a major merger after $z = 0.5$. We have checked that only excluding major mergers after $z = 0.25$ leads to very similar results. A major merger here is defined as a merger of halos with mass ratio larger than $0.3$. It is clear that the profiles in the two panels are very similar. In fact, the profiles of halos without major mergers reach somewhat steeper slopes at $r \approx \rtom$, which may be due to variations in the outer profiles produced by mergers that smooth out features in the median profile. The similarity of the samples with and without major mergers implies that the primary factor in defining the shape of the outer profiles is mass accretion rate, rather than major mergers. In an additional experiment, we verified that selecting halos by the time of their last major merger does not preferentially select profiles with steep outer slopes.

These results highlight an important point: significant growth of halos, in particular in observational analyses of groups and clusters, is often identified with apparent disturbances, such as asymmetries, substructure, deviations from hydrostatic equilibrium, etc. However, real halos grow by a combination of major mergers and the accretion of many low-mass halos. The latter mode of accretion actually dominates at most epochs. An object that appears quite relaxed in its inner regions can thus still be in the process of accreting mass at a high rate because the accretion of many small halos from different directions will not produce strong disturbances typically associated with unrelaxed clusters, for example. 

Additional evidence for the connection between the mass accretion rate and the shape of the outer density profiles is provided by the infall velocity profiles of halos. The top panel of Figure \ref{fig:prof_vr} shows the median radial velocity profiles of the same $\nu$ bins as in Figure \ref{fig:prof_by_nu}, rescaled by $v_{\rm 200m} \equiv (G \mtom / \rtom)^{1/2}$. As could be expected, the high-$\nu$ halos have much more negative (corresponding to infall) average radial velocities than low-$\nu$ halos, even when rescaled to $v_{\rm 200m}$. In fact, the lowest-$\nu$ bin appears to experience no average infall in any radial shell \citep[see also][]{diemand_07_haloevolution, cuesta_08_infall}. The bottom panel of Figure \ref{fig:prof_vr} shows the velocity profiles of the same $1.5 < \nu < 2$ sample as in Figure \ref{fig:prof_accrate}, again split by the mass accretion rate, $\Gamma$. It is clear that for a given mass the halos with the highest $\Gamma$ have a more pronounced infall region compared to the low-$\Gamma$ halos. Interestingly, the maximum infall velocity is reached at radii about a factor of 1.5 larger than the radius where the profiles reach their steepest slope. The latter radius appears to correspond to the radius where the median radial infall velocity approaches zero. Furthermore, the radius of the largest infall velocity shows a similar dependence on $\Gamma$ as the radius where the profiles reach the steepest slope (compare the bottom panels of Figures \ref{fig:prof_accrate} and \ref{fig:prof_vr}).

Finally, Figure \ref{fig:prof_accrate2} shows the median profiles of halos of different peak heights but with a similar accretion rate $\Gamma$. The highest-$\nu$ bin is omitted as it contains too few halos to be split into subsamples. The figure shows that the profiles of halos with a given accretion rate show little variation with $\nu$, except for those samples with the lowest accretion rates and peak heights. For these samples, a significant fraction of systems are located next to bigger systems, and their profiles thus do not reflect the intrinsic shape of the halo profile itself but the contribution from the profiles of their massive neighbors. On the other hand, the higher-$\nu$ systems are relatively isolated on average, and the profiles of halos with $\nu > 1.5$ are independent of $\nu$ for a given range in $\Gamma$. 

The results presented in this section clearly demonstrate that the outer ($0.5 \lesssim r / \rtom \lesssim 2$) density profiles of halos forming in the \LCDM cosmology depend on the halo's mass accretion rate. The profiles are sensitive to the overall mass accretion rate rather than the mass accreted via major mergers. This result opens an interesting possibility of using observational signatures of the mass distribution in galaxies, groups, and clusters to estimate their mass accretion rate (see the discussion in Section \ref{sec:discussion:observable}).

\subsection{Fitting Formula}
\label{sec:results:formula}

Several analytic fitting formulae for the outer halo density profiles have been proposed in the recent literature \citep{prada_06_outerregions, tavio_08, hayashi_08, oguri_11}. However, we find that these forms are not sufficiently flexible to accurately fit the variations of the outer profiles discussed in the previous sections (see Appendix~\ref{sec:app:previous} and Figure \ref{fig:fitfuncs}).  Thus, we developed a new fitting formula to account for the trends and features we observe,
\begin{align}
\label{eq:formula}
\rho(r) &= \rho_{\rm inner} \times f_{\rm trans} + \rho_{\rm outer} \nonumber \\
\rho_{\rm inner} &= \rho_{\rm Einasto} = \rho_{\rm s} \exp \left( -\frac{2}{\alpha} \left[ \left( \frac{r}{r_{\rm s}} \right)^\alpha -1 \right] \right) \nonumber \\
f_{\rm trans} &= \left[ 1 + \left( \frac{r}{r_{\rm t}}\right)^\beta \right]^{-\frac{\gamma}{\beta}} \nonumber \\ 
\rho_{\rm outer} &= \rho_{\rm m} \left[ b_{\rm e} \left( \frac{r}{5\, \rtom} \right)^{-s_{\rm e}} + 1\right] \,.
\end{align}
The inner part of the halo is described by the Einasto profile, which is characterized by three parameters. The transition term, $f_{\rm trans}$, captures the steepening of the profile around a truncation radius, $r_{\rm t}$. The parameters $\gamma$ and $\beta$ define the steepness of the profile at $r\sim\rtom$ and how quickly the slope changes, respectively. Finally, the outermost profile is described by a power law, plus the mean density of the universe, $\rhom$. Our choice of the pivot radius at $5 \rtom$ is somewhat arbitrary, but we have checked that our results are not sensitive to the exact choice in the range of $1-5\rtom$. Profiles with a power law that decreases with radius ($s_{\rm e} > 0$) approach $\rhom$ at sufficiently large radii. Note, however, that the power-law function is only a convenient approximation for the range of radii we are considering here. At larger radii, the profile is not expected to follow a power law, or to reach the mean density until much larger radii. Instead, the profile at $r \gtrsim 9 \rvir$ will follow a shape proportional to the matter correlation function. We discuss alternative ways to parameterize the outer profile based on the 2-halo term in Appendix \ref{sec:app:outer}. For the purposes of describing the profiles at radii $\rvir \lesssim r < 9 \rvir$, we find that a simple power law is accurate, and therefore we adopt it as our fiducial choice due to its relative simplicity.

We first consider the trends of the best-fit parameters as a function of peak height, $\nu$. When varying all eight of the free parameters in Equation (\ref{eq:formula}), the analytic profile fits both the mean and median profiles as a function of $\nu$, for all peak height bins considered in this paper, at all redshifts up to $z = 6$, and at radii between $0.1 \rvir$ and $9 \rvir$, with fractional errors of  $\lesssim 5\%$. 

However, we note that some of the parameters are correlated, indicating that the number of free parameters can be reduced. For example, we can fix the Einasto parameter $\alpha$ to the relation with $\nu$ calibrated by \citet{gao_08},
\begin{equation}
\label{eq:gao}
\alpha(\nu) = 0.155 + 0.0095 \nu^2 \,.
\end{equation}
Furthermore, we find that fixing  $\beta = 4$ and $\gamma = 8$ in the $f_{\rm trans}$ term provides an accurate fit if the truncation radius is related to $\nu$ and $\rtom$ as
\begin{equation}
r_{\rm t}=(1.9 - 0.18 \nu) \times \rtom,
\end{equation}
so that
\begin{equation}
\label{eq:rt_lin}
f_{\rm trans} = \left[ 1 + \left( \frac{r}{(1.9 - 0.18 \nu) \times \rtom}\right)^4 \right]^{-2} \,.
\end{equation}
We find that equally good fits can be obtained by fixing $r_{\rm t}$ but varying $\gamma$ with $\nu$, setting $\gamma = 4 \nu$ and $r_{\rm t} = 1.495 \rtom$. The transition term then takes on the form
\begin{equation}
\label{eq:gamma_lin}
f_{\rm trans} = \left[ 1 + \frac{1}{5} \left( \frac{r}{\rtom}\right)^4 \right]^{-\nu} \,.
\end{equation}
In either case, we vary only four parameters in a fit: the remaining Einasto parameters $\rho_{\rm s}$ and $r_{\rm s}$, and two parameters for the outer profile, $b_{\rm e}$ and $s_{\rm e}$. The shape of the transition region is fixed, with a mild dependence on $\nu$ and no dependence on redshift. This modified fitting function fits the mean and median at all peak heights, redshifts, and radii with fractional errors of $\lesssim 10\%$ (Figure \ref{fig:fits1} in Appendix \ref{sec:app:previous}). 

We note that the transition term has virtually no influence on the best-fit parameters of the Einasto part of the profile. We have compared the concentrations obtained by fitting the Einasto profile to the inner part of the profile only with the concentrations derived from the full fit, and find that the differences are negligible. Thus, one can safely fix $r_{\rm s}$ using a concentration--mass relation without influencing the fits to the outer profiles. Likewise, modifying the dependence of $\alpha$ on peak height (e.g., varying between the \citet{gao_08} and \citet{duffy_08} relations) has very little influence on the best-fit parameters for the outer profile.

\begin{figure}
\centering
\includegraphics[trim = 3mm 3mm 0mm 3mm, clip, scale=0.8]{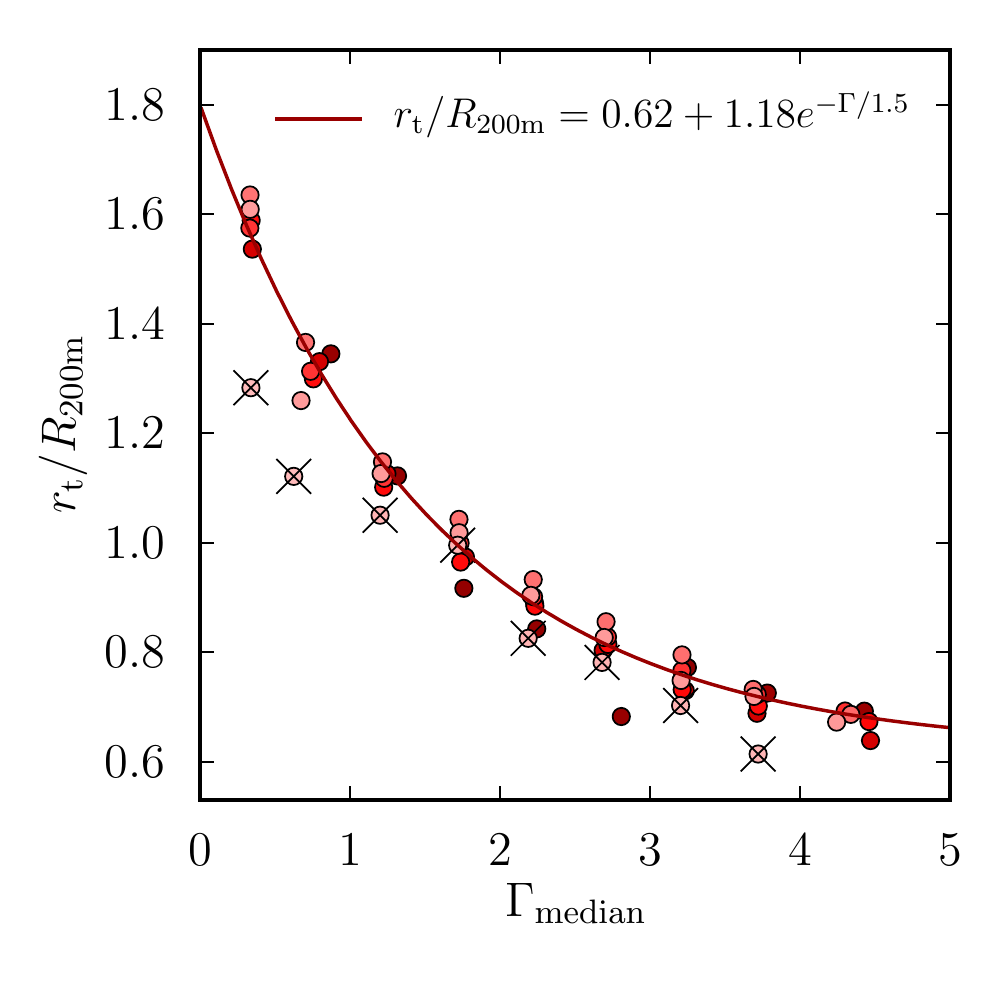}
\caption{Correlation between the median accretion rate of a halo sample, $\Gamma$, and the best-fit truncation radius, $r_{\rm t}$, derived from fits to the median density profile of the sample.  Darker points correspond to higher-$\nu$ samples. The profiles of the lowest-$\nu$ bin (lightest data points, highlighted with black crosses) deviate from the relation somewhat, which could be due to the larger scatter in the outer profiles due to neighboring massive halos.}
\label{fig:gamma_rt}
\end{figure}

We now consider samples of halos binned by their mass accretion rate, $\Gamma$, as well as $\nu$. The median profiles of the $\Gamma$-selected samples show distinct features, such as shallow inner profiles and a sharp downturn at a radius that depends on $\Gamma$ (see Figure \ref{fig:prof_accrate2}). The sharpness of the turnover indicates that $\beta$ may be larger than for the $\nu$-selected samples. Indeed, we obtain accurate fits by fixing $\beta = 6$, $\gamma = 4$, and $\alpha$ according to Equation (\ref{eq:gao}). With these constraints, the fit quality is slightly worse than for the $\nu$-selected samples, but fractional deviations for all samples are still within $\approx 15\%$. Most importantly, the dependence of the radius where the profile steepens on the accretion rate is reflected in the best-fit values for $r_{\rm t}$, which follow a relation with $\Gamma$,
\begin{equation}
\label{eq:gamma_rt}
r_{\rm t} = \left(0.62 + 1.18\, e^{-\Gamma / 1.5} \right) \times \rtom \,.
\end{equation}
This remarkably tight correlation is shown in Figure \ref{fig:gamma_rt}. The corresponding relation between the mean $\Gamma$ of a halo sample and the best-fit $r_{\rm t}$ to the mean profile exhibits slightly more scatter than the median but is well fit by the same relation. Thus, Equation (\ref{eq:gamma_rt}) allows us to infer the accretion rate of a halo sample from a fit to its density profile.

%%%%%%%%%%%%%%%%%%%%%%%%%%%%%%%%%%%%%%%%%%%%%%%%%%%%%%%%%%%%%%%%%%%%%%%%%%
% DISCUSSION
%%%%%%%%%%%%%%%%%%%%%%%%%%%%%%%%%%%%%%%%%%%%%%%%%%%%%%%%%%%%%%%%%%%%%%%%%%

\section{Discussion}
\label{sec:discussion}

The inner density profiles of halos have been the focus of a large number of studies over the past two decades. In this paper, we investigate the outer profile using a large suite of cosmological simulations and show that the density profile at $r \approx 0.5-1\rtom$ exhibits a strong dependence on a halo's mass accretion rate. In particular, the logarithmic slope of the density profile at these radii sharply steepens, and the steepening becomes more pronounced with increasing mass accretion rate. This dependence is not described by the analytic profiles previously proposed in the literature, such as the NFW and Einasto profiles. We propose a new fitting formula to describe the outer profiles and present best-fit parameters as a function of a halo's peak height and mass accretion rate.

Importantly, we find that the outermost density profiles at $r \gtrsim \rtom$ are remarkably self-similar when radii are scaled by $\rtom$ (or, more generally, by any radius around $\rtom$ that is defined with respect to the mean density). This self-similarity indicates that radii defined with respect to the mean density are preferred to describe the structure and evolution of the outer profiles. However, the inner density profiles at smaller radii are most self-similar when radii are scaled by $\rtoc$. In this section, we further discuss some of these findings.

\subsection{The Origin of the Outer Profile Variation}

As we discussed in Section \ref{sec:intro}, secondary infall models predict power-law density profiles with a slope that depends on the slope of the initial density perturbation \citep{fillmore_84}. However, the actual profile resulting from the collapse of peaks is different. Even in the purely self-similar case, the radial orbit instability establishes a break in the power-law profile at a radius of $\approx 0.1 r_{\rm ta}$, where $r_{\rm ta}$ is the turnaround radius \citep{vogelsberger_11}. In a realistic collapse of triaxial density peaks,  adiabatic contraction due to the deepening of the potential during successively collapsing shells further modifies the inner density profile from a power law to an NFW-like form \citep{lithwick_11} {\it with a maximum outer slope of $-3$}. Despite these differences, a common feature of secondary infall solutions to gravitational collapse is the presence of a sharp density jump at an outer radius. This jump corresponds to the apocenter of the most recently accreted particles that have passed through the pericenter of their orbit once since their infall. The density jump is infinitely steep in the spherical collapse case, but has a finite slope in the collapse of triaxial peaks \citep{lithwick_11}. In simulations, such a caustic can be further smoothed out by relaxation due to mergers, interactions with subhalos, etc. Nevertheless, signatures of caustics are detected in cosmological simulations \citep{diemand_08, vogelsberger_09}. 

The position of the caustic in the spherical collapse model depends on the slope of the initial density profile, varying from $r_{\rm caustic} / r_{\rm ta} \approx 0.15$ to $\approx 0.36$ when the slope of the initial density profile varies from flat to steep \citep{vogelsberger_11}. The Gaussian peaks in realistic cosmological initial conditions have shallow inner slopes and steep outer slopes, corresponding to the fast and slow mass accretion regimes of halos \citep{dalal_08}. We can thus expect the radius of the caustic in units of the turnaround radius to increase with decreasing accretion rate. 

We now predict at what radius the caustic should appear at $z = 0$. The turnaround radius for halos in our simulations, defined operationally as the outermost radius where the radial velocity becomes zero, varies from  $\approx 2 \rtom$ to $\approx 2.8 \rtom$, depending on the median $\nu$ of a halo sample. This estimate gives roughly the same physical turnaround radius as shown in Figure 2 of \citet{busha_05} for clusters at $z = 0$. However, for low-$\nu$ halos, the turnaround is not well defined because the velocity profile joins smoothly into the Hubble flow. For $\nu > 1$ halos, on the other hand, the turnaround radius occurs at $\approx 2.8 \rtom$ for the median profile, regardless of $\nu$ (see Figure~\ref{fig:prof_vr}). We can thus expect a caustic at $\approx 0.5-1 \rtom$. This estimate is consistent with the variation of the radius at which the density profile sharply steepens for samples with different accretion rates shown in Figure~\ref{fig:prof_accrate}. This agreement strongly suggests that the steepening of the outer density profile corresponds to the region where particles accumulate on the first apocenter passage after their infall onto a halo. The dependence of this radius on the accretion rate provides motivation for the choice of fits in which we keep the slope parameters in the truncation term of Equation (\ref{eq:formula}), $\beta$ and $\gamma$, fixed but allow the truncation radius $r_{\rm t}$ to vary with peak height and mass accretion rate. 

We note that a similar steepening can also be expected for gas profiles. Indeed, simulations show that in objects that can sustain hot halos, there are two virial shocks: one strong shock at $r\approx 3 \rvir$ and a second, weaker shock at $r \approx \rvir \approx 0.7-0.8 \rtom$ \citep{molnar_09}. This inner shock is thus expected to occur at the same radii where we observe a steepening in the dark matter profiles. Preliminary tests using $N$--body$+$hydrodynamics simulations of individual galaxy clusters indicate that gas density profiles do indeed exhibit such a steepening if the halo is strongly accreting (E. Lau, private communication). 

\subsection{Origin and Implications of the Self-similarity of Profiles}
\label{sec:discussion:selfsimilarity}

The relation between the steepening of the outer halo profile and the caustic associated with the first orbit apocenter, as discussed above, can explain the self-similarity of the outer ($r \gtrsim \rtom$) profile shape when radii are rescaled by $\rtom$. The dynamics of the infall region and the density profile are expected to be universal in units of the turnaround radius, according to self-similar models. In these models, the turnaround radius is a fixed multiple of the radius enclosing a given fixed overdensity with respect to the mean density of the universe. The profile can thus be expected to be self-similar in $r / \rtom$.

The reason for the self-similarity of the {\it inner} density profile when rescaled with $\rho_c$ and $\rtoc$ is less clear, and needs to be investigated further in future studies. This remarkable self-similarity nevertheless has immediate practical implications. First, it justifies using radii defined with respect to the critical density in defining cluster masses and observable properties, as is often done in galaxy cluster studies \citep[e.g.,][]{evrard_08}. Second, it implies that halo concentrations defined using a radius tied to the critical density (e.g., $c_{\rm 200c} = \rtoc / r_s$) should be remarkably universal at fixed $\nu$, as confirmed by \citet{prada_12} and \citet{ludlow_13_cm}.

Finally, Figure \ref{fig:prof_physical} showed that scaling with $\rtom$ or $\rtoc$ absorbs large differences in the actual physical density profiles of halos corresponding to a given $\nu$ at different redshifts. This fact demonstrates that such radii are physically motivated and useful. Abandoning them altogether, as suggested recently by \citet{zemp_14} to remove the pseudo-evolution of mass \citep{diemer_13_pe}, is not warranted, expect perhaps for the smallest mass halos at low redshifts. 

\subsection{Observational Signatures of Halo Mass Accretion Rate}
\label{sec:discussion:observable}

In Section \ref{sec:results:accrate} we showed that the outer halo density profile depends on the mass accretion rate experienced by a halo over the past few billion years. Moreover, the profiles are sensitive to the overall amount of accretion rather than just to major mergers. This correlation potentially opens a new avenue for assessing the dynamical evolution state of halos, not accessible to other commonly used indicators of halo growth, such as structural signatures of mergers and interactions. The important question is thus whether there are signatures of the mass accretion rate that can be detected in observations. 

\begin{figure}
\centering
\includegraphics[trim = 8mm 6mm 0mm 2mm, clip, scale=0.72]{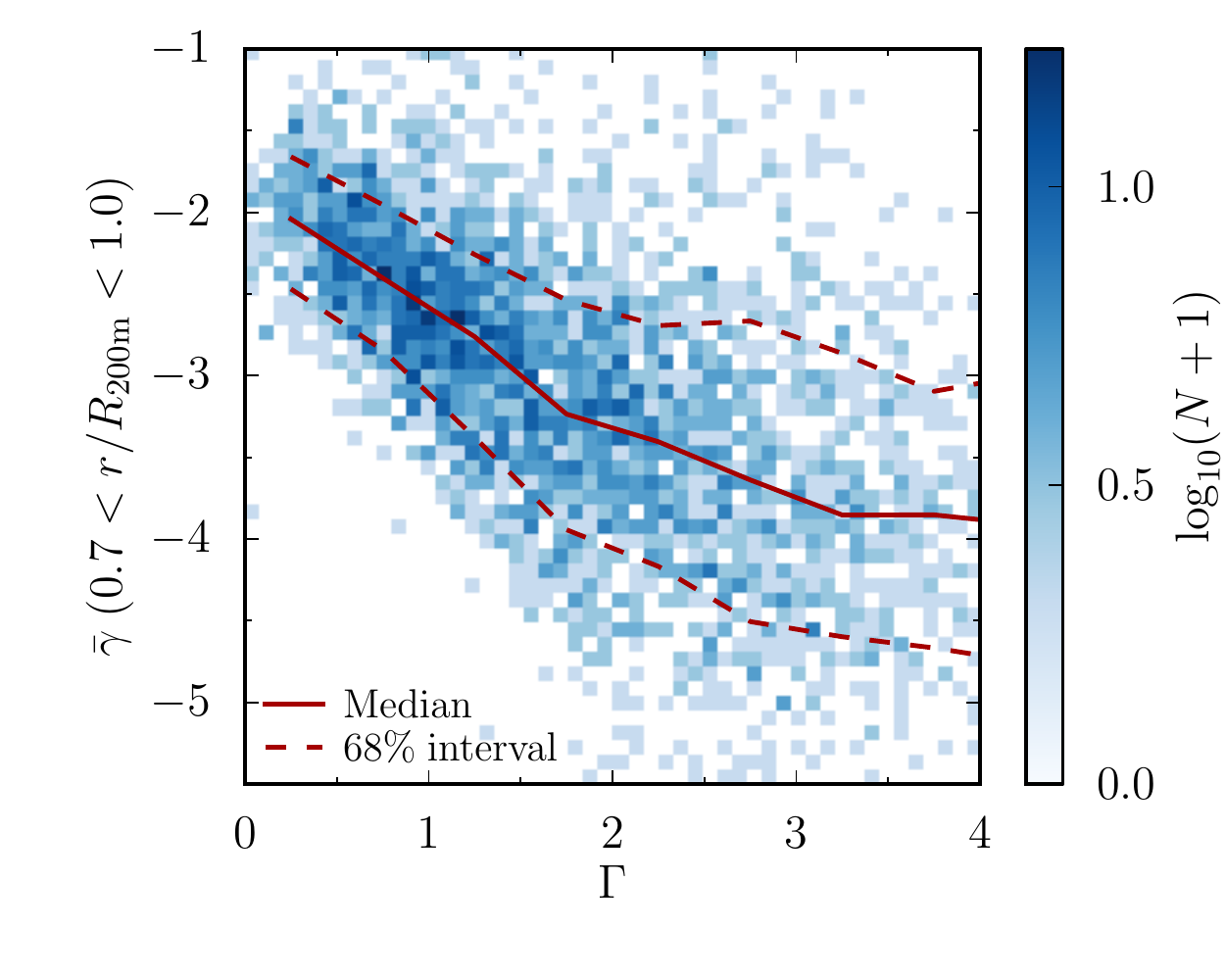}
\caption{Correlation between the accretion rate of a halo, $\Gamma$, and the slope of its density profile for a sample of about $3000$ individual, massive halos ($\mvir > 3 \times 10^{14} \msunh$). The slope is averaged over the radial range $0.7 < r/\rtom < 1$. The slopes follow the same trend observed in the median profiles in Figure \ref{fig:prof_accrate2}, namely, a steepening slope around $\rtom$ with increasing $\Gamma$. The majority of large accretion rates, $\Gamma > 2$, are due to recent major mergers, but excluding such halos barely changes the median trend.}
\label{fig:Gamma_slope}
\end{figure}

First, we must assess whether the trends with $\Gamma$ observed in the median density profiles (Figure \ref{fig:prof_accrate2}) hold for individual halos as well. Figure \ref{fig:Gamma_slope} shows the distribution of slopes around $\rtom$ as a function of $\Gamma$. We focus on the cluster-sized halos, which have the best near-term prospects for measurements of the outer density profiles via X-ray or weak-lensing observations. As with the median profiles in Figure \ref{fig:prof_accrate2}, the slope around $\rtom$ steepens with increasing $\Gamma$, but the relation is subject to significant scatter. In order to reduce the scatter, we plot the average slope in the radial range $0.7 < r / \rtom < 1$. Averaging, however, means that the slopes are somewhat less steep than the steepest slopes in Figure \ref{fig:prof_accrate2}. Although the scatter makes it unlikely that the slope of individual objects can be used for an accurate estimate of their accretion rate, observational measurements of a steep slope (e.g., $\bar{\gamma} \lesssim -4$) would be a distinct signature of a high accretion rate ($\Gamma \gsim 1.5$).

\begin{figure}
\centering
\includegraphics[trim = 12mm 7mm 3mm 2mm, clip, scale=\panelsize]{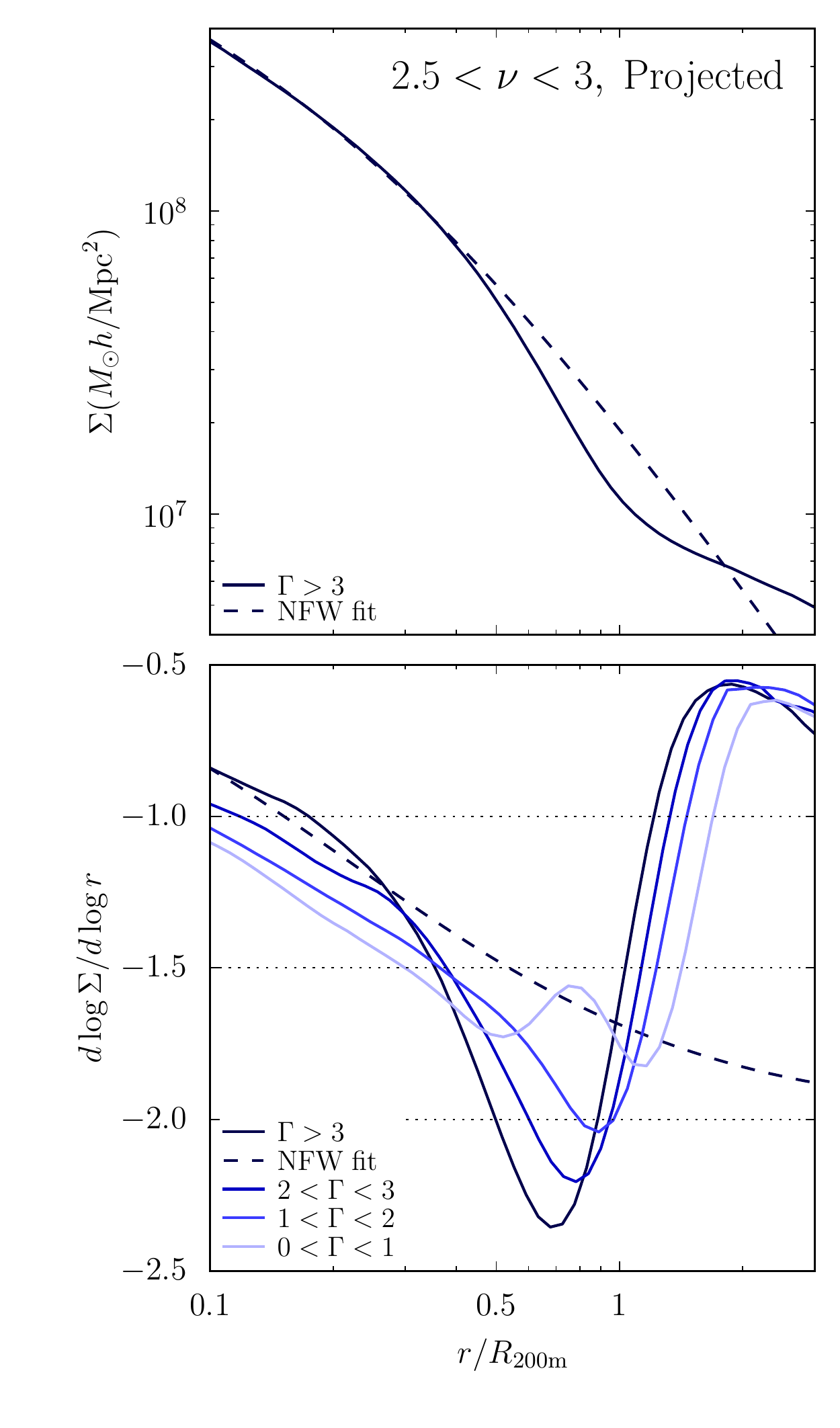}
\caption{Impact of accretion rate on the projected density profiles of halos. We investigate the sample with $2.5 < \nu < 3$ as an example. Top panel: projected surface density profile of the highest-$\Gamma$ bin (solid line) and an NFW fit (dashed line). An analytical expression for the projected NFW profile \citep{bartelmann_96} was fit at $r < 0.5 \rvir$. The $\Gamma > 3$ profile is easily distinguishable from the NFW fit. Bottom panel: logarithmic slopes of the projected density profiles of four $\Gamma$ bins, with the same NFW fit as in the left panel for comparison. While the slopes are shallower in projection than in three dimensions, the effect of the accretion rate is still easily discernible.}
\label{fig:projected}
\end{figure}

Another question is whether the steepening of the profiles would be detectable using weak gravitational lensing, which probes the projected mass distribution. Figure \ref{fig:projected} shows the projected surface density profiles of the median profiles of halo samples with different accretion rates. The three-dimensional profiles were integrated out to $10 \rvir$. For clarity, only the highest-$\Gamma$ bin is shown in the top panel along with the best-fit projected NFW profile. We note that there is no analytical expression for the projected Einasto profile \citep[e.g.,][]{retanamontenegro_12}. However, such a projected profile will be very close to the S\'ersic profile \citep[e.g.,][]{limaneto_99}, and thus an analytic projected profile, equivalent to the density profile of Equation (\ref{eq:formula}), can be constructed.

Even in projection, the profile steepening is clearly visible at radii $r > 0.5 \rtom$. The bottom panel of Figure \ref{fig:projected} shows the slopes of the profile in the top panel, the NFW fit, and the bins with a lower accretion rate for comparison. The dependence of the profile shape on the mass accretion rate is clearly discernible in the projected mass profiles. 
 
\begin{figure}
\centering
\includegraphics[trim = 120mm 2mm 2mm 0mm, clip, scale=\panelsize]{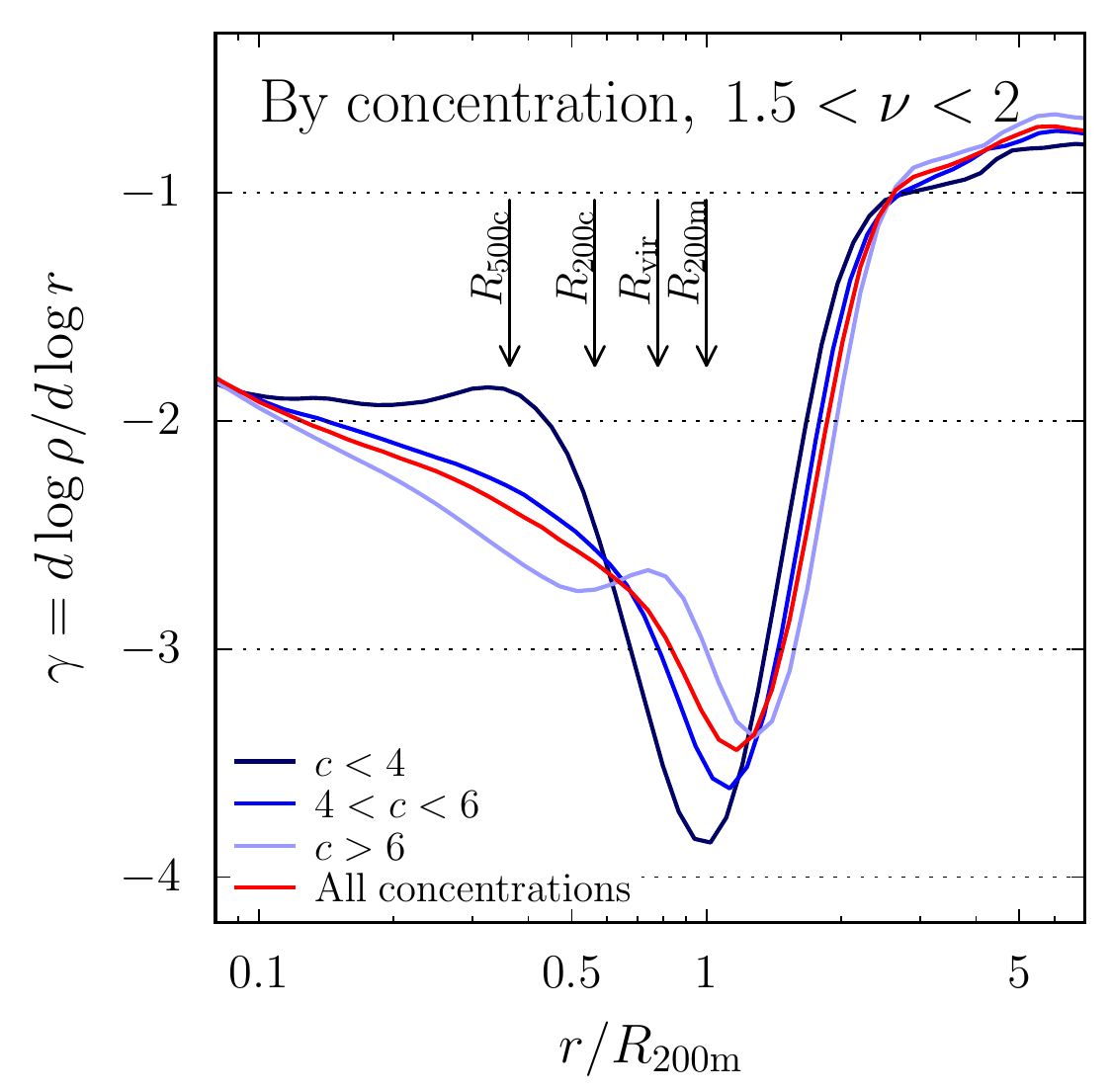}
\caption{Same as Figure \ref{fig:prof_accrate}, but for samples split by concentration. The red line shows the slope of the median density profile of all halos in the $1.5 < \nu < 2$ sample. As concentration correlates with accretion rate, low-concentration halos have steeper outer density profiles. The steepening is not quite as pronounced as when selecting the sample by accretion rate directly, probably due to scatter in the relation between concentration and accretion rate.}
\label{fig:conc}
\end{figure}

There are several other halo properties known to correlate with the mass accretion history of a halo. For example, concentration strongly correlates with the formation epoch of halos \citep{wechsler_02_halo_assembly}. Halos tend to accrete mass at a high rate while they are in the fast accretion regime \citep{zhao_03_mah}, but their accretion slows down at an epoch that can be identified as a halo's formation redshift \citep{wechsler_02_halo_assembly}. While  a halo is in the fast accretion regime, its concentration is approximately constant, $\cvir \approx 4$ \citep{zhao_09_mah}. During the slow accretion regime, halos mostly pseudo-evolve and their scale radius stays constant \citep{bullock_01_halo_profiles}, but the pseudo-evolution of the virial radius leads to an increase of concentration with time \citep{diemer_13_pe}. Thus, concentration is an indicator of how long a halo has been in the slow accretion regime and thus should also correlate with the mass accretion rate (see Figure 7 of \citealt{wechsler_02_halo_assembly}). In fact, this correlation can be inferred from the slope profiles in Figure \ref{fig:prof_accrate}. The profiles of fast-accreting halos exhibit slopes shallower than $-2$ out to large radii and thus have a larger scale radius. All samples share roughly the same $\rvir$ due to the fixed mass bin, meaning that the larger scale radii translate to smaller concentrations. 

We now test whether we can invert this correlation and select halos with steep outer profiles using concentration. Figure \ref{fig:conc} shows the same $\nu$ bin as in Figure \ref{fig:prof_accrate}, but split into three bins in concentration (using $\cvir$ as measured by the halo finder). The lowest concentration sample exhibits a steepening of the outer profile almost as pronounced as the highest-$\Gamma$ bin in Figure \ref{fig:prof_accrate}, whereas the highest concentration bin resembles the slowly accreting halos. The trend with concentration shown in Figure \ref{fig:conc} is also apparent in the mass profiles in Figure 7 of \citet{ludlow_13_mah}. Their paper, however, focuses on the inner regions ($r < \rtom$) and the connection between the shape of the inner profile and a halo's mass accretion history \citep[see also][]{wechsler_02_halo_assembly, zhao_03_mah, zhao_09_mah}.

The steepening around $\rtom$ may lead to a slightly different $\rvir$ at fixed scale radius and thus slightly influence the concentration. However, it is clear from Figures \ref{fig:prof_accrate} and \ref{fig:conc} that the correlation between accretion rate and concentration is mostly driven by the different shape of the inner profile and the resulting differences in scale radius at fixed mass. We thus conclude that concentration as derived from a fit to the inner density profile provides a more or less independent estimate of the mass accretion rate. Both low concentration and steep outer profiles would thus be indications of the high mass accretion rate, and their combination could be used for consistency checks.

%%%%%%%%%%%%%%%%%%%%%%%%%%%%%%%%%%%%%%%%%%%%%%%%%%%%%%%%%%%%%%%%%%%%%%%%%%
% CONCLUSION
%%%%%%%%%%%%%%%%%%%%%%%%%%%%%%%%%%%%%%%%%%%%%%%%%%%%%%%%%%%%%%%%%%%%%%%%%%

\section{Conclusions}
\label{sec:conclusion}

We have presented a detailed study of the outer density profiles of \LCDM halos and shown that they exhibit a strong dependence on the recent mass accretion rate of halos, which we operationally define as $\Gamma = \Delta\log M/\Delta\log a$, measured between $z = 0.5$ and $z = 0$. This dependence means that the density profiles of halos over the entire radial range out to $\rvir$ are not accurately described by a simple universal function, such as the NFW or Einasto profile.  Specifically, our main conclusions are as follows:
\begin{enumerate}
\item The median density profiles of halos exhibit a steepening at $\approx 0.5-1 \rtom$ that becomes more pronounced with increasing peak height, $\nu$, or with increasing mass accretion rate, $\Gamma$, at a fixed $\nu$. The median profiles of halo samples of different $\nu$ or $\Gamma$ reach their steepest slope at $r \approx \rtom$ with values of the logarithmic slope of $\approx -3$ for $\nu \lesssim 1$ or $\Gamma \lesssim 1$, and $\approx -4$ for $\nu \gtrsim 3.5$ and $\Gamma \gtrsim 3$. The steeper the slope at $\rtom$, the larger are the deviations from the NFW and Einasto profiles at $r \gsim 0.5 \rvir$.

\item We find that the slope profiles of the outermost density profiles of halos of a given $\nu$ at $r \gtrsim \rtom$ are remarkably similar at different redshifts, when radii are scaled by $\rtom$. This indicates that radii defined using a fixed overdensity with respect to the mean density are preferred to describe the structure and evolution of the outer density profiles over radii defined using a variable overdensity, such as  $\rvir$, or radii defined with respect to the critical density, such as $\rtoc$. At the same time, we find that the inner density profiles are most self-similar when radii are scaled by $\rtoc$. 

\item We show that the slope of the outer halo profile at $r\approx\rtom$ depends primarily on $\Gamma$, and becomes steeper with increasing mass accretion rate. This dependence induces a corresponding trend with $\nu$, because higher-$\nu$ objects accrete at a higher rate, on average. In addition, higher-$\nu$ halos tend to be more isolated and their outer profiles are less affected by the presence of massive neighbors, and thus exhibit less scatter.

\item We propose a new fitting formula (Equation (\ref{eq:formula})) to describe the outer profiles and present best-fit parameters as a function of $\nu$ and $\Gamma$. We show that this formula provides fits with a fractional accuracy of $\lsim 10-15\%$ for the median and mean profiles of all halo samples and all redshifts we studied.

\item We show that the steepening of the outer profile can, in principle, be observable in the outer mass surface density profiles derived using future weak-lensing analyses. The steep outer profiles at $r\gtrsim 0.5\rtom$ should be accompanied by lower concentrations of the inner profiles at $r\lesssim 0.5\rtom$. 
\end{enumerate}

Our results motivate further work to explore the observational signatures of the steepening slope of the outer profiles. They also indicate possible avenues for improving models of the halo--matter correlation function in the transition region between the virialized portion of a halo's volume and large radii that still evolve in the linear or quasi-linear regime. We leave the exploration of these directions for future work.  

%%%%%%%%%%%%%%%%%%%%%%%%%%%%%%%%%%%%%%%%%%%%%%%%%%%%%%%%%%%%%%%%%%%%%%%%%%
% ACKNOWLEDGMENTS
%%%%%%%%%%%%%%%%%%%%%%%%%%%%%%%%%%%%%%%%%%%%%%%%%%%%%%%%%%%%%%%%%%%%%%%%%%

%\section*{Acknowledgments}
\vspace{0.5cm}

We are grateful to Matt Becker for his assistance with setting up some of our simulations, his analyses using the Rockstar halo finder, and the extraction of the density profiles. A.K. would like to thank Alexey Vikhlinin for fruitful discussions of the outer density profiles of galaxy clusters that prompted this study. We would also like to thank Neal Dalal, Oleg Gnedin, Andrew Hearin, Doug Watson and Surhud More for useful discussions and comments. We thank Peter Behroozi for making his Rockstar halo finder code publicly available. Finally, we thank the referee for a detailed and constructive report. This work was supported by NASA ATP grant NNH12ZDA001N and by the Kavli Institute for Cosmological Physics at the University of Chicago through grants NSF PHY-0551142 and PHY-1125897 and an endowment from the Kavli Foundation and its founder Fred Kavli. We have made extensive use of the NASA Astrophysics Data System and {\tt arXiv.org} preprint server. The simulations used in this study have been carried out using the {\tt midway} computing cluster supported by the University of Chicago Research Computing Center.

%%%%%%%%%%%%%%%%%%%%%%%%%%%%%%%%%%%%%%%%%%%%%%%%%%%%%%%%%%%%%%%%%%%%%%%%%%
% APPENDIX
%%%%%%%%%%%%%%%%%%%%%%%%%%%%%%%%%%%%%%%%%%%%%%%%%%%%%%%%%%%%%%%%%%%%%%%%%%

\appendix

\section{Fitting Formula}
\label{sec:app:formula}

In this Appendix, we briefly review a number of fitting formulae for halo density profiles previously proposed in the literature, and we illustrate why they fail to reproduce the steepening of the outer profiles discussed in this paper. We discuss the new fitting formula introduced in Section \ref{sec:results:formula}, quantify its accuracy, and discuss alternative parameterizations of the outermost profile. 

\subsection{Comparison with Previous Work}
\label{sec:app:previous}

For the inner part of halo density profiles, $r \lsim \rvir$, the NFW \citep{navarro_97_nfw} and \citet{einasto_65} profiles are most commonly used. \citet{prada_06_outerregions} proposed to improve the Einasto profile by adding the mean matter density $\rhom$ to account for the outer parts. \citet{tavio_08} extended this idea by using the NFW profile, $\rhom$, and two more terms to describe a cutoff around $\rvir$ and the transition to $\rhom$. 

In principle, in the framework of the halo model, the outermost profile should be related to the 2-halo term of the halo--matter correlation function \citep[e.g.,][]{smith_03_powerspec, hayashi_08}, as
\begin{equation}
\label{eq:2halo}
\rho_{\rm 2h}(r) = \left[ b(\nu) \xi_{\rm lin}(r) + 1 \right] \rhom
\end{equation}
where $b(\nu)$ is the peak-height-dependent bias \citep[e.g.,][]{sheth_99_bias, tinker_10_bias}. $\xi_{\rm lin}$ is the linear matter--matter correlation function, which can be computed from the linear power spectrum as 
\begin{equation}
\xi_{\rm lin}(r) = \frac{1}{2 \pi^2} \int_0^{\infty} k^2 P(k) \frac{\sin(kr)}{kr} dk \,.
\end{equation}
The mean profile is, by definition, guaranteed to approach the 2-halo term at some radius where the 1-halo term becomes negligible.

\begin{figure}
\centering
\includegraphics[trim = 3mm 8mm 120mm 0mm, clip, scale=0.72]{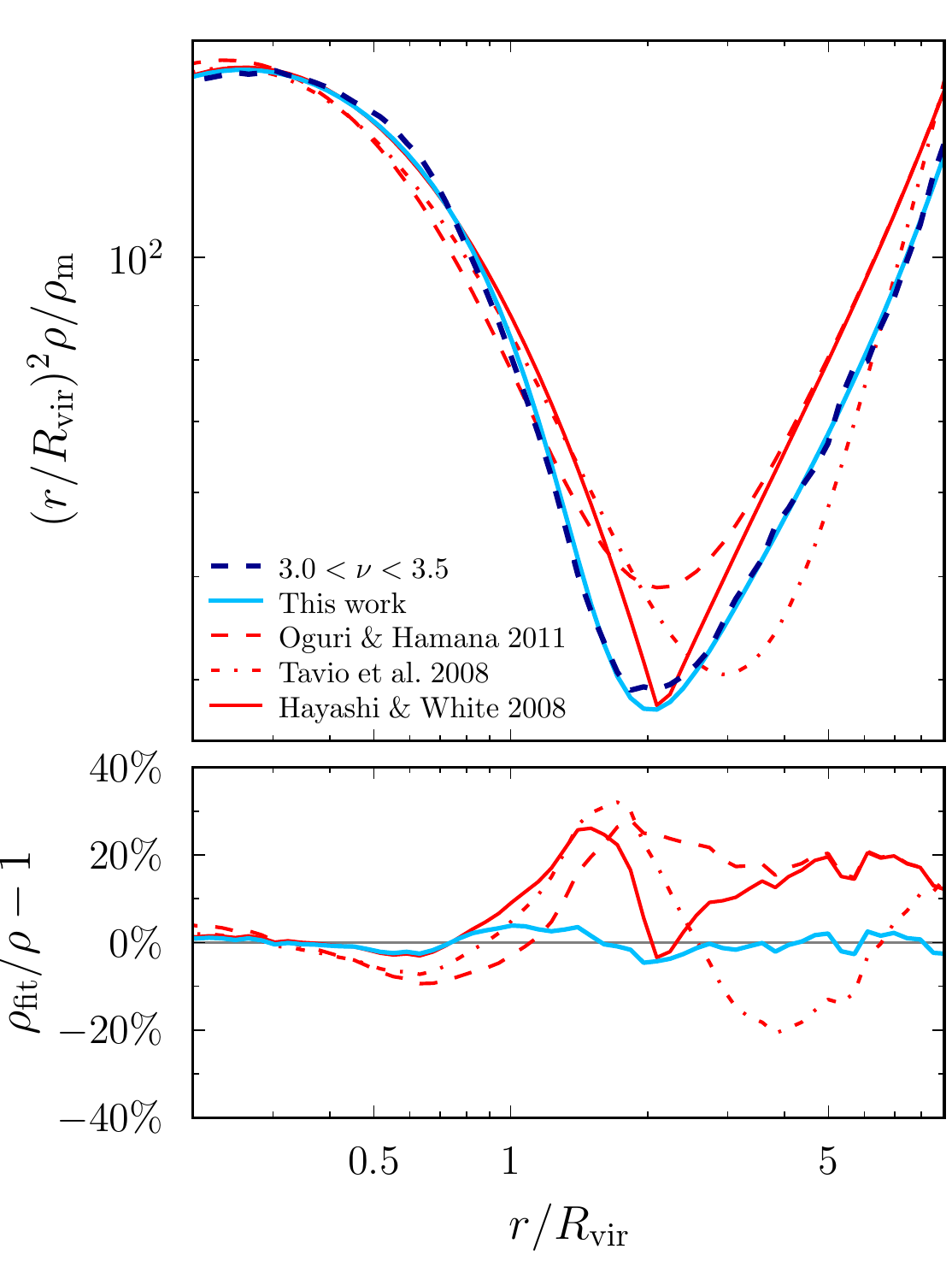}
\caption{Fits to the mean profile of halos with $3 < \nu < 3.5$. The plot shows only fitting functions that were designed to fit the outer halo density profile. The \citet[][red dot-dashed]{tavio_08} function captures the steepening but does not have the freedom to shift it to the correct radius. The fit was performed with four free parameters, rather than the one-parameter version also described in their paper. The functions of \citet[][red solid]{hayashi_08} and \citet[][red dashed]{oguri_11} match the overall shape well but rely on the 2-halo term for the outer regions, which overestimates this particular mean profile significantly. The fitting formula proposed in this paper alleviates these issues (light blue). The fit shown here was performed with fixed $\alpha$, $\beta = 4$, $\gamma = 8$, and $r_{\rm t}$ according to Equation (\ref{eq:rt_lin}) and using Equation (\ref{eq:2h1}) for the outer profile.}
\label{fig:fitfuncs}
\end{figure}

\begin{figure*}
\centering
\includegraphics[trim = 2mm 25mm 0mm 3mm, clip, scale=0.5]{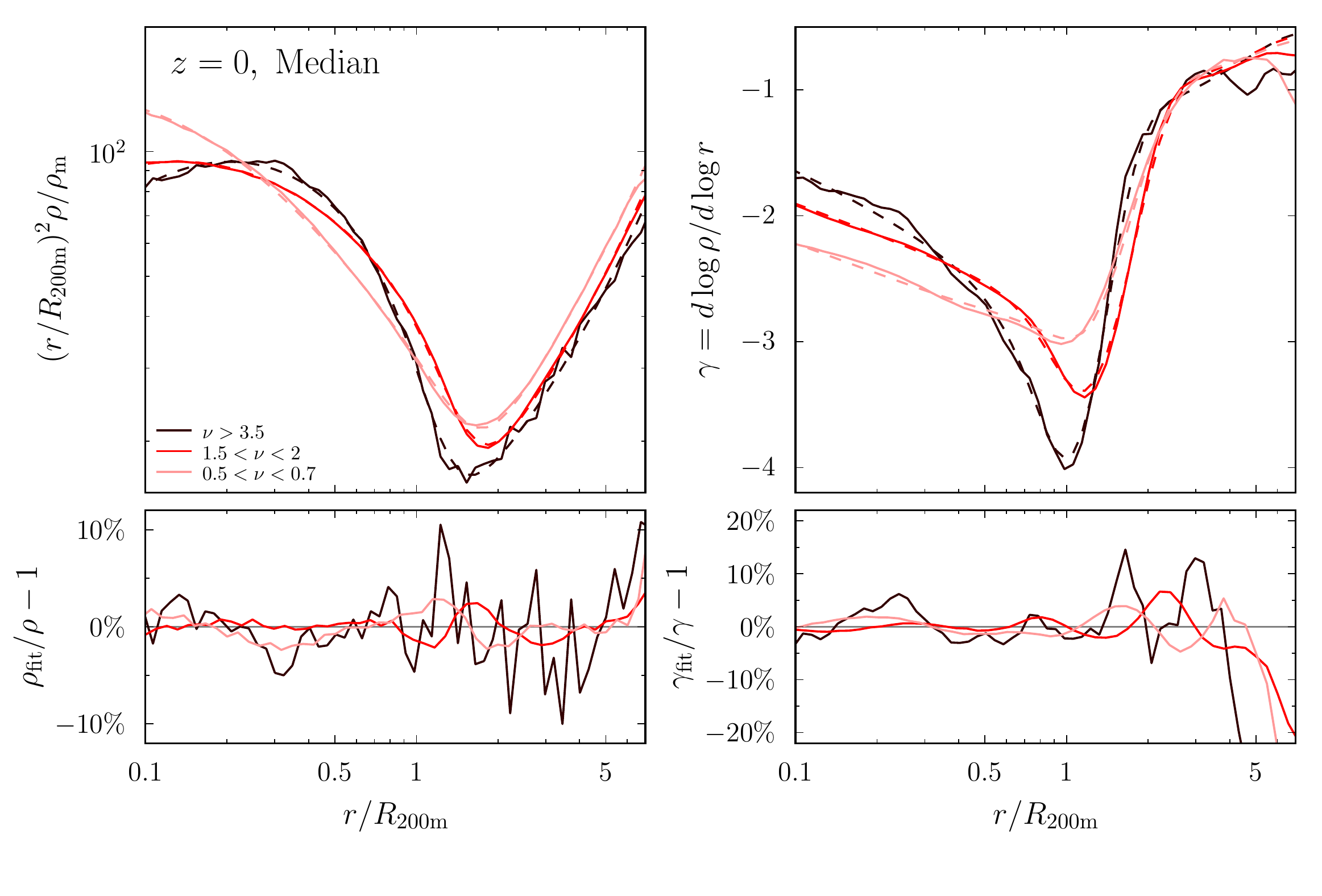}
\includegraphics[trim = 2mm 25mm 120mm 3mm, clip, scale=0.5]{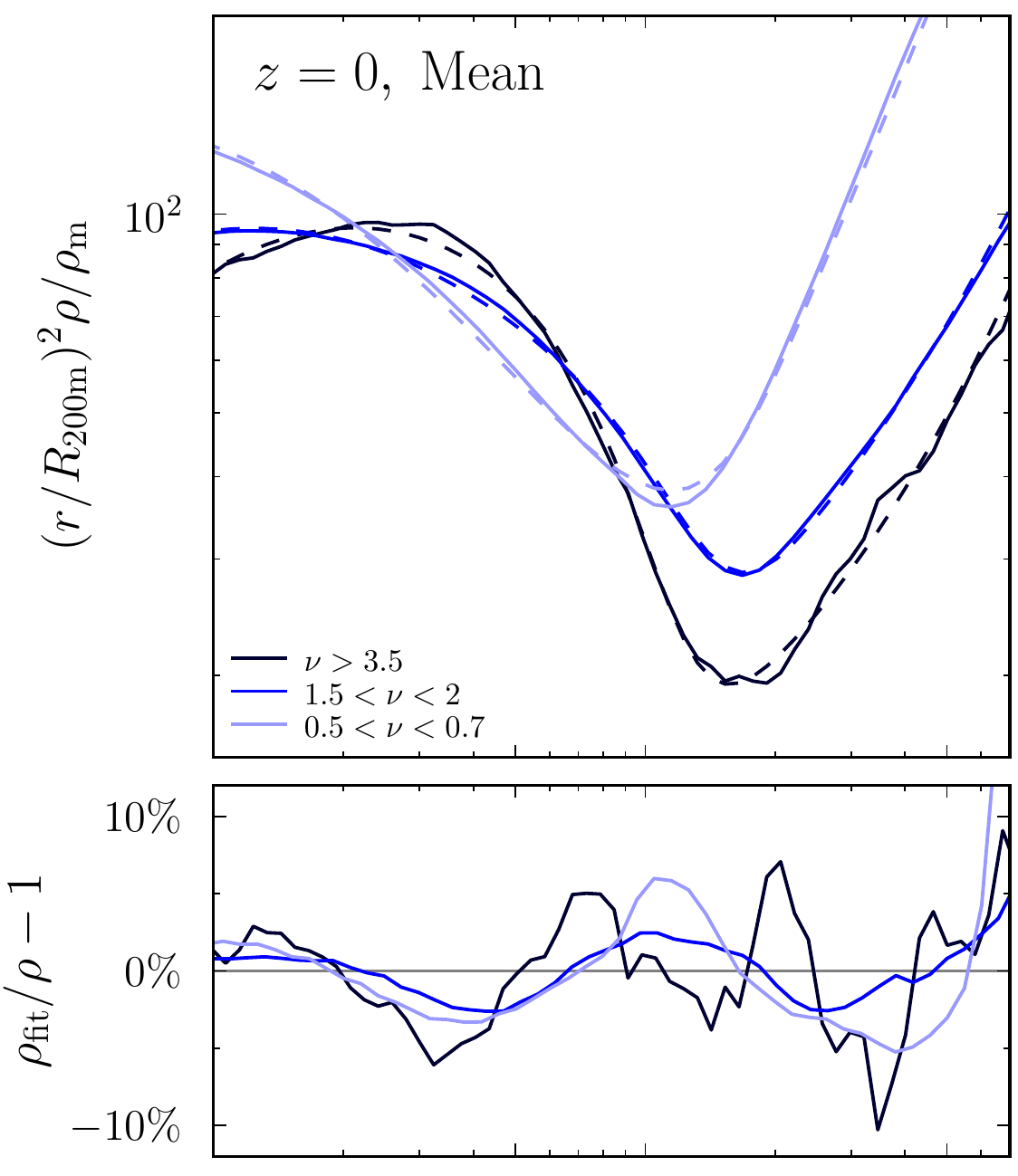}
\includegraphics[trim = 2mm 25mm 0mm 91mm, clip, scale=0.5]{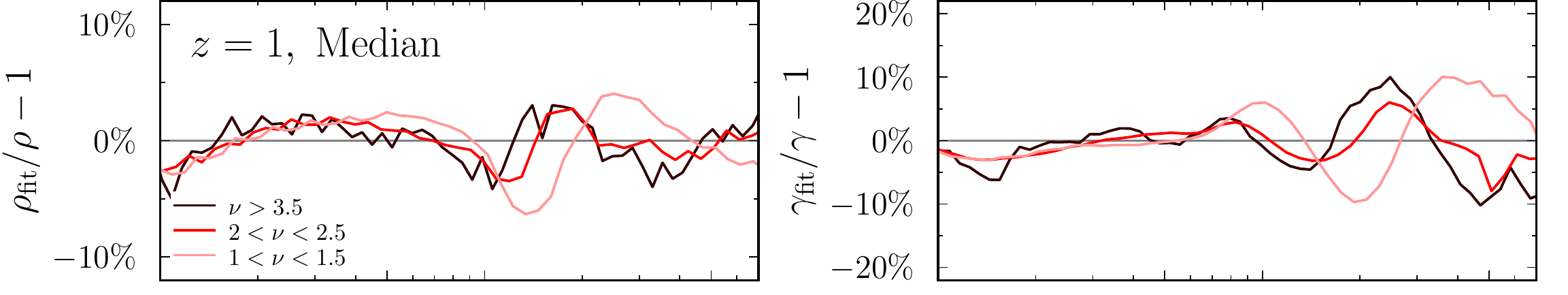}
\includegraphics[trim = 2mm 25mm 120mm 91mm, clip, scale=0.5]{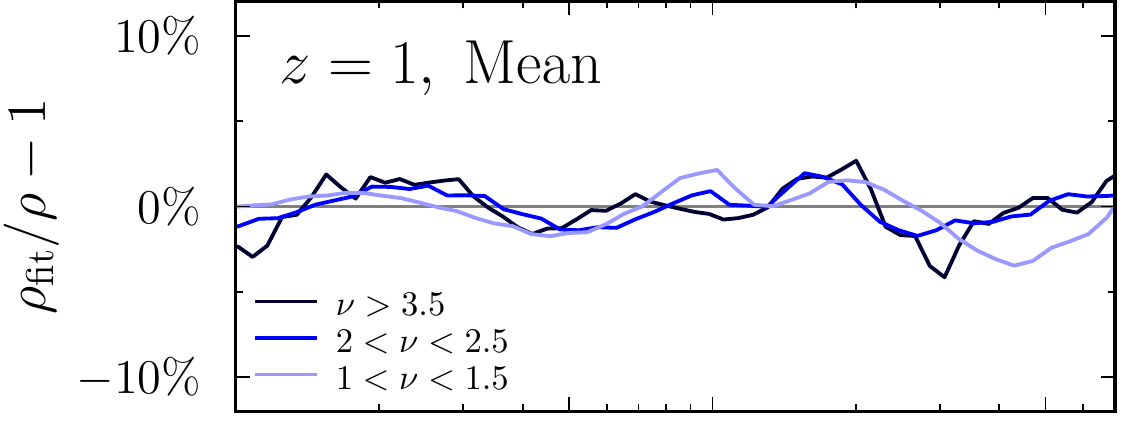}
\includegraphics[trim = 2mm 25mm 0mm 91mm, clip, scale=0.5]{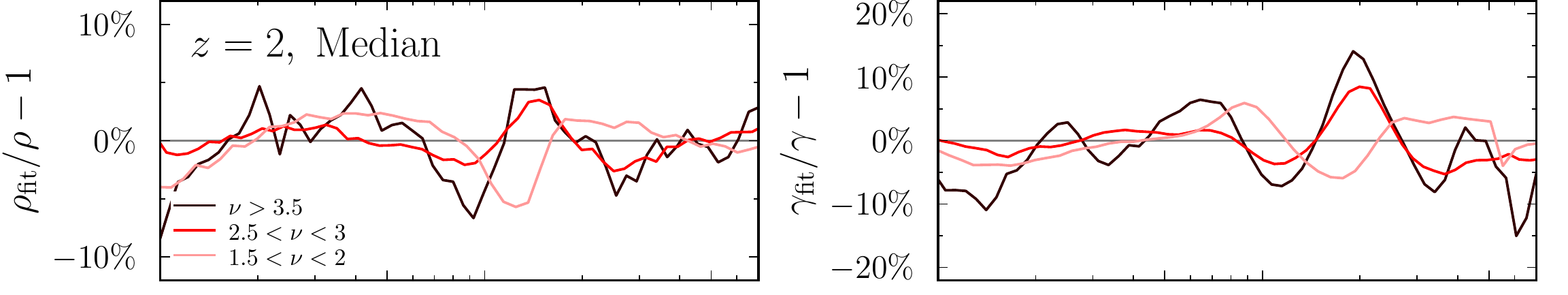}
\includegraphics[trim = 2mm 25mm 120mm 91mm, clip, scale=0.5]{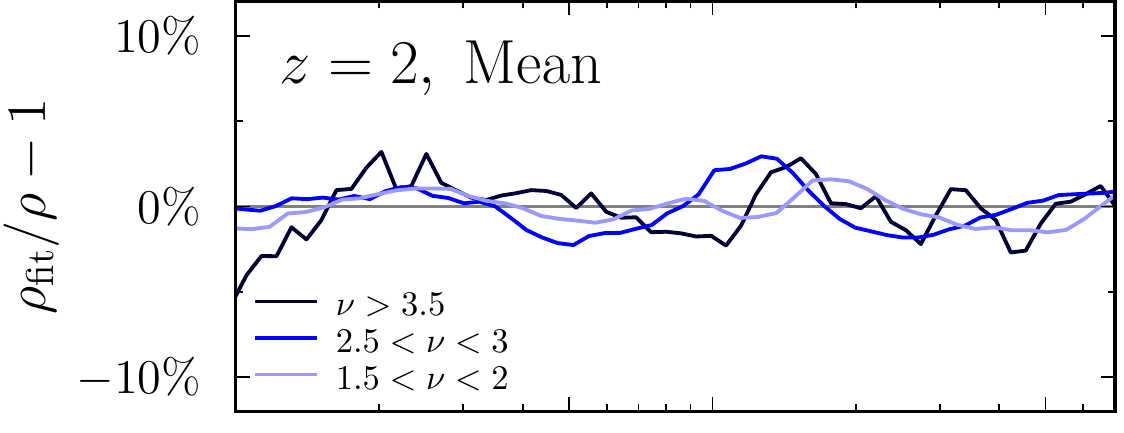}
\includegraphics[trim = 2mm 8mm 0mm 91mm, clip, scale=0.5]{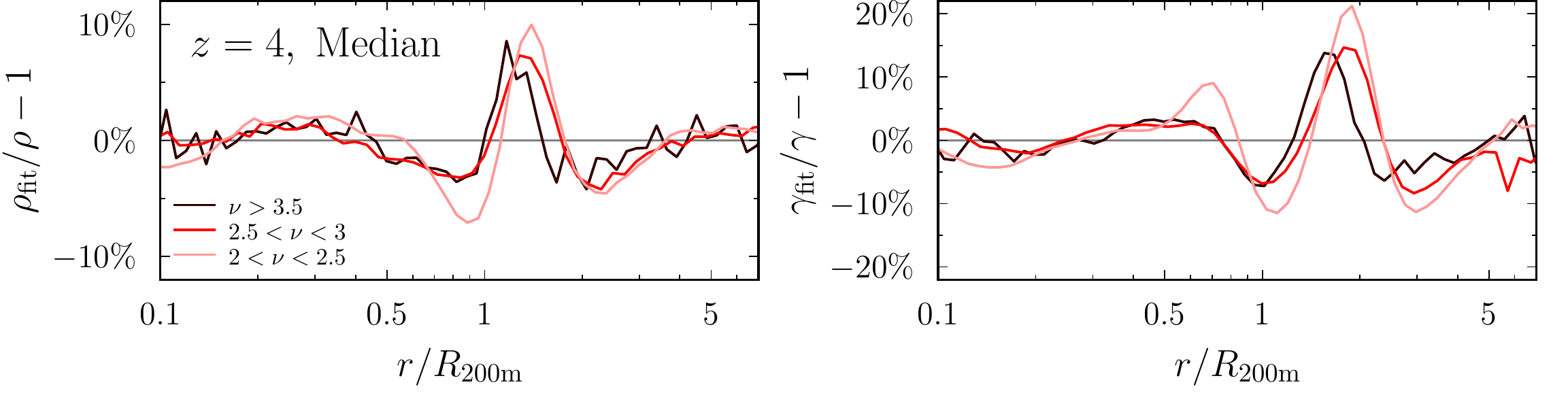}
\includegraphics[trim = 2mm 8mm 120mm 91mm, clip, scale=0.5]{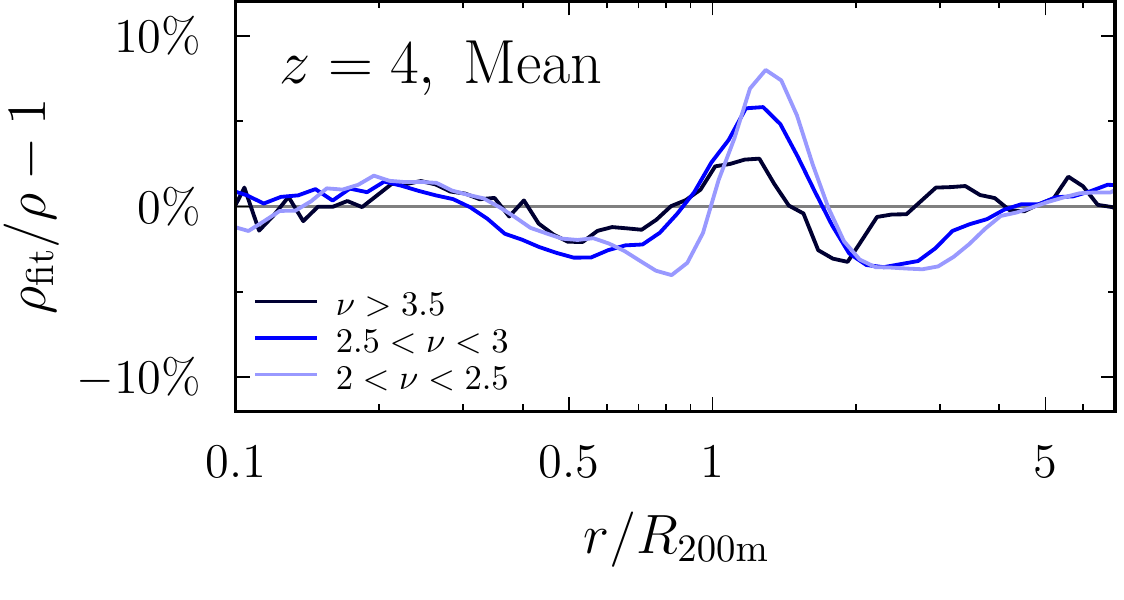}
\caption{Fits of our fitting function (Equation (\ref{eq:formula}), dashed lines) to the median (left column) and mean (right column) profiles of various peak height bins at redshifts 0, 1, 2 and 4. The center column shows the slopes of the median profiles, the slopes of the fits, and the difference between them (the slope profiles were not used in the fit, however). Only $\rho_{\rm s}$, $r_{\rm s}$, $b_{\rm e}$, and $s_{\rm e}$ are varied, while $\alpha$, $\beta$, $\gamma$, and $r_{\rm t}$ are fixed according to Equation (\ref{eq:rt_lin}). The actual profiles are only shown for $z = 0$. Note the larger scale of the slope difference panels in the center column.}
\label{fig:fits1}
\end{figure*}

\begin{figure*}
\centering
\includegraphics[trim = 2mm 25mm 120mm 91mm, clip, scale=0.5]{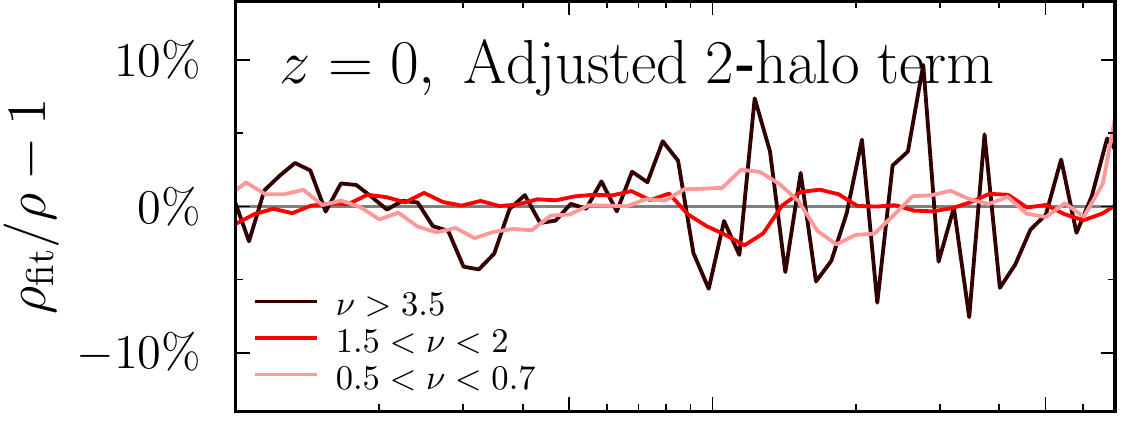}
\includegraphics[trim = 2mm 25mm 120mm 91mm, clip, scale=0.5]{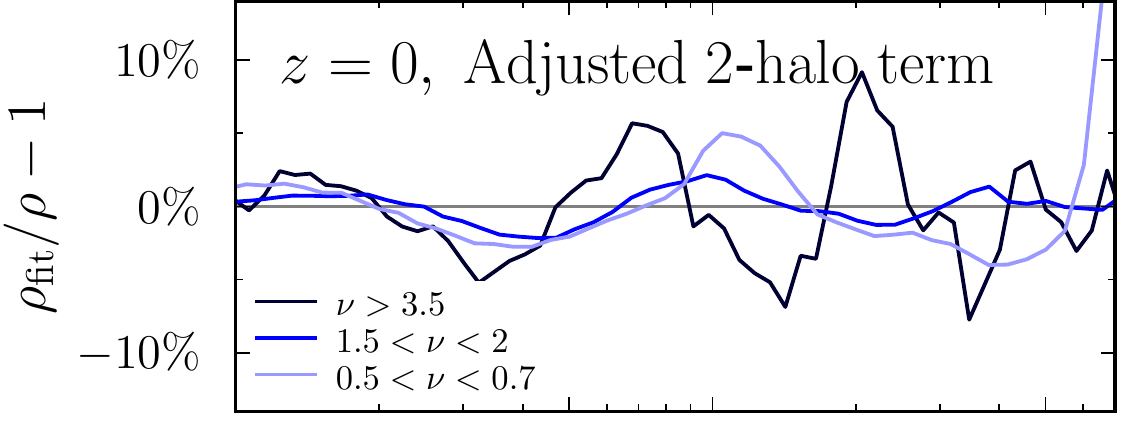}
\includegraphics[trim = 2mm 25mm 120mm 91mm, clip, scale=0.5]{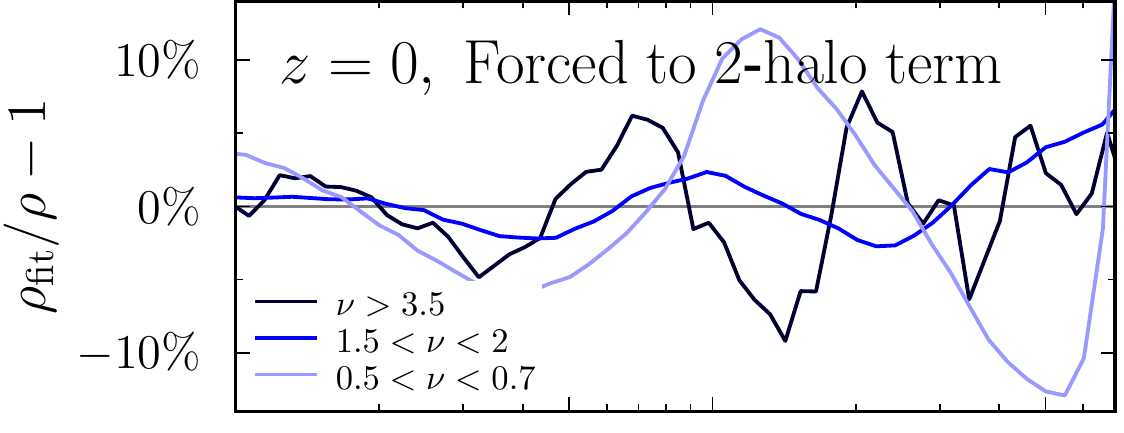}
\includegraphics[trim = 2mm 8mm 120mm 91mm, clip, scale=0.5]{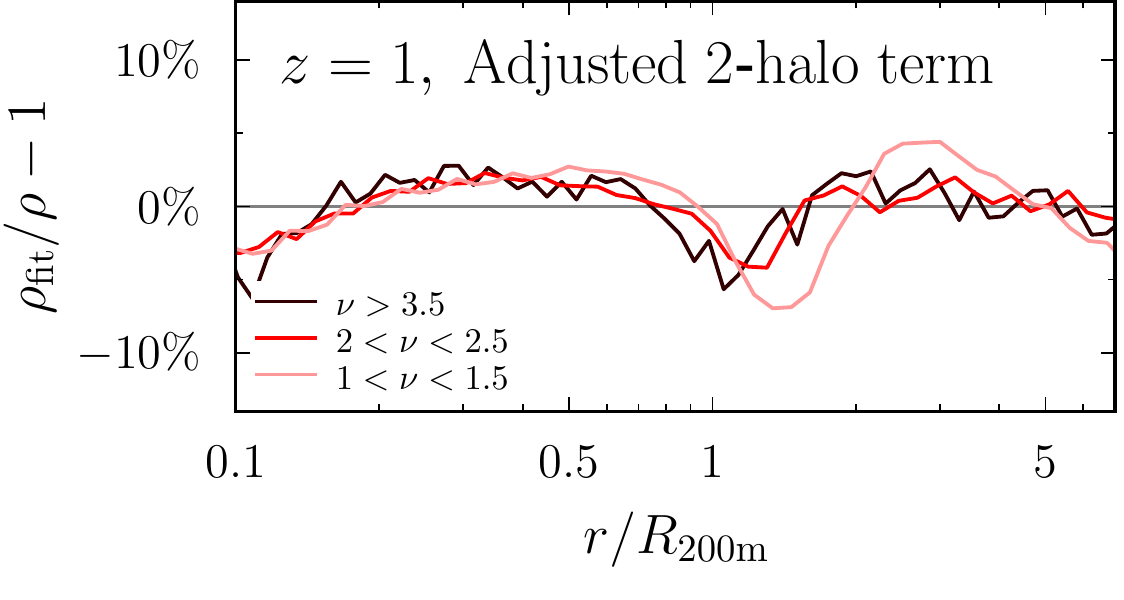}
\includegraphics[trim = 2mm 8mm 120mm 91mm, clip, scale=0.5]{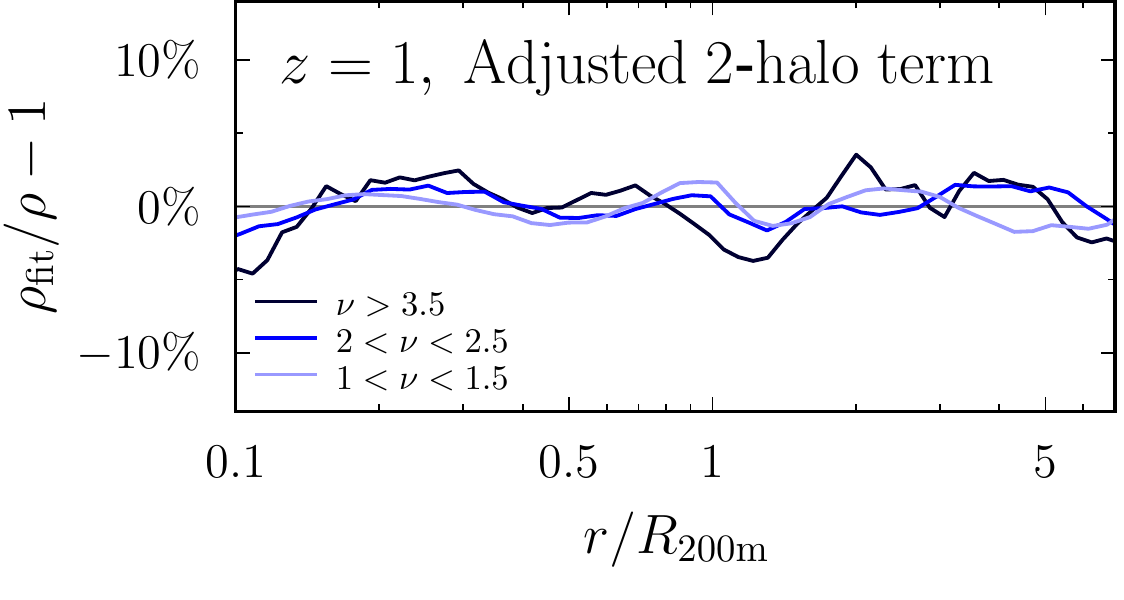}
\includegraphics[trim = 2mm 8mm 120mm 91mm, clip, scale=0.5]{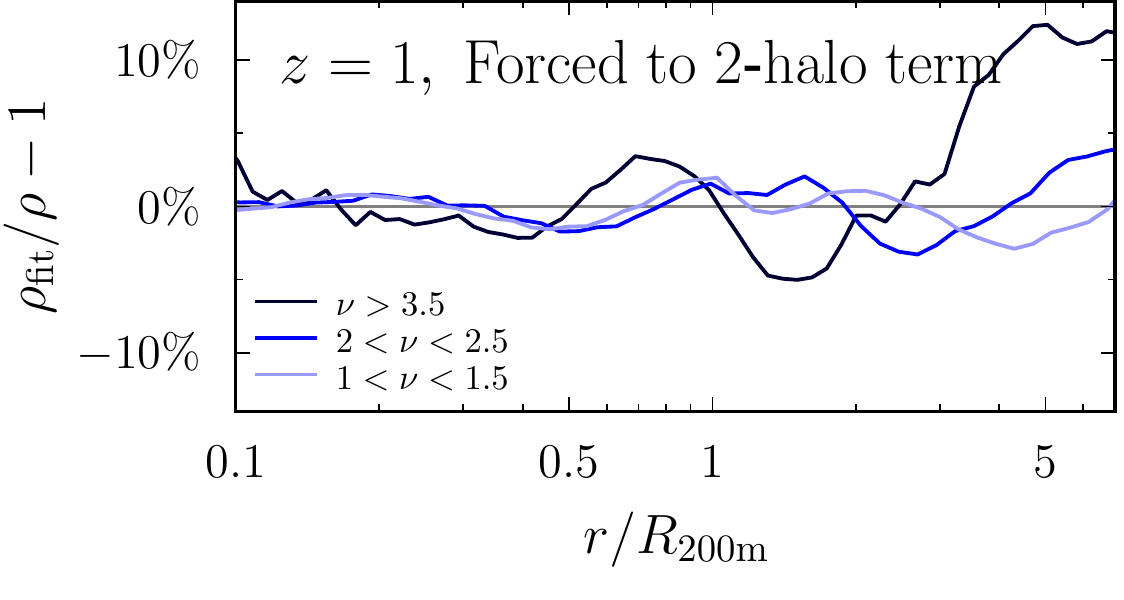}
\caption{Same as Figure \ref{fig:fits1}, but for different parameterizations on the outer profile. Only the fractional differences of the profiles are shown. Left column: median, fitted with the adjusted 2-halo term (Equation (\ref{eq:2h1})) as the outer profile. The fits are slightly better than using a pure power law (compared to Figure \ref{fig:fits1}). Center column: same as the left column, but for the mean. Right column: mean, fitted with an outer profile that is forced to converge to the 2-halo term at large radii (Equation (\ref{eq:2h2})). Particularly the lowest-$\nu$ bin at $z = 0$ is poorly fit by the 2-halo term as its profile diverges from this term with radius at $r < 9 \rvir$, possibly due to sample variance. We do not show fits of Equation (\ref{eq:2h2}) to the median profiles, as the 2-halo term is not expected to be a good description of the median.}
\label{fig:fits2}
\end{figure*}

\citet{hayashi_08} proposed to fit density profiles with the maximum of the 1-halo and 2-halo terms, where the 1-halo term can be represented by either the NFW or Einasto profile. \citet{oguri_11} continued in this spirit but proposed adding the 1- and 2-halo terms rather than taking their maximum, rendering the function differentiable, which is desirable when computing the weak-lensing properties of a profile. They model the 1-halo term using the profile of \citet{baltz_09}, given by an NFW profile multiplied by a truncation term. We note that the truncation term used in \citet{oguri_11} corresponds to the transition term in Equation (\ref{eq:formula}), but with fixed values of $\beta = 2$ and $\gamma = 4$. 

Figure \ref{fig:fitfuncs} shows fits of some of the functional forms described above to the mean density profile of halos with $3.0 < \nu < 3.5$ at $z = 0$ (the fit results for the median profile of this sample are similar). While the functions can fit the inner regions of the density profile well, none of them have sufficient flexibility to accurately match the shape of the outer profile. For example, the function advocated by \citet{tavio_08} reaches errors of up to $\approx 35\%$ at $r\sim 1-2\rtom$. The function of \citet[][not plotted in Figure \ref{fig:fitfuncs}]{prada_06_outerregions} approaches the mean density at large radii, but $\rhom$ underestimates the true profile dramatically. Finally, the 2-halo term as computed from Equation (\ref{eq:2halo}) describes the particular mean halo profile in Figure \ref{fig:fitfuncs} with an accuracy of only $\approx 20\%$. In many cases, the 2-halo term overestimates the median profiles by even larger margins.

All fits shown in Figure \ref{fig:fitfuncs}, as well as those shown in the following figures, were performed over the radial range $0.1 \rvir < r < 9 \rvir$. The fits in Figure \ref{fig:fitfuncs} were derived by minimizing $\Delta(r^2\rho)$. If we instead minimize $\Delta \rho / \rho$, the previously proposed fitting functions fit the outer profile slightly better, but at the expense of accuracy in the inner regions. For the functional profile form we propose in this study, the difference between fits with different merit functions is small, but using $\Delta \rho / \rho$ places somewhat more emphasis on the transition region at $\sim 1-3\rtom$, where $r^2\rho$ is smaller by an order of magnitude compared to the inner and outer radii. As we are particularly interested in this region, all fits except for those shown in Figure \ref{fig:fitfuncs} are performed by minimizing $\Delta \rho / \rho$. We have verified that minimizing $\Delta(r^2\rho)$ does not systematically change any of the best-fit parameters or conclusions.

With its eight free parameters, our function can fit the density profiles to better than 5\% error at almost all radii, with some deviations to about 5\% around the steepest part at higher redshift. However, in Section \ref{sec:results:formula} we claimed that the number of free parameters can be reduced to four without a significant loss in fit quality. Figure \ref{fig:fits1} shows fits to the mean and median samples of halos with various peak heights, using our fitting function (Equation (\ref{eq:formula})) and the linear relation between $\nu$ and $r_{\rm t}$ (Equation (\ref{eq:rt_lin})). The fits match the true profiles to better than $\approx 10\%$ at virtually all radii, redshifts, and peak heights. At $z=6$, we observe deviations slightly larger than $10\%$ at the radius of the steepest slope.

\subsection{The Outer Profile and the 2-halo Term}
\label{sec:app:outer}

\begin{figure}
\centering
\includegraphics[trim = 0mm 2mm 2mm 0mm, clip, scale=0.6]{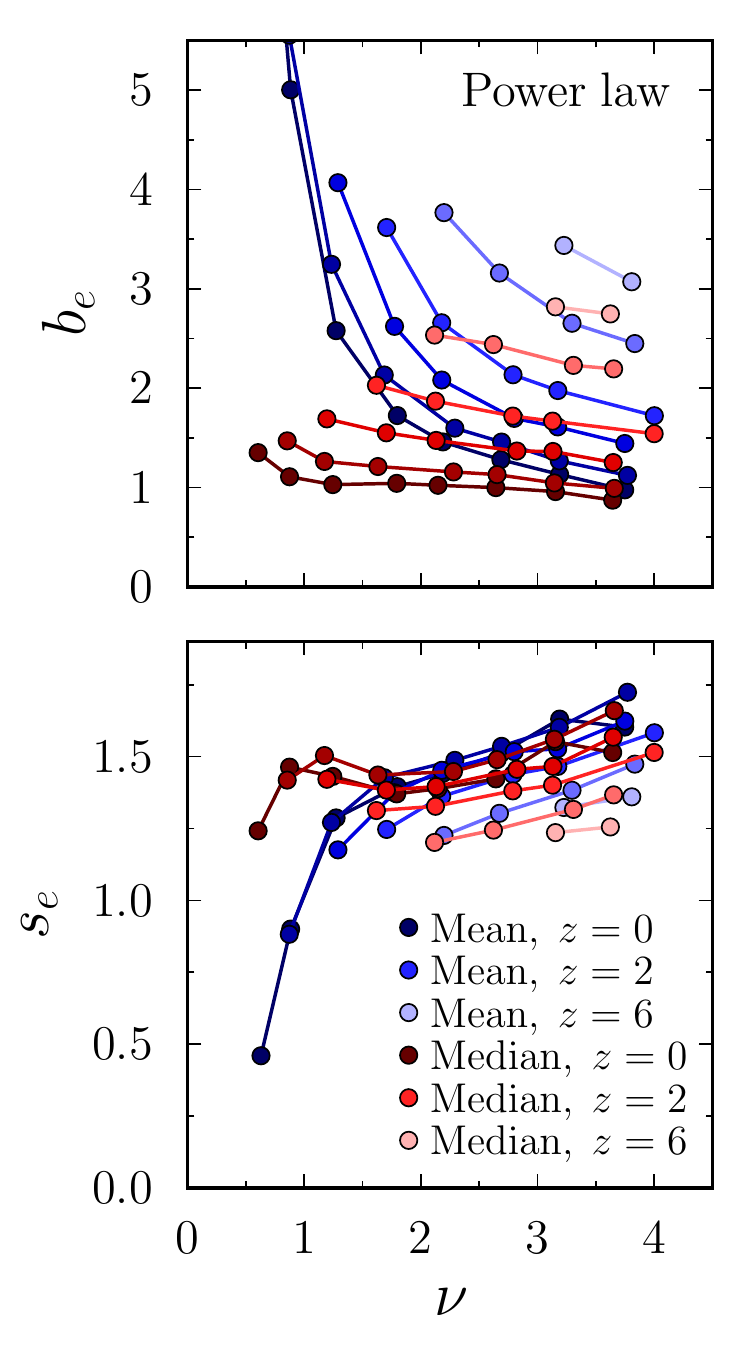}
\includegraphics[trim = 8mm 2mm 2mm 0mm, clip, scale=0.6]{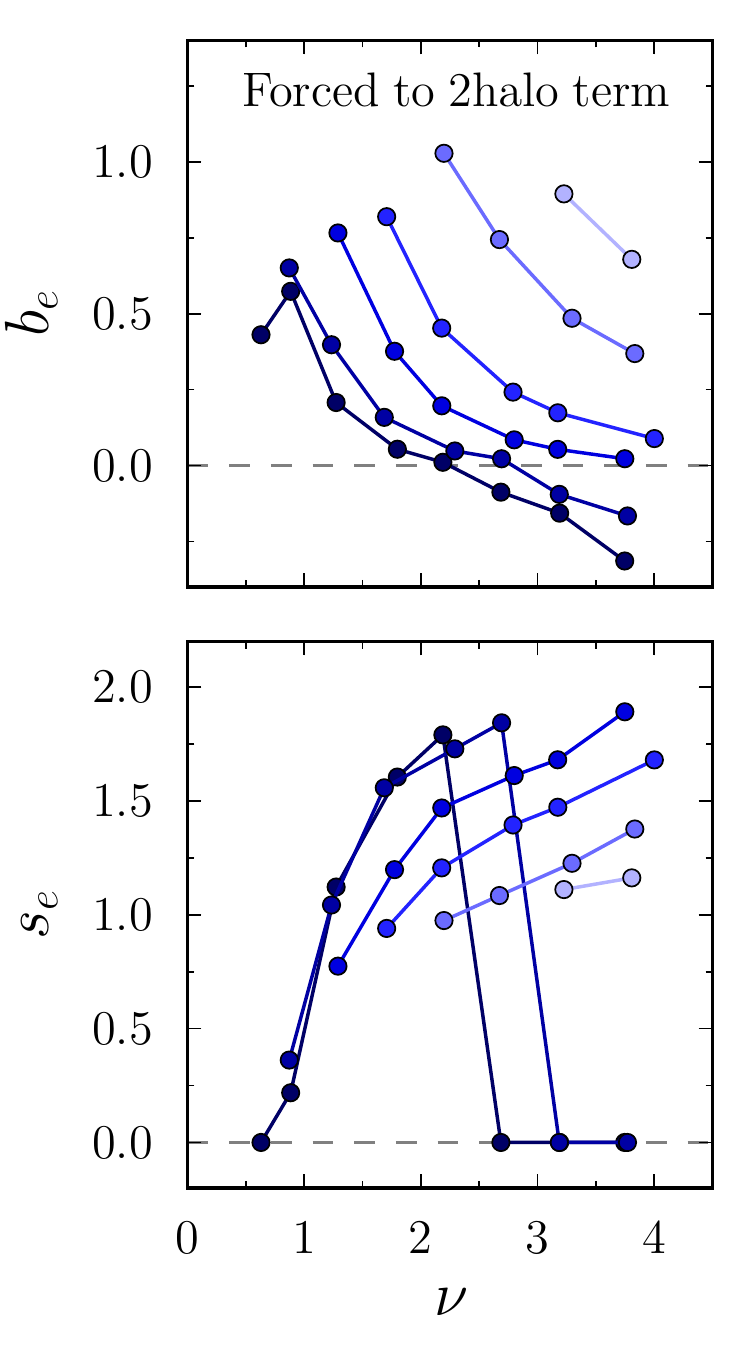}
\caption{Best-fit values for the excess bias and excess slope parameters, $b_{\rm e}$ and $s_{\rm e}$, for two different parameterizations of the outer profile. Red colors indicate results for the median profiles, blue for the mean, with darker colors indicating lower redshifts ($z = [0, 0.5, 1, 2, 4, 6]$). Left column: results for the power-law parameterization (Equation (\ref{eq:formula})), where $b_{\rm e}$ signifies the normalization of the profile in units of $\rhom$ at $5 \rtom$. Right column: results for a fit that is forced to converge to the 2-halo term at large radii (Equation (\ref{eq:2h2})). $b_{\rm e} < 0$ means that the profile lies below the 2-halo term, and $s_{\rm e} = 0$ indicates that the convergence occurs at an infinite radius. This situation occurs when the actual profile runs parallel to, or away from the 2-halo term.}
\label{fig:fitresouter}
\end{figure}

In the fitting formula presented in Equation (\ref{eq:formula}), we simply parameterized the outer profile with a power law instead of attempting to describe the shape of the profile with the 2-halo term. We have experimented with fits based on this term and found that its complexity is not warranted in general.

As demonstrated in Figure \ref{fig:fitfuncs}, the 2-halo term based on the matter correlation function provides an inferior fit to the density profiles at $r < 9 \rvir$ compared to our fiducial choice of the power law. To explore this issue more generally, we have used fits in which we parameterize the outer profile as the 2-halo term with a power-law correction,
\begin{equation}
\label{eq:2h1}
\rho_{\rm outer} = \rho_{\rm m} \left[b_{\rm e} \left( \frac{r}{5\, \rtom} \right)^{-s_{\rm e}} b_{\rm h}(\nu) \xi_{\rm lin}(r) + 1\right]  \,,
\end{equation}
where we use the bias model of \citet{tinker_10_bias} to calculate $b_{\rm h}(\nu)$. The results of such fits are shown in the left and center columns of Figure \ref{fig:fits2}. We note that the 2-halo term is itself close to a power law at $r < 9\rvir$, meaning that the fits do not differ greatly from the simple power law of Equation (\ref{eq:formula}). 

For some applications, it might be desirable for the fitting function to converge to the 2-halo term at large radii, $r >> 9\rvir$. This behavior can be achieved with the following parameterization of the outer profile: 
\begin{equation}
\label{eq:2h2}
\rho_{\rm outer} = \rho_{\rm m} \left[ b_{\rm h}(\nu) \xi_{\rm lin}(r) \left( 1 + b_{\rm e}  \left( \frac{r}{5\, \rtom} \right)^{-s_{\rm e}} \right)  + 1\right] \,.
\end{equation}
The resulting fits are somewhat worse and exhibit fractional deviations of up to $\approx 15\%$, as shown in the right column of Figure \ref{fig:fits2}. Moreover, the 2-halo term can only be expected to be an accurate description of the outer part of the mean profiles, but not the median profiles. Indeed, we find that the 2-halo term overestimates the median profiles significantly, and that the ratio of the mean to the median profiles varies with radius.

These considerations have motivated our use of a simple power-law form for the outermost density profile in our fiducial fits. Figure \ref{fig:fitresouter} shows the best-fit values for the two parameters for the outer profile, $b_{\rm e}$ and $s_{\rm e}$, from fits to halo samples of different $\nu$, at different redshifts, and for both the mean and median profiles. The best-fit values show a weak variation with $\nu$, but a stronger variation with  redshift. Overall, the variations are relatively mild though, which is particularly true for the slope $s_{\rm e}$, as can also be visually seen in Figures~\ref{fig:prof_by_nu} and \ref{fig:prof_accrate}. The best-fit values are also somewhat different for the mean and median profiles of the samples. The strongest deviations for low-$\nu$ samples are likely due to sample variance. Indeed, we find that in our smallest simulation, L0063, the outermost density profiles of halos of a given mass are systematically higher compared to their counterparts from the L0125 box.

Figure \ref{fig:fitresouter} also shows $b_{\rm e}$ and $s_{\rm e}$ derived from fits using Equation (\ref{eq:2h2}), i.e., forcing the outer profile to converge to the 2-halo term at large radii. In this case, $s_{\rm e}$ indicates how quickly the mean profiles approach the 2-halo term. $s_{\rm e}=0$ means that the profile runs either parallel to the 2-halo term, or even away from it.

Finally, we have verified that using the parameterizations of Equation (\ref{eq:2h1}) or Equation (\ref{eq:2h2}) does not change the best-fit parameters for $\beta$, $\gamma$, or $r_{\rm t}$, or the relation between $r_{\rm t}$ and $\Gamma$ when fitting $\Gamma$-selected samples.

%%%%%%%%%%%%%%%%%%%%%%%%%%%%%%%%%%%%%%%%%%%%%%%%%%%%%%%%%%%%%%%%%%%%%%%%%%
% BIBLIOGRAPHY
%%%%%%%%%%%%%%%%%%%%%%%%%%%%%%%%%%%%%%%%%%%%%%%%%%%%%%%%%%%%%%%%%%%%%%%%%%

\bibliographystyle{apj}
\bibliography{sf}

\end{document}